\documentclass[a4paper,11pt]{article}
\pdfoutput=1 

\usepackage{silence}
\usepackage{jheppub}

\usepackage{float}
\usepackage{relsize}
\usepackage{fontenc}
\usepackage{amsmath,amsfonts,amssymb,bm}
\usepackage{tabularx}
\usepackage{tabularray}
\usepackage{array}
\usepackage{makecell}
\usepackage{fontawesome}
\usepackage{cellspace} % add in preamble
\usepackage{nicematrix}
\usepackage{placeins}
\usepackage{adjustbox}
\usepackage{tikz}
\usepackage{slashed}
\usepackage{pgfplots}
\usepackage{mathtools}
\usepackage{siunitx}
\usepackage{physics}
\usepackage[compat=1.1.0]{tikz-feynman}
\usepackage[T1]{fontenc}
\usepackage{multirow}
\usepackage{listings}
\usepackage{tabularx}
\usepackage{enumitem} 
\usepackage[ruled,vlined]{algorithm2e}
\usepackage{booktabs}
\usepackage{cleveref}
  \tikzfeynmanset{
  double line/.style={
    draw,
    double distance=0.7pt, % separation between the two lines
    line width=0.4pt
  }
}

\def\({\left(} 
\def\){\right)}

\newcommand\rmd { {\rm d} }
\newcommand\imagewidth{0.99\linewidth}

\newcommand{\Amp}{\mathcal{M}}

\tikzset{graviton/.style={decorate, decoration={snake, amplitude=.6mm, segment length=1.5mm, pre length=.3mm, post length=.3mm}, double}}

\usepackage{xcolor}
\xdefinecolor{redviolet}{RGB}{228,38,207}
\xdefinecolor{brickred}{RGB}{220,60,66}
\xdefinecolor{bluscuro}{RGB}{51,51,179}

\usepackage{hyperref}

\title{\boldmath Radiative 
return at NLOPS accuracy}

\author[a,b]{Ettore Budassi,}
\author[b]{Carlo M. Carloni Calame,}
\author[a,b]{Marco Ghilardi,}
\author[c,d]{Andrea Gurgone,}
\author[a,b]{Guido Montagna,}
\author[e,f]{Mauro Moretti,}
\author[b]{Oreste Nicrosini,}
\author[b,g]{Fulvio Piccinini,}
\author[a,b]{and Francesco P. Ucci}

\affiliation[a]{Dipartimento di Fisica ``Alessandro Volta'', Universit\`a di Pavia, Via A. Bassi 6, 27100 Pavia, Italy}
\affiliation[b]{INFN, Sezione di Pavia, Via A. Bassi 6, 27100 Pavia, Italy}
\affiliation[c]{Dipartimento di Fisica, Universit\`a di Pisa, Largo Bruno Pontecorvo 3, 56127 Pisa, Italy}
\affiliation[d]{INFN, Sezione di Pisa, Largo Bruno Pontecorvo 3, 56127 Pisa, Italy}
\affiliation[e]{Dipartimento di Fisica e Scienze della Terra, Universit\`a di Ferrara, Via Saragat 1, 44122 Ferrara, Italy}
\affiliation[f]{INFN, Sezione di Ferrara, Via Saragat 1, 44122 Ferrara, Italy}
\affiliation[g]{INFN, Galileo Galilei Institute for Theoretical Physics, Largo E. Fermi 2, 50125 Firenze, Italy}

\emailAdd{ettore.budassi01@universitadipavia.it}
\emailAdd{carlo.carloni.calame@pv.infn.it}
\emailAdd{marco.ghilardi01@universitadipavia.it}
\emailAdd{andrea.gurgone@df.unipi.it}
\emailAdd{guido.montagna@unipv.it}
\emailAdd{mauro.moretti@unife.it}
\emailAdd{oreste.nicrosini@pv.infn.it}
\emailAdd{fulvio.piccinini@pv.infn.it}
\emailAdd{francesco.ucci@pv.infn.it}

\abstract{The radiative return, together with the energy scan, is the method used at 
flavour factories
to measure the pion form factor, which is a crucial input for the data-driven dispersive computation of the leading-order hadronic contribution to the muon anomalous magnetic moment. We consider the radiative hadronic
and leptonic channels of main experimental interest, namely the processes $e^+e^-\to X^+X^-\gamma$, with 
$X = \{\pi \, , \mu \}$. For such processes, we compute the exact next-to-leading order (NLO) corrections matched to a Parton Shower (PS) 
to describe exclusive multiple photon emission. All sources of radiative corrections from initial-state and final-state radiation, as well as their interference, are considered according to QED for $e^+e^-\to\mu^+\mu^-\gamma$ and QED$\oplus$F$\times$sQED 
(Factorised scalar QED) for $e^+e^-\to\pi^+\pi^-\gamma$. We describe in detail the novel features of our PS approach to compute the fixed-order corrections in association with higher-order contributions to $2\to3$ processes, with a 
hard photon in the final state. We present validation tests and
comparisons with 
NLO predictions available in the literature to 
cross-check various ingredients of our formulation. We also show numerical results at NLOPS accuracy according to realistic event selection criteria for precision 
measurements at flavour factories. Our calculation is implemented in an updated version of the
Monte Carlo event generator
\textsc{BabaYaga@NLO}, which can be used for fully exclusive simulations and data analysis in radiative return experiments.
}


\begin{document} 
\maketitle

\flushbottom

\section{Introduction}
\label{sec:intro}
The anomalous magnetic moment of the muon~\cite{Jegerlehner:2009ry,Jegerlehner:2017gek,Gabrielse:2025jep,Hertzog:2025ssc}, $a_\mu = (g-2)_\mu/2$, is currently 
measured to a precision of about one hundred parts per billion~\cite{Muong-2:2025xyk} and, as such, it represents a highly sensitive probe of the Standard Model (SM) of particle physics.

On the theoretical side, the main bottleneck to obtain comparatively accurate predictions for $a_\mu$ is the evaluation of the leading-order hadronic vacuum polarization (HVP) correction, which cannot be calculated using perturbation theory. To compute the latter, two methods are presently available~\cite{Aliberti:2025beg,Aoyama:2020ynm}: the traditional data-driven dispersive approach~\cite{Jegerlehner:2017gek,Keshavarzi:2018mgv,Davier:2019can,Benayoun:2019zwh,Keshavarzi:2024wow} and the more recent lattice QCD calculations~\cite{Borsanyi:2021,Boccaletti:2024guq,Djukanovic:2024cmq,RBC:2018dos,RBC:2023pvn,RBC:2024fic,Meyer:2018til}\footnote{In~\cite{Aliberti:2025beg}, also the dispersive 
approach based on $\tau$ decays for the computation of the HVP correction 
to $a_\mu$ has been revisited, according to 
the recent results obtained in 
\cite{Davier:2023fpl,Davier:2010fmf,Castro:2024prg,Colangelo:2022prz,Hoferichter:2023sli}.}. A more complete discussion on all the lattice QCD contributions to $a_\mu^{\rm HVP}$ is reported in \cite{Aliberti:2025beg}.

According to the latest theoretical update on the muon anomalous magnetic moment, as compiled by the Muon $g-2$ Theory Initiative in 2025~\cite{Aliberti:2025beg}, there is no tension between the SM prediction and the experimental world average, which is dominated by the measurements by the Muon $g-2$ Collaboration at Fermilab~\cite{Muong-2:2025xyk,Muong-2:2023cdq,Muong-2:2021ojo}. This data-theory agreement represents a significant change compared to the recent past and is the consequence of a major shift in the estimate of the HVP contribution to $a_\mu$. Actually, in~\cite{Aliberti:2025beg} the evaluation of the HVP contribution is based on an average of lattice QCD calculations rather than the data-driven approach.
 
Despite the significant progress in lattice simulations, the dispersive approach remains a key and sound tool to determine the HVP contribution to $a_\mu$. In particular, understanding the origin of the discrepancy between the lattice and dispersive calculations, which is dubbed as the ''new muon $g-2$ puzzle'', is essential for a completely satisfactory solution of the muon anomaly. In turn, this requires a better insight of the tensions between the measurements of the $e^+e^- \to hadrons$ cross sections by different experiments operating at low-energy colliders, known as flavour factories.

Actually, the dispersive theoretical predictions are based on the optical theorem and make use of the measurements of the hadron production cross sections in $e^+ e^-$ annihilation at flavour factories as input data. In this context, the $e^+ e^- \to \pi^+ \pi^-$ process is the most relevant channel, as it is responsible for more than $70\%$ of the HVP contribution and is the main source of uncertainty of the data-driven approach~\cite{Aliberti:2025beg,Aoyama:2020ynm}. The measurement of the $e^+ e^- \to \pi^+ \pi^-$ cross section is closely related to the extraction from the data of the time-like pion form factor $F_\pi (Q^2)$, which accounts for the non-perturbative internal structure of the pion.

At flavour factories, there are two methods to measure the two-pion production cross section and the corresponding form factor: energy scan and radiative return. In this paper, we focus on the latter method, in order to provide improved theoretical predictions compared to the existing literature, as detailed in the following. Similar results for energy scan experiments were recently presented by us in \cite{Budassi:2024whw}.

Radiative return is a technique used at GeV-scale $e^+ e^-$ colliders to measure the hadronic cross sections by analysing events where one of the colliding particles emits a hard photon before the collision. In this way, the nominal centre of mass (c.m.) energy of the collider is effectively reduced, allowing for cross section measurements in a range of continuous energies at a fixed beam energy. The high luminosity of flavour factories, such as $\phi$, $\tau$-charm and $B$ factories, compensates for the suppression factor of the electromagnetic coupling constant associated with photon emission. When applied to the $e^+ e^- \to \pi^+ \pi^- \gamma$ process, the radiative return enables to extract the energy dependence of the $e^+ e^- \to \pi^+ \pi^-$ cross section and to measure the pion form factor at $Q^2 = s^{\prime} = M^2_{\pi\pi}$, where $\sqrt{s^\prime}$ is the reduced c.m. energy and $M_{\pi\pi}$ is the two-pion invariant mass.

Calculations of hard photon emission processes in 
$e^+ e^-$ collisions date back to the sixties to eighties of the past century~\cite{Baier:1965skb,
Baier:1965jz,Baier:1966jf,Berends:1973tz,Berends:1973jb,PhysRevD.11.58,Kuraev:1977rp,Eidelman:1978rw,Baier:1980kx,Berends:1981rb}. The application of the radiative return method to the 
high-luminosity flavour factories materialised in the late 
nineties~\cite{Spagnolo:1998mt,Arbuzov:1998te,
Benayoun:1999hm,Binner:1999bt,Khoze:2000fs},
along with first 
computations of the radiative corrections to 
$e^+ e^- \to \pi^+ \pi^-\gamma$. From the early 2000s, the method was applied to the measurement of $F_\pi (Q^2)$ by BABAR~\cite{BaBar:2012bdw}, BESIII~\cite{BESIII:2015equ} and KLOE~\cite{KLOE:2004lnj,KLOE:2008fmq,KLOE:2010qei,KLOE:2012anl,KLOE-2:2017fda} collaborations, by taking data at colliders with different c.m. energies\footnote{The method was also adopted in \cite{Xiao:2017dqv} by using data collected with the CLEO-c detector at the CESR collider.}. The resulting pion form factors show some discrepancy~\cite{Davier:2023fpl} and all of them significantly disagree with the recent CMD-3 measurement\footnote{In \cite{SND_orsay2025} a recent analysis of the SND data has been presented and found to be in agreement with the CMD-3 measurement.} via energy scan~\cite{CMD-3:2023alj,CMD-3:2023rfe} . This experimental situation has increased the tensions among data-driven dispersive evaluations of the HVP contribution to $a_\mu$ to a level that makes it impossible to combine the results in a meaningful way~\cite{Muong-2:2025xyk}. Improved analyses of radiative return data are in progress\footnote{The preliminary result by the BABAR collaboration, using the full set of data, can be found in \cite{Zhang:2026jvb,Polat:2026ysh}. } and more accurate measurements of the pion form factor 
with sub-percent precision are expected in the upcoming years.

In radiative return experiments, different event selection criteria are applied, in particular by requiring one hard photon at large or small angles with respect to the beams. Moreover, the leptonic process $e^+ e^- \to \mu^+ \mu^- \gamma$ plays a crucial role in this kind of measurements for a twofold reason. First, it is the main background to radiative pion pair production. Secondly, in some measurements, the hadronic $N_{\pi\pi\gamma}$ data are normalized to the measured number of $\mu^+ \mu^- \gamma$ events, $N_{\mu\mu\gamma}$, in order to cancel in the $N_{\pi\pi\gamma} / N_{\mu\mu\gamma}$ ratio relevant experimental and theoretical systematics. For both reasons, a QED test using this leptonic process is performed in data analysis, by comparing the yield of events of the $\mu^+ \mu^- \gamma$ channel with a precise QED prediction as a function of the two-muon invariant mass.

Precision Monte Carlo (MC) event generators are essential for simulating and analysing radiative return events. By its nature, the method relies on a detailed knowledge of the radiative corrections to $e^+ e^- \to X^+ X^- \gamma$, $X = \left\{ \pi, \mu \right\}$, in the presence of realistic cuts and, in particular, on a high-precision modelling of the contribution due to initial-state radiation (ISR). The standard tool for the simulation of radiative processes is 
\textsc{Phokhara}, which is mainly based on the calculations described in~\cite{Rodrigo:2001jr,Rodrigo:2001kf,Kuhn:2002xg,Czyz:2002np,Czyz:2003ue}\footnote{A first MC generator developed for radiative return was 
\textsc{EVA}~\cite{Binner:1999bt}, which was based on a leading-order (LO) treatment of ISR and final-state radiation (FSR) for one photon emission, supplemented by the inclusion of additional ISR radiation via collinear structure functions~\cite{Kuraev:1985hb,AltarelliMartinelli,
Nicrosini:1986sm,Skrzypek:1990qs,
Skrzypek:1992vk,Cacciari:1992pz,Arbuzov:2010}.}. The most recent version of the \textsc{Phokhara} generator incorporates the complete set of NLO radiative corrections for both the  $\mu^+\mu^-\gamma$ and the $\pi^+\pi^-\gamma$ channel\footnote{Gauge-invariant subsets of radiative corrections to $e^+ e^- \to X^+ X^- \gamma$, $X =\left\{ \pi, \mu \right\}$, that had previously been omitted, were computed in~\cite{Campanario:2013uea,Campanario:2019mjh} and added in \textsc{Phokhara}.}. Because of the lack of higher-order contributions due to multiple photon emission, the theoretical accuracy of \textsc{Phokhara} is estimated to be 0.5\%~\cite{Rodrigo:2001kf,WorkingGrouponRadiativeCorrections:2010bjp}, which is the value quoted by the experiments as a source of systematic error. In \cite{BaBar:2012bdw}, also the MC generator \textsc{AfkQED} was used by the BABAR collaboration for the theoretical modelling of radiative processes. In this code, the signal photon can be generated within the detector acceptance at LO. Additional ISR contributions are simulated through the collinear structure function method, while FSR photons are generated with the \textsc{Photos} package~\cite{Barberio:1990ms,Barberio:1993qi}. Therefore, an event generator for radiative processes including both NLO corrections and exclusive exponentiation is still missing. This is at variance with the situation of $2 \to 2$ processes, where codes with both features, such as \textsc{BabaYaga@NLO}~\cite{Budassi:2024whw,CarloniCalame:2000pz,CarloniCalame:2001ny,Balossini:2006wc,Balossini:2008xr}, \textsc{Bhwide}~\cite{Jadach:1995nk}, \textsc{Kkmc}~\cite{Jadach:2000ir,Jadach:2022mbe} 
and \textsc{Sherpa}~\cite{Sherpa:2024mfk,Price:2025fiu}
are available.
 
Concerning recent progress in the field, it is worth mentioning that NLO predictions for $2 \to 3$ processes, with a hard photon in the final state, can be obtained by the MC integrator \textsc{McMule}~\cite{Banerjee:2020rww}. The latter includes the full set of NLO corrections for the $\mu^+\mu^-\gamma$ channel, while for the $\pi^+\pi^-\gamma$ final state only ISR corrections are taken into account. Recent calculations of the two-loop corrections to $e^+ e^- \to \gamma\gamma^*$ can be found in \cite{Fadin:2023phc,Badger:2023xtl}. Attempts to compute the full set of next-to-next-to-leading order (NNLO) corrections to radiative processes are underway~\cite{Aliberti:2024fpq,PetitRosas:2025xhm}. A dedicated measurement of additional radiation in radiative return measurements at BABAR was published in~\cite{BaBar:2023xiy}. 
A new method to deal with the inclusion of the pion form factor in the predictions for 
pion pair production processes has been recently proposed in~\cite{Fang:2025mhn}. 

In order to fill the gap in the literature and provide state-of-the-art predictions with an improved accuracy, we compute in this paper the exact NLO corrections matched to a Parton Shower (PS) to the processes $e^+ e^- \to X^+ X^- \gamma$, $X = \left\{ \pi, \mu \right\}$. All the sources of radiative corrections due to ISR, FSR and IFI are included. For pion pair production, FSR and IFI are computed according to the F$\times$sQED approach~\cite{Budassi:2024whw,Aliberti:2024fpq}, in which the scalar QED (sQED) 
point-like amplitudes are multiplied by the pion form factor at appropriate virtualities. A more refined treatment of this issue, that is possible in other theoretical frameworks~\cite{Budassi:2024whw,Ignatov:2022iou,Colangelo:2022lzg}, will be addressed elsewhere. The contribution of multiple photon emission is simulated by a PS algorithm in QED for $e^+ e^- \to \mu^+ \mu^- \gamma$ and in QED$\oplus$sQED for $e^+ e^- \to \pi^+ \pi^- \gamma$. Our calculation is implemented in a new release of the MC event generator 
\textsc{BabaYaga@NLO}, which is available for fully exclusive simulations and data analysis in radiative return experiments.

The rest of the paper is organised as follows.
In Sec.~\ref{sec:fix-ord} we describe the NLO calculation of the leptonic and hadronic processes considered in our work. 
In Sec.~\ref{sec:PS} we focus on the main and novel features of our PS approach to radiative processes, 
as well as on its matching with NLO corrections.
In Sec.~\ref{sec:numerical-results} we present a number of validation tests to cross-check various elements of our MC implementation and we perform an analysis of the impact of LO and NLO gauge-invariant subsets associated to ISR, FSR and IFI in different experimental setups. In the same Section, we also show phenomenological results at NLOPS according to various 
experimental scenarios. In Sec.~\ref{sec:conc} we summarise 
the main conclusions of our work. In Appendix~\ref{sec:app1} we provide further numerical results for experimentally relevant observables, while in Appendix~\ref{sec:app2} we show NLOPS predictions for $2\to2$ processes according to the new formulation here presented. In Appendix~\ref{sec:app_ps_and_is} and Appendix~\ref{sec:mapping} we describe the multi-photon phase-space parametrisation, the technique of multi-channel importance sampling and the implementation of the clustering and mapping algorithms in the updated version v2.0 of \textsc{BabaYaga@NLO}, which is publicly available on \href{https://github.com/cm-cc/BabaYagaNLO}{GitHub \faGithub}. 

\section{Fixed-order calculation}
\label{sec:fix-ord}
\begin{figure}
    \centering
\begin{equation*}
    \begin{tikzpicture}
  
  \begin{feynman}[small]
    \vertex (a) ;
    \vertex[right=1cm of a,dot] (b) {};
    \vertex[above left=0.75cm and 0.75cm of a] (c);
    \vertex[below left=0.75cm and 0.75cm of a] (d);
    \vertex[above right=0.75cm and 0.75cm of b] (e);
    \vertex[below right=0.75cm and 0.75cm of b] (f);
    \vertex[above left =0.5 and 0.5cm of a] (g);
    \vertex[below left =0.5 and 0.5cm of a] (h);
    \vertex[above left=0.35cm and 0.35cm of a](i);
    \vertex[above right=0.4cm and 0.6cm of i](j);
    \diagram* {
      (a) -- [photon] (b),
      (d) -- [fermion] (a),
      (a) -- [fermion] (c),
      (e) -- [double line] (b) --[double line] (f),
      (i) -- [photon] (j),
    };
  \end{feynman}
\end{tikzpicture}
\quad
    \begin{tikzpicture}
  \begin{feynman}[small]
    \vertex (a) ;
    \vertex[right=1cm of a,dot] (b) {};
    \vertex[above left=0.75cm and 0.75cm of a] (c);
    \vertex[below left=0.75cm and 0.75cm of a] (d);
    \vertex[above right=0.75cm and 0.75cm of b] (e);
    \vertex[below right=0.75cm and 0.75cm of b] (f);
    \vertex[above left =0.5 and 0.5cm of a] (g);
    \vertex[below left =0.5 and 0.5cm of a] (h);
    \vertex[below left=0.35cm and 0.35cm of a](i);
    \vertex[below right=0.4cm and 0.6cm of i](j);
    \diagram* {
      (a) -- [photon] (b),
      (d) -- [fermion] (a),
      (a) -- [fermion] (c),
      (e) -- [double line] (b) --[double line] (f),
      (i) -- [photon] (j),
    };
  \end{feynman}
\end{tikzpicture}
 \quad
 \begin{tikzpicture}
  \begin{feynman}[small]
    \vertex (a) ;
    \vertex[right=1cm of a,dot] (b) {};
    \vertex[above left=0.75cm and 0.75cm of a] (c);
    \vertex[below left=0.75cm and 0.75cm of a] (d);
    \vertex[above right=0.75cm and 0.75cm of b] (e);
    \vertex[below right=0.75cm and 0.75cm of b] (f);
    \vertex[above left =0.5 and 0.5cm of a] (g);
    \vertex[below left =0.5 and 0.5cm of a] (h);
    \vertex[above right=0.35cm and 0.35cm of b](i);
    \vertex[below right=0.4cm and 0.6cm of i](j);
    \diagram* {
      (a) -- [photon] (b),
      (d) -- [fermion] (a),
      (a) -- [fermion] (c),
      (e) -- [double line] (b) --[double line] (f),
      (i) -- [photon] (j),
    };
  \end{feynman}
\end{tikzpicture}\quad
    \begin{tikzpicture}
  \begin{feynman}[small]
    \vertex (a) ;
    \vertex[right=1cm of a,dot] (b) {};
    \vertex[above left=0.75cm and 0.75cm of a] (c);
    \vertex[below left=0.75cm and 0.75cm of a] (d);
    \vertex[above right=0.75cm and 0.75cm of b] (e);
    \vertex[below right=0.75cm and 0.75cm of b] (f);
    \vertex[above left =0.5 and 0.5cm of a] (g);
    \vertex[below left =0.5 and 0.5cm of a] (h);
    \vertex[below right=0.35cm and 0.35cm of b](i);
    \vertex[above right=0.4cm and 0.6cm of i](j);
    \diagram* {
      (a) -- [photon] (b),
      (d) -- [fermion] (a),
      (a) -- [fermion] (c),
      (e) -- [double line] (b) --[double line] (f),
      (i) -- [photon] (j),
    };
  \end{feynman}
\end{tikzpicture}
\qquad
    \begin{tikzpicture}
  \begin{feynman}[small]
    \vertex (a) ;
    \vertex[right=1cm of a,style=blob] (b) {};
    \vertex[above left=0.75cm and 0.75cm of a] (c);
    \vertex[below left=0.75cm and 0.75cm of a] (d);
    \vertex[above right=0.75cm and 0.75cm of b] (e);
    \vertex[below right=0.75cm and 0.75cm of b] (f);
    \vertex[above left =0.5 and 0.5cm of a] (g);
    \vertex[below left =0.5 and 0.5cm of a] (h);
    \vertex[above left=0.35cm and 0.35cm of a](i);
    \vertex[above right=0.4cm and 0.6cm of i](j);
    \vertex[right=0.8cm of b](k);
    \diagram* {
      (a) -- [photon] (b),
      (d) -- [fermion] (a),
      (a) -- [fermion] (c),
      (e) -- [scalar] (b) --[scalar] (f),
      % (i) -- [photon] (j),
      (b) --[photon] (k)
    };
  \end{feynman}
\end{tikzpicture}
\end{equation*}
\begin{equation*}
\begin{gathered}
     \begin{tikzpicture}
  \begin{feynman}[small]
    \vertex (a) ;
    \vertex[right=1cm of a, dot] (b) {};
    \vertex[above right=1cm and 1cm of b] (e) {$X^-$};
    \vertex[below right=1cm and 1cm of b] (f) {$X^+$};
    \vertex[above left =0.5 and 0.5cm of a] (g);
    \vertex[below left =0.5 and 0.5cm of a] (h);
    \vertex[above left=0.35cm and 0.35cm of a](i);
    \vertex[above right=0.4cm and 0.6cm of i](j);
    \vertex[right=0.8cm of b](k);
    \diagram* {
      (e) -- [double line] (b) --[double line] (f),
    };
  \end{feynman}
\end{tikzpicture}
\end{gathered}=
\begin{gathered}
     \begin{tikzpicture}
  \begin{feynman}[small]
    \vertex (a) ;
    \vertex[right=1cm of a] (b);
    \vertex[above right=0.7cm and 0.7cm of b] (e) {$\mu^-$};
    \vertex[below right=0.7cm and 0.7cm of b] (f) {$\mu^+$};
    \diagram* {
      (f) -- [fermion] (b) --[fermion] (e),
    };
  \end{feynman}
\end{tikzpicture}
\end{gathered}     ,\quad
\begin{gathered}
     \begin{tikzpicture}
  \begin{feynman}[small]
    \vertex (a) ;
    \vertex[right=1cm of a,style=blob] (b) {};
    \vertex[above right=1cm and 1cm of b] (e) {$\pi^-$};
    \vertex[below right=1cm and 1cm of b] (f) {$\pi^+$};
    \vertex[above left =0.5 and 0.5cm of a] (g);
    \vertex[below left =0.5 and 0.5cm of a] (h);
    \vertex[above left=0.35cm and 0.35cm of a](i);
    \vertex[above right=0.4cm and 0.6cm of i](j);
    \vertex[right=0.8cm of b](k);
    \diagram* {
      (f) -- [scalar] (b) --[scalar] (e),
    };
  \end{feynman}
\end{tikzpicture}
\end{gathered}     
\end{equation*}
    \caption{Tree-level diagrams contributing to the $e^+e^-\to X^+X^-\gamma$ process: the first two are ISR diagrams, while the latter three represent FSR. The first four contribute to the signatures with $X=\left\{\mu,\pi\right\}$, while the last one contributes  to the pion channel only. The shaded blob 
    represents the pion form factor.}
    \label{fig:LOdiags}
\end{figure}
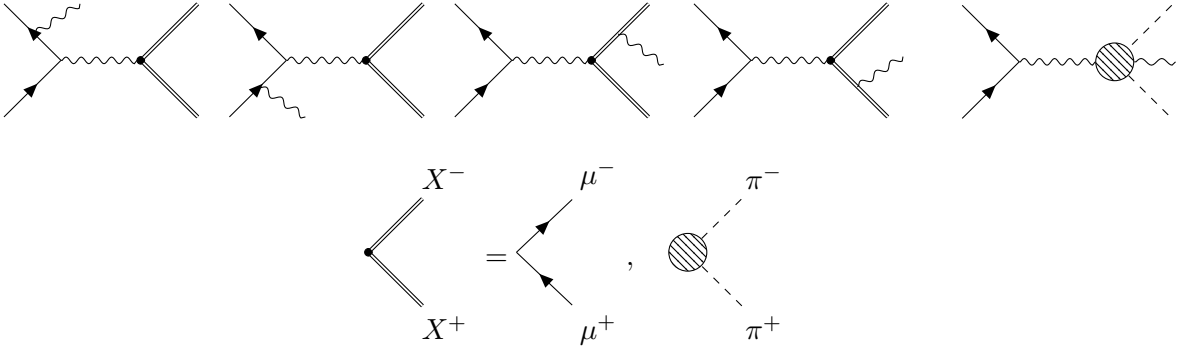
 We define the radiative process as follows 
\begin{equation}
\label{eq:momenta_born}
    e^- (p_1)  + e^+(p_2) \to X^-(p_3) + X^+(p_4) + \gamma (k), \qquad X=\{\mu,\pi\}
\end{equation}
and the generalised Mandelstam variables as 
\begin{equation}
    \begin{aligned}
        s&= (p_1+p_2)^2 = 4E^2\\
        s'&=(p_3+p_4)^2=M_{XX}^2\\
            p_{ij}&=p_i\cdot p_j\\
        s_{ij}&=(p_i-p_j)^2 = m_i^2+m_j^2 - 2 p_{ij}\qquad i=1,2;\, j=3,4
    \end{aligned}
\end{equation} 
where $E$ is the beam energy and $M_{XX}$ is the invariant mass of the final-state charged particles.
 At the tree level, the diagrams shown in Fig.~\ref{fig:LOdiags} contribute to the matrix element, as the photon can be either emitted by the initial or final charged legs. Therefore, the LO matrix-element can be decomposed according to ISR, FSR and IFI contributions as
\begin{equation}
    |\Amp_{\rm LO}|^2=\left|\Amp^{\rm ISR}_1\right|^2+2\text{Re}\left\{\Amp^{{\rm ISR},\dagger}_1\Amp_1^{\rm FSR}\right\}+\left|\Amp^{\rm FSR}_1\right|^2\,,
\end{equation}
where the interference is odd 
under the exchange of $\pi^+$ and $\pi^-$ momenta (or electron and positron momenta), hence it does not contribute to the cross section integrated over symmetric angular ranges.

For the $X=\pi$ final state, in the simplest case of photons emitted only by the initial state and in the limit of $m_e\to0$, the LO cross section reads~\cite{Rodrigo:2001jr,WorkingGrouponRadiativeCorrections:2010bjp}
\begin{equation}
    s'\dv{\sigma}{s'}=\frac{\alpha^3\beta_\pi^3}{3s}|F_\pi(s')|^2\left\{\frac{s^2+s'^2}{s(s-s')}\left(\log\frac{s}{m_e^2}-1\right)\right\}\,,
\end{equation}
after integrating over the photon phase space, where $\beta_\pi=\sqrt{1-4m_\pi^2/s'}$ is the pion velocity. We stress that, 
analogously to the fully inclusive LO cross section of 
$e^+ e^- \to \gamma\gamma$~\cite{Berends:1973tm,Berends:1980px}, a collinear logarithm $L\equiv \log (s/m_e^2)$ is present already in the tree-level cross section.

At NLO, the fixed order calculation receives $\order{\alpha^4}$ virtual and real contributions from the insertion of an additional photon. Therefore, the NLO cross section can be written with the phase-space slicing as
\begin{equation}
\sigma_{\text{NLO}}=\int_{\Omega_\text{sig}}  (1+\mathcal{C}_\alpha)|\mathcal{M}_1|^2\dd\Phi_1+ \int_{\Omega_\text{sig}\times\Omega_\text{hard}}   |\mathcal{M}_2|^2\dd \Phi_2 \,,
\label{eq:sigmaNLO}
\end{equation}
where the pedex $n$ counts the number of photons, so that $\mathcal{M}_n$ is the exact matrix element for the process $e^+e^-\to X^+X^-n\gamma$ and $\dd \Phi_n$ is the exact $(2+n)$-body phase space. The integration domain for the charged particles is the same in both terms, therefore we focus on the photonic phase space. The first term of Eq.~\eqref{eq:sigmaNLO} represents the Born cross section plus the soft+virtual (SV) photon contribution and it is integrated over the region $\Omega_\text{sig}$, which encodes the kinematical cuts.
The second term of Eq.~\eqref{eq:sigmaNLO} accounts for the emission of an additional hard photon, with energy integrated in the range $\Omega_\text{hard}=[\varepsilon\sqrt{s}/2,(s-M_{XX\gamma_{\rm sig}}^2)/(2\sqrt{s})]$, given $M_{XX\gamma_{\rm \rm sig}}^2=\left(p_3+p_4+k_{\rm sig}\right)^2$, while the tagged photon is integrated over $\Omega_\text{sig}$. We have introduced the  phase-space slicing parameter $\varepsilon\ll 1$, which is an unphysical separator whose dependence cancels in $\sigma_\text{NLO}$, as shown in our previous work~\cite{Budassi:2024whw}. The soft+virtual and real corrections appearing in Eq.~\eqref{eq:sigmaNLO} are described in the following, while the non-perturbative effects are treated in the last part of this Section.

The exact $n$-photon phase space, divided by the incoming flux and including a factor $\left(2\pi\right)^4$, is given by
\begin{equation}
    \dd \Phi_n (\{p\},\{k\})= \frac{\left(2   \pi\right)^4}{\mathcal{F}\,[2(2\pi)^3]^{n+2}}\frac{\dd^3\vec p_3}{E_3}\, \frac{\dd^3\vec p_4}{E_4}
    \left (
    \prod_{i=1}^n \frac{\dd^3\vec k_i} {\omega_i} 
    \right )
    \delta^{\left(4\right)}\left(p_1+p_2-p_3-p_4-\sum_{i=1}^n k_i\right) \,,
    \label{eq:phsp}
\end{equation}
with $\mathcal{F}=2s\sqrt{1-4m_e^2/s}$ being the flux factor. In order to sample the soft-collinear enhancements present in the matrix elements, as well as to generate phase-space points returning to a reduced c.m. energy $\sqrt{s'}<\sqrt{s}$ via the radiation of a hard photon, a multichannel sampling procedure is applied to the phase-space integral, which is detailed in 
Appendix~\ref{sec:app_ps_and_is}.

The coefficient $\mathcal{C}_\alpha$ appearing in Eq.~\eqref{eq:sigmaNLO} is the exact SV correction at NLO, given by 
\begin{equation}
    \mathcal{C}_\alpha=\frac{2\text{Re}\{\mathcal{M}_1^\dagger \mathcal{M}_V\}}{|\mathcal{M}_1|^2} +\frac{\alpha}{4\pi^2}\int_{\Omega_\text{soft}}\frac{\dd k_0}{k_0}\int_{4\pi} \dd \Omega_k\,\mathcal{I}\left(k,\left\{p\right\}\right) \,,
    \label{eq:calpha}
\end{equation}
where the first term contains virtual photonic correction, while the second term accounts for the emission of soft photons in the region $\Omega_\text{soft}=\left\{\lambda\le k_0\le\varepsilon\sqrt{s}/2\right\}$, the photon energy ranging from the fictitious photon mass $\lambda$ -- used to regularise infrared divergences -- up to $\varepsilon \sqrt{s}/2$.  The eikonal function entering Eq.~(\ref{eq:calpha}) reads
\begin{equation}
  \mathcal{I}\left(k,\left\{p\right\}\right)=  -\sum_{i,j=1}^4 \mathcal{I}_{ij}\left(k,\left\{p\right\}\right)=- \sum_{i,j=1}^4 \eta_i\eta_j \frac{\left(p_i\cdot p_j\right)}{\left(p_i\cdot k\right)\left(p_j\cdot k\right)}k_0^{2}\,,
  \label{eq:eik}
\end{equation}
 where $k$ is the momentum of the real photon with energy $k_0$ and $\eta_i$ represents the physical charge in positron units flowing into the process from the $i$-th particle.
 
\begin{figure}
    \centering
\begin{equation*}
\begin{gathered}
        \begin{tikzpicture}
  \begin{feynman}[small]
    \vertex (a) ;
    \vertex[right=1cm of a,dot] (b) {};
    \vertex[above left=0.75cm and 0.75cm of a] (c);
    \vertex[below left=0.75cm and 0.75cm of a] (d);
     \vertex[above left=0.6cm and 0.6cm of a] (v);
    \vertex[below left=0.6cm and 0.6cm of a] (z);
    \vertex[above right=0.75cm and 0.75cm of b] (e);
    \vertex[below right=0.75cm and 0.75cm of b] (f);
    \vertex[above left =0.5 and 0.5cm of a] (g);
    \vertex[below left =0.5 and 0.5cm of a] (h);
    \vertex[above left=0.65cm and 0.65cm of a](i);
    \vertex[above right=0.4cm and 0.6cm of i](j);
      \vertex[above left=0.1cm and 0.1cm of a](k);
        \vertex[above left=0.55cm and 0.55cm of a](l);
    \diagram* {
      (a) -- [photon] (b),
      (d) -- [fermion] (a),
      (a) -- [fermion] (c),
      (e) -- [double line] (b) --[double line] (f),
      (i) -- [photon] (j),
      (k)--[photon, half left, looseness=1.3] (l),
    };
  \end{feynman}
\end{tikzpicture}
\end{gathered}
\qquad
\begin{gathered}
     \begin{tikzpicture}
  \begin{feynman}[small]
    \vertex (a) ;
    \vertex[right=1cm of a,dot] (b) {};
    \vertex[above left=0.75cm and 0.75cm of a] (c);
    \vertex[below left=0.75cm and 0.75cm of a] (d);
     \vertex[above left=0.5cm and 0.5cm of a] (v);
    \vertex[below left=0.5cm and 0.5cm of a] (z);
    \vertex[above right=0.75cm and 0.75cm of b] (e);
    \vertex[below right=0.75cm and 0.75cm of b] (f);
    \vertex[above left =0.5 and 0.5cm of a] (g);
    \vertex[below left =0.5 and 0.5cm of a] (h);
    \vertex[above left=0.65cm and 0.65cm of a](i);
    \vertex[above right=0.4cm and 0.6cm of i](j);
      \vertex[below left=0.35cm and 0.35cm of a](k);
    \vertex[below right=0.4cm and 0.6cm of k](l);
    \diagram* {
      (a) -- [photon] (b),
      (d) -- [fermion] (a),
      (a) -- [fermion] (c),
      (e) -- [double line] (b) --[double line] (f),
      (i) -- [photon] (j),
      (v) -- [photon,half right, looseness=1,] (z),
      % (k) -- [photon] (l),
    };
  \end{feynman}
\end{tikzpicture}
\end{gathered}
\qquad 
  \begin{gathered}
           \begin{tikzpicture}
  \begin{feynman}[small]
    \vertex (a);
    \vertex[right=1cm of a,dot] (b){};
    \vertex[below=1cm of a] (i);
    \vertex[below=1cm of b,dot] (j) {};
    \vertex[above left=0.75cm and 0.75cm of a] (c);
    \vertex[below left=0.75cm and 0.75cm of i] (d);
    \vertex[above right=0.75cm and 0.75cm of b] (e);
    \vertex[below right=0.75cm and 0.75cm of j] (f);
      \vertex[above left=0.35cm and 0.35cm of a](k);
    \vertex[above right=0.4cm and 0.6cm of k](l);
    \diagram* {
      (a) -- [photon] (b),
      (d) -- [fermion] (i) -- [fermion] (a) -- [fermion] (c),
      (e) -- [double line] (b) --[double line] (j) -- [double line] (f),
      (i) -- [photon] (j),
      (k) -- [photon] (l),
    };
  \end{feynman}
\end{tikzpicture}      
  \end{gathered}
 \qquad
\begin{gathered}
          \begin{tikzpicture}
  \begin{feynman}[small]
    \vertex (a);
    \vertex[below right=0.3cm and 1cm of a,dot] (b){};
    \vertex[below=1.6cm of a] (i);
    \vertex[below left=0.8cm and 0.6cm of a] (v);
    \vertex[left=1cm of v](z);
    \vertex[below=1cm of b,dot] (j) {};
    \vertex[above left=0.75cm and 0.75cm of a] (c);
    \vertex[below left=0.75cm and 0.75cm of i] (d);
    \vertex[above right=0.75cm and 0.75cm of b] (e);
    \vertex[below right=0.75cm and 0.75cm of j] (f);
      \vertex[above left=0.35cm and 0.35cm of a](k);
    \vertex[above right=0.4cm and 0.6cm of k](l);
    \diagram* {
      (a) -- [photon, momentum=$q$] (b),
      (d) -- [fermion] (i) -- [fermion](v)--[fermion] (a) -- [fermion] (c),
      (e) -- [double line] (b) --[double line] (j) -- [double line] (f),
      (i) -- [photon] (j),
      (v) --[photon](z),
      % (k) -- [photon] (l),
    };
  \end{feynman}
\end{tikzpicture}  
\end{gathered}
\end{equation*}
    \caption{Examples of virtual diagrams where the photon is emitted by the initial state, divided by classes.}
    \label{fig:NLOvirtDiags}
\end{figure}
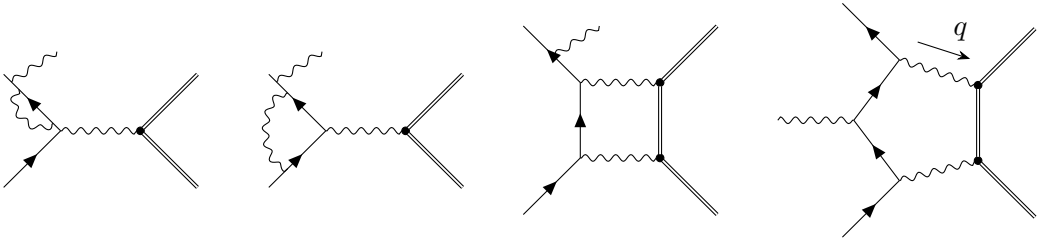

 The virtual amplitude $\Amp_V$ in
  Eq.~(\ref{eq:calpha}) is given by the sum of many contributions, that are schematised in Fig.~\ref{fig:NLOvirtDiags}. The counterterm diagrams are not shown, while the photon vacuum polarisation is discussed later on. For the treatment of ultraviolet divergences,
  we employ the on-shell renormalisation scheme with dimensional regularisation. In order to define an interference probability at NLO, we write 
\begin{equation}
     \mathcal{P}_V^\text{IFI} = 2\,\text{Re}\left\{\Amp_1^\dagger\Amp_{V}\right\}-2\,\text{Re}\left\{\Amp_1^{{\rm ISR},\dagger}\Amp_{V}^{\rm ISR}\right\}-2\,\text{Re}\left\{\Amp_1^{{\rm FSR},\dagger}\Amp_{V}^{\rm FSR}\right\}\,.
\end{equation}
Here, ISR and FSR refer to virtual corrections exclusively on the initial and final state, respectively, in which also the LO photon has been emitted by the corresponding current. 
This definition is naturally gauge invariant, since the ISR and FSR contributions are separately alike. Therefore, the interference contains not only virtual diagrams connecting the initial- and final-state currents, but also the interference of ISR (FSR) virtual corrections on initial legs with FSR (ISR) LO diagrams. For this reason, unlike in the $e^+e^-\to\pi^+\pi^-$ process, the IFI does contain also a charge-even contribution that does not vanish under angular-symmetric integration. The amplitudes have been generated using the \textsc{FeynArts}~\cite{Hahn:1998yk,Hahn:2000kx,Hahn:2010zi}$\to$\textsc{FeynCalc}~\cite{Shtabovenko:2016sxi,Shtabovenko:2020gxv,Shtabovenko:2023idz} chain and the loop functions are evaluated using the $\textsc{Collier}$~\cite{Denner:2014gla} library. The calculation of the virtual contributions to $\pi^+\pi^-\gamma$ has been cross-checked with an independent computation implemented in \textsc{Form}~\cite{Vermaseren:2000nd,Kuipers:2013pba,Ruijl:2017dtg} with \textsc{LoopTools}~\cite{Hahn:1998yk} to evaluate loop functions, finding perfect agreement. We also verified that the virtual corrections for the $\mu^+\mu^-\gamma$ channel are under control, by comparing our prediction with the \textsc{Recola}~\cite{Actis:2016mpe} automatic package. Particular care is needed for the correct numerical evaluation of the 5-point loop functions needed for the pentagon diagrams shown in the last panel of Fig.~\ref{fig:NLOvirtDiags}. At first, the routing of the momenta, $i.e.$ all the permutations of the loop momentum $q$ that can be placed in any of the internal lines, has been chosen in order to reproduce the best agreement between \textsc{LoopTools} and \textsc{Collier}. Moreover, due to the presence of electron propagators $[(p_i-k)^2-m_e^2]^{-1}$ that can become nearly on shell, the pentagons are evaluated in quadruple precision when $p_i\cdot k<5 \cdot10^{-8}~\rm{GeV}^{2}$. With this caution, we do not observe significant numerical instabilities.

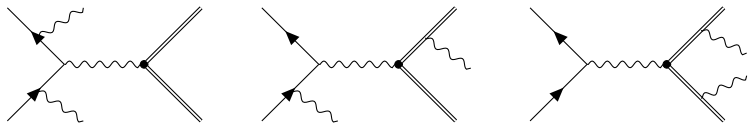
\begin{figure}
    \centering
\begin{equation*}
    \begin{tikzpicture}
  \begin{feynman}[small]
    \vertex (a) ;
    \vertex[right=1cm of a,dot] (b) {};
    \vertex[above left=0.75cm and 0.75cm of a] (c);
    \vertex[below left=0.75cm and 0.75cm of a] (d);
    \vertex[above right=0.75cm and 0.75cm of b] (e);
    \vertex[below right=0.75cm and 0.75cm of b] (f);
    \vertex[above left =0.5 and 0.5cm of a] (g);
    \vertex[below left =0.5 and 0.5cm of a] (h);
    \vertex[above left=0.35cm and 0.35cm of a](i);
    \vertex[above right=0.4cm and 0.6cm of i](j);
      \vertex[below left=0.35cm and 0.35cm of a](k);
    \vertex[below right=0.4cm and 0.6cm of k](l);
    \diagram* {
      (a) -- [photon] (b),
      (d) -- [fermion] (a),
      (a) -- [fermion] (c),
      (e) -- [double line] (b) --[double line] (f),
      (i) -- [photon] (j),
      (k) -- [photon] (l),
    };
  \end{feynman}
\end{tikzpicture}
\qquad
    \begin{tikzpicture}
  \begin{feynman}[small]
    \vertex (a) ;
    \vertex[right=1cm of a,dot] (b) {};
    \vertex[above left=0.75cm and 0.75cm of a] (c);
    \vertex[below left=0.75cm and 0.75cm of a] (d);
    \vertex[above right=0.75cm and 0.75cm of b] (e);
    \vertex[below right=0.75cm and 0.75cm of b] (f);
    \vertex[above left =0.5 and 0.5cm of a] (g);
    \vertex[below left =0.5 and 0.5cm of a] (h);
    \vertex[below left=0.35cm and 0.35cm of a](i);
    \vertex[below right=0.4cm and 0.6cm of i](j);
        \vertex[above right=0.35cm and 0.35cm of b](k);
    \vertex[below right=0.4cm and 0.6cm of k](l);
    \diagram* {
      (a) -- [photon] (b),
      (d) -- [fermion] (a),
      (a) -- [fermion] (c),
      (e) -- [double line] (b) --[double line] (f),
      (i) -- [photon] (j),
          (k) -- [photon] (l),
    };
  \end{feynman}
\end{tikzpicture}
 \qquad
 \begin{tikzpicture}
  \begin{feynman}[small]
    \vertex (a) ;
    \vertex[right=1cm of a,dot] (b) {};
    \vertex[above left=0.75cm and 0.75cm of a] (c);
    \vertex[below left=0.75cm and 0.75cm of a] (d);
    \vertex[above right=0.75cm and 0.75cm of b] (e);
    \vertex[below right=0.75cm and 0.75cm of b] (f);
    \vertex[above left =0.5 and 0.5cm of a] (g);
    \vertex[below left =0.5 and 0.5cm of a] (h);
    \vertex[above right=0.45cm and 0.45cm of b](i);
    \vertex[below right=0.3cm and 0.6cm of i](j);
     \vertex[below right=0.45cm and 0.45cm of b](k);
    \vertex[above right=0.3cm and 0.6cm of k](l);
    \diagram* {
      (a) -- [photon] (b),
      (d) -- [fermion] (a),
      (a) -- [fermion] (c),
      (e) -- [double line] (b) --[double line] (f),
      (i) -- [photon] (j),
      (k) -- [photon] (l),
    };
  \end{feynman}
\end{tikzpicture}
\end{equation*}
    \caption{Examples of classes of diagrams for the $2\to2+2\gamma$ process. The first diagram accounts for ISR photon emission, the central diagram exemplifies the set of diagrams in which one photon is emitted from the initial current and one from the final one. The rightmost diagram represents the double FSR emission.}
    \label{fig:2gammaDiags}
\end{figure}

At $\order{\alpha^4}$, one has to consider also the process $e^+e^-\to X^+X^-\gamma\gamma$, entering the second term of the NLO cross section of Eq.~\eqref{eq:sigmaNLO}. The matrix element is given by three separately gauge-invariant contributions, associated with two ISR/FSR photon emissions and the IFI subset, as schematised in Fig.~\ref{fig:2gammaDiags}.
The IFI is given by the square of all diagrams in which an emission occurs both from the initial and final state particles, as well as the interference of the different classes of diagrams. Again, the matrix elements have been obtained via the usual tools, but in order to simplify the algebraic structure of the amplitudes an intermediate step using the \textsc{Form} package was necessary, as to speed up the numerical evaluation. Several tests have been performed, finding perfect agreement with the automatic calculation provided by \textsc{Recola} for the $\mu^+\mu^-\gamma$ final state and with a modified version of \textsc{Alpha}~\cite{Caravaglios:1995cd} for the $\pi^+\pi^-\gamma$ channel.

\subsection{Non-perturbative contributions}
We complete the description of the radiative return processes by addressing the treatment of the non-perturbative ingredients of our calculation. The following discussion applies both for the virtual and real corrections, with the necessary adaptation that will be clarified when not obvious. Let us define a generic function $\Lambda(Q^2)=\left\{\alpha(Q^2),F_\pi(Q^2)\right\}$, where 
\begin{equation}
    \alpha(Q^2)=\frac{1}{1-\Delta \alpha(Q^2)}\,,
\end{equation}
is the Dyson-resummed electromagnetic coupling, which parametrises leptonic and hadronic vacuum polarisation effects in the $X=\mu$ final state; $F_\pi(Q^2)$ is the electromagnetic form factor encoding the non-perturbative internal structure of the pion. The latter is introduced for the $X=\pi$ final state on top of the point-like calculation, in the so-called  F$\times$sQED approach~\cite{Aliberti:2024fpq,Budassi:2024whw}, and it is intended to be the \textit{dressed} form factor, \textit{i.e.} including vacuum polarisation effects.

In general, $\Lambda(Q^2)$ is a complex valued function inserted as a multiplicative modification of the photon propagator acting as
\begin{equation}
D_{\mu\nu}(Q^2)=\frac{-ig_{\mu\nu}}{Q^2} \quad \longrightarrow \quad D_{\mu\nu}(Q^2)\Lambda(Q^2)\,,
\end{equation}
and then the amplitude is squared accordingly. For pure ISR and FSR virtual and real contributions, this amounts to the combinations
\begin{equation}
  \left| \Amp^\text{ISR}\right|^2\left|\Lambda(M_{XX}^2)\right|^2, \qquad 
  \left| \Amp^\text{FSR}\right|^2\left|\Lambda(s)\right|^2\,.
  \end{equation}
For what concerns the IFI contributions, the treatment is slightly different. For virtual IFI corrections, the function $\Lambda(Q^2)$ is evaluated at $M_{XX}^2 $(or  $s$) if the emission occurs from the initial (or final) state. On the other hand, the momentum transfer for real IFI matrix elements with two photons depend on how many photons $i\in \text{FS}$ have been emitted from the final state, therefore giving
\begin{equation}
  \sum_{I,J} 2\text{Re}\left\{\Amp_I^\dagger  \Amp_J \,\,\Lambda_I^\dagger\left[\left(p_3+p_4+{\textstyle\sum_{i\in \text{FS}} k_i}\right)^2\right]\Lambda_J\left[\left(p_3+p_4+{\textstyle\sum_{j\in \text{FS}} k_j}\right)^2\right]\right\}\,,
\end{equation}
where the indices $I,J$ are looped over all the amplitudes contributing to the IFI subset. We stress that in this contribution the scales entering the non-perturbative functions $\Lambda_I(Q^2_i)$ are in general different. Therefore, a proper extraction of $F_\pi(Q^2)$ from radiative events should take into account such differences.

\section{Beyond the fixed order: Parton Shower}
\label{sec:PS}
In this Section, we discuss how the exponentiation of the leading logarithmic  
corrections due to multi-photon emission is performed by means of a PS method. 

\subsection{The pure  Parton Shower approach}
\label{subsec:PurePS}

We focus on the key and novel features of the PS algorithm used in our simulation for radiative processes. The PS approach, as already outlined in~\cite{Budassi:2024whw,CarloniCalame:2000pz,CarloniCalame:2001ny,Balossini:2006wc,Balossini:2008xr}, accounts for the emission of an arbitrary number of additional photons on top of the underlying process in the leading logarithmic (LL) approximation, going beyond the strictly collinear limit and including the transverse momenta of both the emitting particle and photon at each branching.

The master formula for the merging of the PS on top of an exact LO matrix element reads as follows
\begin{equation}
    \rmd\sigma^{\rm LL} \, = \,  \sum_{n=1}^\infty \, \Pi(\varepsilon, \{p\}) \,
  \frac{1}{n!}\;\left|{\cal M}_{n}^{\rm J}\right|^2\;\rmd\Phi_{n}(\{p\},\{k\}) \,.
  \label{eq:PS}
\end{equation}
In Eq.~(\ref{eq:PS}), $\{p\}$ and $\{k\}$ represent the sets of final-state momenta and $\rmd\Phi_{n}(\{p\},\{k\})$ is the exact phase space of the 
$e^+ e^- \to X^+ X^- n \gamma$ process, as given 
by Eq.~(\ref{eq:phsp}) in 
Sec.~\ref{sec:fix-ord}. The term ${\cal M}_{n}^{\rm J}$ stands for 
the exact matrix element ${\cal M}^{\rm ex}_{n}$ for $n = 1,2$ and for the LL approximation ${\cal M}^{\rm LL}_{n}$ for $n \geq 3$, as motivated in the following.

The Sudakov form factor $\Pi(\varepsilon, \{p\})$
in Eq.~(\ref{eq:PS}) is defined as follows
    {\small \begin{equation}
    \Pi (\varepsilon,\{p\} ) \, = \, {\rm exp} \left\{ - \frac{\alpha}{ \pi} \,\int 
    \rmd \Omega_{k} \, {\cal I}\left(k,\left\{p\right\}\right)\left[ \int_0^{1-\varepsilon} \rmd z \, P_r(z)+ \int_0^{1-\varepsilon} \rmd z \, P_f(z)\right]\right\} \quad
\begin{array}{ll}
&\text{$r=f$ for muons}\\
&\text{$r=s$ for pions}
\end{array}\,,
    \label{eq:SFF}
\end{equation}}
and it represents the probability of \textit{non-resolved} emission, $i.e.$ a real (soft) photon with energy less than $\varepsilon\sqrt{s}/2$, with $\varepsilon\ll 1$, or a virtual photon.

The non-infrared eikonal function ${\cal I}\left(k,\left\{p\right\}\right)$, defined in Eq.~\eqref{eq:eik}, derives from the Yennie--Frautschi--Suura (YFS) exponentiation of soft photon contributions \cite{Yennie:1961ad} and keeps the information about the angular distribution of soft and collinear photons. The integration over the photon angular variables $\rmd \Omega_k$ of the diagonal terms in Eq.~(\ref{eq:eik}) yields constant contributions, while the off-diagonal terms lead to collinear logarithms.

In Eq.~(\ref{eq:SFF}), $P(z)$ is the Altarelli-Parisi unregularised splitting function for the $p\to p\,\gamma$ branching in massless approximation, with $p=f,s$ for fermions in QED and scalars in sQED, respectively. The variable $z$ represents the energy fraction of the daughter charged particle involved in the splitting with respect to the parent one. The explicit expression of $P(z)$ describes the way the energy is shared after the splitting and depends on the spin of the emitter. The formulae of interest in our computation are given by~\cite{CZ:1975,Altarelli:1977zs,Leenaarts:2019,Chen:2016wkt}
\begin{equation}
    P_f(z)=\frac{1+z^2}{1-z}\, ,\qquad\qquad 
    P_s(z)=\frac{2z}{1-z}\, ,
\end{equation}
that refer to photon emission from a fermion and a scalar leg, respectively. Note that, according to Eq.~(\ref{eq:SFF}) and the above ingredients, 
in our approach the resummation of soft and collinear logarithms applies, by construction, to the contributions due to ISR, FSR and IFI, both for the hadronic and leptonic final states.

 Among the $n$ photons, one of them is the tagged photon of $e^+e^- \to X^+X^-\gamma$, while the other ones are generated with energy greater than $\varepsilon\sqrt{s}/2$. Thanks to factorization theorems of soft and collinear 
singularities, the $\rm LL$ approximation of the squared matrix element, that holds for $n\ge3$, is written as
\begin{equation}
    \left|{\cal M}_{n}^{\rm LL}\right|^2=
    \prod_{i=1}^{n-2} \left[\frac{\alpha}{2\pi} P(z_{i}) \, {\cal I}_{\rm SF}\left(k_{i},z_{i}\right) \, \frac{4\pi^{2}\left(1-z_i\right)}{z_i\omega_i^2}\right] \left|{\cal M}_{2}^{\rm ex}(\{\tilde{p}\},\{\tilde{k}\})\right|^2\,.
    \label{eq:llapprox}
\end{equation}
The function $ \mathcal{I}_{\rm SF}$ appearing in Eq.~(\ref{eq:llapprox}) is given by
\begin{equation}
    {\cal I}_{\rm SF}(k,z)=-\sum_{ij}{\cal I}_{ij}\left(k,\left\{p\right\}\right)\, F_{ij}(z)\,.
    \label{eq:eikmod}
\end{equation}
The eikonal function $\mathcal{I}_{ij}$ is given in Eq.~\eqref{eq:eik}, where $i,j$ label the radiating legs and the correction factor $F_{ij}$ is
\begin{align}
\begin{cases}
    F_{ij}(z)=1-\delta_{ij}\displaystyle\frac{\left(1-z\right)^2}{1+z^{2}} \, &\text{for fermions}\,\,\\
    F_{ij}(z)=1 \, &\text{for scalars}\,
\end{cases}
\label{eq:fijeik}
\end{align}
which accounts for the \textit{spin-flip} of massive spin-1/2 particles. As a matter of fact, the modification of the YFS-like eikonal structure as in Eq.~(\ref{eq:eik}) through the factors $F_{ij}$ allows to reproduce in the collinear limit, \textit{i.e.} for $\left(k\cdot p\right)\to 0$ while keeping the photon energy $k^0$
 fixed, the unpolarised massive splitting function as in Eq.~(2.5) of \cite{Dittmaier:1999mb}. It is important to emphasise that this modification does not affect the soft limit: indeed, in the limit $z\to 1$, the correction factors $F_{ij}$ go to 1, recovering the YFS eikonal function as in Eq.~(\ref{eq:eik}).

In Eq.~(\ref{eq:llapprox}), the 
contribution $ \left|{\cal M}_{2}^{\rm ex}(\{\tilde{p}\},\{\tilde{k}\})\right|^2$ is the exact LO squared matrix element 
of the two photon emission process $e^+ e^- \to X^+ X^- \gamma\gamma$,
on top of which the iteration of the LL-approximated matrix elements is built. 
We choose to construct the LL approximation on top of the exact matrix element associated to double hard bremsstrahlung for the following reasons. First, 
we want to maintain in our approach as much information as possible about the dynamics of exclusive events, that is given in our computation by the matrix element for two-photon emission. Secondly, the samples with two photons are an important component of the inclusive radiative return signature $e^+ e^- \to X^+ X^- \gamma \, (\gamma)$ and these samples are also used by the experiments for validation of MC codes~\cite{BaBar:2023xiy}. Furthermore, this choice enables to simulate with sufficiently good accuracy the exclusive signature with three hard photons in the final state, even in the presence of a resonant behaviour\footnote{Note that this is not the case if the mapping is 
applied on the zero or one-photon kinematics, as we 
explicitly checked, hence our choice. This aspect is further discussed in the following.}, as shown 
in Sec.~\ref{sec:numerical-results}.
However, the radiative nature of the underlying process introduces an ambiguity, namely the choice of the photons to be used in the evaluation of $\left|{\cal M}_{2}^{\rm ex}(\{\tilde{p}\}, \{\tilde{k} \})\right|^2$. In our approach, the selection of the two photons follows a CKKW-like criterion~\cite{Catani:2001cc}, 
adapted from QCD to QED, as it is infrared and collinear safe. 
For each photon $i$, we define the CKKW-like distance as
\begin{equation}
    y_{ij} = \min\left\{k_{0,i}^2, p_{0,j}^2\right\}\,\left(1 - \cos \theta_{ij}\right), 
    \label{eq:CKKW}
\end{equation}
looping over the charged legs $j=1\dots 4$, where $\theta_{ij}$ is the relative angle between $\vec{k}_i$ and $\vec{p}_j$. We define $d(k_i) = \min_j\{y_{ij}\}$ and select the two photons with indices $l$ and $m$ such that $d(k_l) \ge d(k_m)\ge\dots\ge d(k_n)$. Furthermore, to evaluate the exact matrix element $ \left|{\cal M}_{2}^{\rm ex}(\{\tilde{p}\},\{\tilde{k} \})\right|^2$ in Eq.~(\ref{eq:llapprox}), we introduce a procedure mapping the momenta $\{p\},\{k\}$ of the 
$2 \to (4 + n)$-kinematics into reduced momenta
$\{\tilde{p}\}, \{\tilde{k} \}$ of the 
$2 \to 4$ kinematics. More details about the clustering and mapping algorithms can be found in Appendix~\ref{sec:mapping}.

The above theoretical description allows to simulate radiative events in the LL approximation by exponentiating the universal soft and collinear contributions to all orders. The dependence of observables from the cut-off parameter $\varepsilon$, introduced as a soft-hard photon separator, cancels between the Sudakov factor and the integration of the hard-photon matrix element in the phase space region where the photon energy is larger than $\varepsilon \sqrt{s}/2$. 

\subsection{Matching NLO corrections to Parton Shower}\label{subsec:PS-match}

As in the case of non-radiative processes \cite{Budassi:2024whw,Balossini:2006wc,Balossini:2008xr}, the expansion at ${\cal O}\left(\alpha\right)$ of the LL cross section in Eq.~(\ref{eq:PS}) does not coincide with the exact NLO cross section as in Eq.~(\ref{eq:sigmaNLO}) of Sec.~\ref{sec:fix-ord}.
The explicit $\mathcal{O}\left(\alpha\right)$ expansion of Eq.~(\ref{eq:PS}) is given by
\begin{equation}
    \rmd\sigma^{\rm LL}_{\alpha} \, = \, \left[1+C_{\alpha}^{\rm LL}\right]\left|{\cal M}_{1}^{\rm ex}\right|^2\rmd\Phi_{1}+\left|{\cal M}_{2}^{\rm ex}\right|^2\rmd\Phi_{2}\,,
    \label{eq:ll-ordalphaexp}
\end{equation}
where $C_{\alpha}^{\rm LL}$ is the argument of the exponential in Eq.~(\ref{eq:PS}). The terms $|{\cal M}_{1}^{\rm ex} |^2$ and $|{\cal M}_{2}^{\rm ex} |^2$ are the exact squared
matrix elements of the processes $e^+ e^- \to X^+ X^- \gamma$ and $e^+ e^- \to X^+ X^- \gamma\gamma$, respectively. On the other hand, the exact NLO calculation can be written as
\begin{equation}
    \rmd\sigma_{\alpha} \, = \, \left[1+C_{\alpha}\right]\left|{\cal M}_{1}^{\rm ex}\right|^2\rmd\Phi_{1}+\left|{\cal M}_{2}^{\rm ex}\right|^2\rmd\Phi_{2}\,\, ,
    \label{eq:ex-ordalphaexp}
\end{equation}
where the term $C_{\alpha}$ is the exact SV correction given in Eq.~(\ref{eq:calpha}).

In order to consistently match the LL exponentiation on top of the NLO calculation, we introduce appropriate correction factors which carry the information about the exact $\mathcal{O}\left(\alpha\right)$ calculation. 
 To this end, the master formula for the cross section calculated at NLO and matched to the PS (NLOPS) reads as follows
\begin{equation}
    \rmd\sigma_{\rm NLOPS} \, = \,  \sum_{n=1}^\infty \, \Pi(\varepsilon, \{p\}) \,F_{\rm SV} \,
  \frac{1}{n!}\;\left|{\cal M}_{n}^{\rm J}\right|^2\;\rmd\Phi_{n}(\{p\},\{k\}) \,.
  \label{eq:PSMATCH}
\end{equation}
where the factor $F_{\rm SV}$ reproduces the exact 
SV correction and ${\cal M}_{n}^{\rm J}$ stands for 
${\cal M}^{\rm ex}_{n}$ for $n = 1,2$ and ${\cal M}^{\rm LL}_{n}$ for $n \geq 3$.
The above matrix elements contain  the running of $\alpha$ for the di-muon final state and the pion form factor for the hadronic channel, with $\Lambda(Q^2)=\left\{\alpha (Q^2),F_{\pi} (Q^2)\right\}$ computed at the appropriate 
virtualities, as discussed in Sec.~\ref{sec:fix-ord}.  We stress that the predictions according to Eq.~(\ref{eq:PSMATCH}) are independent of the cut-off parameter $\varepsilon$, in close analogy with the pure PS formulation, as will be shown in Sec.~\ref{sec:numerical-results}.

A comment is in order here. The matching procedure here described 
for radiative processes is somewhat different from the formulation used 
in~\cite{Budassi:2024whw,Balossini:2006wc,Balossini:2008xr} for the modelling and simulation of two pion production and QED processes. In those papers, the LL approximation for real photon emission is constructed on top of the LO matrix elements of the $2 \to 2$ processes under study and the mapping is performed on the 
zero-photon kinematics. Consequently, correction factors are also introduced in the 
NLOPS cross section to exactly reproduce the dynamics of hard photon bremsstrahlung up to NLO. As shown in \cite{Budassi:2024whw,
Balossini:2006wc,Balossini:2008xr,WorkingGrouponRadiativeCorrections:2010bjp}, this approach provides an accurate description of non-radiative processes when considering observables that are sufficiently inclusive over the emitted radiation and tend to isolate the contribution of soft photon emission. However, for radiative processes with hard photons in the final state, we checked that hard 
 correction factors may introduce spurious enhancements in the vicinity of resonances, because of their inability of keeping under control the correct virtualities that are needed for the evaluation of $\Lambda(Q^2)$.
 In the present formulation, 
 this feature is absent as the treatment of resonances is exactly accounted for up to two real photon emissions, as previously noticed, and enables a sufficiently accurate simulation of the exclusive three photon contribution, as shown in the next Section. To improve the modelling of hard bremsstrahlung dynamics in the presence of resonant enhancements, the formulation described in this work has been also implemented in 
 \textsc{BabaYaga@NLO} for the simulation at NLOPS accuracy of the $2 \to 2$ processes 
 $e^+ e^- \to \mu^+ \mu^-$ and 
 $e^+ e^- \to \pi^+ \pi^-$. We discuss the implementation in Appendix~\ref{sec:app2}, where we show that for an energy scan setup the new formulation agrees with the previous version of \textsc{BabaYaga@NLO} at the per mille level, while for radiative return scenarios the results are not adequate for sub-percent accuracy.

In Eq.~(\ref{eq:PSMATCH}), the definition of the $F_{\rm SV}$ factor presents the same ambiguity encountered in Eq.~(\ref{eq:llapprox}) when $n\ge 2$, namely the choice of the photon on which the correction should be evaluated. According to the 
soft-photon approximation, namely when the emitted photon with momentum $k$ becomes soft, the QED virtual correction to a generic radiative process with $n+1$ photons can be cast as the product of a universal eikonal factor and the virtual correction to the corresponding process with $n$ photons. The explicit expression for this correction can be written as follows \cite{Yennie:1961ad, Weinberg:1995mt,Engel:2021ccn}
\begin{equation}
\label{eq:LBK}
    \Re{\mathcal{\mathcal{M}}_{n+1}^{V} \mathcal{\mathcal{M}}_{n+1}^{\dagger}}=4\pi\alpha\left(\frac{\mathcal{I}\left(k,\left\{p\right\}\right)}{k_0^2}\right)\Re{\mathcal{\mathcal{M}}_{n}^{V} \mathcal{\mathcal{M}}_{n}^{\dagger}} \,.
\end{equation}
Therefore, in our approach, the $F_{\rm SV}$ factor is evaluated by using the kinematics obtained by mapping the $(n+1)$-photon configuration onto a single-photon one. Moreover, in order to better reproduce the desired behaviour of Eq.~(\ref{eq:LBK}), the definition of $F_{\rm SV}$ is given by
\begin{equation}
    F_{\rm SV}=1+\left(\mathcal{C}_{\alpha}-\mathcal{C}_{\alpha}^{\rm LL}\right)_h\,.
\end{equation}
where $C_\alpha^{\rm LL}$ and $C_\alpha$ are given in Eqs.~(\ref{eq:ll-ordalphaexp}, \ref{eq:ex-ordalphaexp}), respectively, in which $\mathcal{M}_1^{\rm ex}$ is calculated on the kinematics preserving the momentum of the hardest photon, while mapping the other ones.
This factor allows for the exact NLO SV correction to be reproduced in the one-photon sample. At the same time, when considering the sample with $n+1$ photons, with $n\ge1$, this prescription enables a correct reconstruction of the SV correction in the limit where the $n$ additional photons become soft, and ensures the cancellation of any residual dependence on the soft-hard separator $\varepsilon$.

Two important aspects need to be highlighted. First, the term $F_{\rm SV}$ is, by construction, free of infrared logarithms but it may still include collinear logarithms. This can be seen when comparing the SV contribution of Eq.~(\ref{eq:ll-ordalphaexp}) solely with ISR, with the exact SV contribution of Eq.~(23) of \cite{Czyz:2002np}. This occurs when no 
acceptance angular cuts are applied to the tagged photon and is a consequence of the presence of a collinear logarithm in the fully inclusive LO cross section, as discussed in Sec.~\ref{sec:fix-ord}.
Nevertheless, the matching procedure fully recovers the NLO result  without any double counting. Secondly, Eq.~(\ref{eq:PSMATCH}) preserves the exponentiation of LL contributions, as in a pure PS approach, being therefore well suited for its implementation in a MC code, by enabling exclusive event generation at NLOPS accuracy.

It should also be noted that the factorized matching of NLO corrections with the universal LL contributions from the PS algorithm effectively captures the dominant part of the NNLO corrections. This can be explicitly verified by means of a direct comparison with Eq.~(5.19) of \cite{Fadin:2023phc}, which provides an expression for the leptonic tensor up to two loops for the process $e^+e^- \to \gamma \gamma^*$. This result includes the soft–soft (SS), soft-virtual (SV) and virtual–virtual (VV) contributions on top of the radiative process, accounting for ISR corrections only, which are the dominant contributions due to the presence of large collinear logarithms.
Considering the expansion at $\mathcal{O}\left(\alpha^2\right)$ of Eq.~(\ref{eq:PSMATCH}) accounting for the SS+SV+VV corrections, this leads to
\begin{equation}
    \rmd \sigma_{\rm VV}+\rmd \sigma_{\rm SV}+\rmd \sigma_{\rm SS}=\left\{1+\mathcal{C}_{\alpha}-\frac{\left(\mathcal{C}^{\rm LL}_{\alpha}\right)^2}{2!}+\mathcal{C}_{\alpha}\,\mathcal{C}^{\rm LL}_{\alpha}\right\} \, \rmd \sigma_{1}^{\rm ex} \,,
    \label{eq:twoloopvirt}
\end{equation}
where $\rmd \sigma_{1}^{\rm ex}$ is the exact LO cross section of $e^+ e^- \to X^+ X^- \gamma$.
By directly comparing the gauge-invariant subset of ISR corrections in Eq.~(\ref{eq:twoloopvirt}) with Eq.~(5.19) of \cite{Fadin:2023phc} -- without the fermion loop contributions --  we carefully checked that all the infrared-sensitive structures, namely terms proportional to $\log{\varepsilon}$, as well as the terms of $\mathcal{O}\left(\alpha^2L^2\right)$ with respect to the LO differential cross section, are correctly reproduced. Therefore, the first terms not included exactly are of $\mathcal{O}\left(\alpha^2L\right)$, whose typical size at flavour factories is about $0.1\%$.

\section{Numerical results}
\label{sec:numerical-results}
In this Section, the numerical results for the processes $e^+e^-\to X^+X^-\gamma$ with $X =\{\pi,\mu\}$ obtained with the updated version of \textsc{BabaYaga@NLO} event generator are shown and discussed. We start by describing the input parameters and the event selection criteria used for the simulations in Sec.~\ref{sec:inp-dist-sel}. In Sec.~\ref{sec:validation}, we show some internal consistency checks that serve for the validation of the MC implementation. In Sec.~\ref{sec:subsec-giss} we discuss the size of gauge-invariant subsets contributing at LO and NLO. Finally, in Sec.~\ref{sec:subsec-numerical} a sample of differential cross sections is shown for different event selection criteria of interest for radiative return experiments.

It is worth noticing that our 
calculation includes NLO corrections to the processes with one photon in the final state and, therefore, NLO accuracy holds for any observables related to one photon emission. Thus, the 
inclusive process of main experimental interest in radiative return measurements, $i.e.$ $e^+ e^- \to X^+ X^+ \gamma \, (\gamma)$, is modelled at NLOPS accuracy. For observables corresponding to the exclusive signatures with two or more detected photons, the accuracy of our approach is simply LO with LL resummation.

As mentioned in Sec.~\ref{sec:fix-ord}, in the new release of \textsc{BabaYaga@NLO} for radiative processes, we improved the phase-space treatment and implemented suitable multi-channel importance sampling to provide efficient MC integration and event generation even
when including non-perturbative effects, such as narrow resonances and $F_{\pi}(Q^2)$. More details can be found in Appendix~\ref{sec:app_ps_and_is}.

\subsection{Input values, distributions and event selections}\label{sec:inp-dist-sel}

The input parameters for the simulations are set to
\begin{align*}
    \alpha & = 1/137.03599908\,,\\
    m_e    & = 0.51099895\,\text{MeV}, \\
    m_\mu  & = 105.658375\,\text{MeV},  \\
    m_\pi  & = 139.57039\,\text{MeV}.
\end{align*}
We express the pion form factor $F_\pi(Q^2)$ as a combination of Gounaris-Sakurai (GS) and Breit-Wigner (BW) functions, which model the relevant resonances decaying to $\pi^+\pi^-$. As done in~\cite{Aliberti:2024fpq}, we define
\begin{align}
\begin{aligned}
F_\pi(Q^2) &= \frac{\textrm{BW}_\rho^\textrm{GS}(Q^2) \left[1+ (Q^2/m_\omega^2) \, c_\omega\, \textrm{BW}_\omega(Q^2) + (Q^2/m_\phi^2) \, c_\phi\, \textrm{BW}_\phi(Q^2) \right]}{1+c_{\
\rho'}+c_{\rho''}+c_{\rho'''}} \\[2pt]
&+  \frac{c_{\rho'}\, \textrm{BW}_{\rho'}^\textrm{GS}(Q^2)
+ c_{\rho''}\, \textrm{BW}_{\rho''}^\textrm{GS}(Q^2)
+ c_{\rho'''}\, \textrm{BW}_{\rho'''}^\textrm{GS}(Q^2)}{1+c_{\rho'}+c_{\rho''}+c_{\rho'''}}\,,
\end{aligned}
\label{eq:vff}
\end{align}
where, on top of the $\omega,\phi,\rho,\rho',\rho''$ and $\rho'''$ resonances, we also introduce the $\rho-\omega$ and $\rho-\phi$ interferences. The narrow resonances $\omega$ and $\phi$ are described through $\textrm{BW}_v\left(Q^2\right)$ functions with mass $M_v$ and width $\Gamma_v$, defined as
\begin{eqnarray}
    \textrm{BW}_{v}\left(Q^2\right)= \frac{M_v^2}{M_v^2 - Q^2 - i M_v \Gamma_v} \,.
\end{eqnarray}
\begin{table}
\centering
\begin{tabular}{rcccccc}
\toprule
 & $\rho$ & $\rho'$ & $\rho''$ & $\rho'''$ & $\omega$ & $\phi$ \\[1pt]
\midrule
$M_v$ (MeV) & 774.56 & 1485.9 & 1866.8 & 2264.5 & 782.48 & 1019.47 \\
$\Gamma_v$ (MeV) & 148.32 & 373.60 & 303.34 & 113.27 & 8.55 & 4.25 \\
$|c_v|$ & - & 0.14104 & 0.0614 & 0.0047 & 0.00158 & 0.00045 \\
$\varphi_v$ (rad) & - & 3.7797 & 1.429 & 0.921 & 0.075 & 2.888 \\
\bottomrule
\end{tabular}
\caption{Input parameters for $F_\pi(Q^2)$.}
\label{tab:num_vff}
\end{table}
The broad $\rho$ resonances are described by GS functions as
\begin{eqnarray}
    \textrm{BW}_v^\textrm{GS}(Q^2) = \frac{M_v^2 + d(M_v)\,M_v\,\Gamma_v}{M_v^2 - Q^2 + f(Q^2,M_v,\Gamma_v) - i\, M_v\, \Gamma(Q^2,M_v,\Gamma_v)}\,,
\end{eqnarray}
where the definitions of $d(M_v)$, $f(Q^2,M_v,\Gamma_v)$ and $\Gamma(Q^2,M_v,\Gamma_v)$ can  be found in \cite{Gounaris:1968mw}. Finally, the amplitude resonances $c_v$ are defined as complex numbers $c_v=\left|c_v\right|e^{i\varphi_v}$. The numerical values used in the 
parametrisation of Eq.(\ref{eq:vff}) are reported in Tab.~\ref{tab:num_vff}.
 Our choice aims to
enable us to show reasonably realistic prediction for the observable of interest and it is valid up to $M_{\pi\pi}\lesssim 3~\rm{GeV}$, and should be intended as the \textit{dressed} form factor, $i.e.$ including vacuum polarisation effects. 

Moreover, we emphasize that both the pion form factor and the running of the electromagnetic coupling $\Delta \alpha(Q^2)$ are input for the code and should be provided by the users according to their needs. Nevertheless, we provide pre-implemented parametrisations that can be readily used.

The event selection criteria that are used for the simulations are taken as  representative of four experimental scenarios. Their definition, taken from~\cite{Aliberti:2024fpq} and summarised in Tab.~\ref{tab:scenarios}, aims to resemble the realistic cuts applied by experiments at the generator level. Further steps into the detector simulation, which can affect the shape and size of radiative corrections, are left to the user. The variables $\theta^{\pm}$ and $\phi^{\pm}$ denote the polar and azimuthal scattering angles of the $X^{\pm}$ particles, respectively, $E_\gamma$ and $\theta_\gamma$ are the photon energy and polar angle, respectively. The invariant mass of the charged final-state system is defined as $M_{XX}$ and $M_{XX\gamma}$ is the invariant mass of the $XX\gamma$ system, where $\gamma$ refers to the hardest photon passing the photon selection cuts. Finally, the quantities $p^{\pm}$, $p^{\pm}_z $, $p^{\pm}_\perp$ represent the modulus of the three-momentum $\vec{p}^{\,\pm}$, the modulus of the longitudinal and transverse component of $\vec{p}^{\,\pm}$ with respect to the beam direction in the laboratory frame, respectively.

\begin{table}[t]
\centering
\renewcommand{\arraystretch}{1.5}
\small
\begin{tabular}{p{7cm} p{7cm}}
\toprule
\textbf{(a) KLOE\,I Large-Angle (LA)} &
\textbf{(b) KLOE\,II Small-Angle (SA)} \\
\midrule
% \vspace{1mm}
\makecell[l]{
$\sqrt{s}=1.02~\mathrm{GeV}$\\[4pt]
$50^\circ \le \theta^{\pm} \le 130^\circ$\\[4pt]
$p_z^{\pm} \ge 90~\mathrm{MeV}\ \vee\ p_\perp^{\pm} \ge 160~\mathrm{MeV}$\\[4pt]
$50^\circ \le \theta_\gamma \le 130^\circ\,\wedge\, E_\gamma \ge 20~\mathrm{MeV}$\\[4pt]
$0.1~\mathrm{GeV}^2 \le M^2_{XX} \le 0.85~\mathrm{GeV}^2$
}
% \vspace{3mm}
&
% \vspace{1mm}
\makecell[l]{
$\sqrt{s}=1.02~\mathrm{GeV}$\\[4pt]
$50^\circ \le \theta^{\pm} \le 130^\circ$\\[4pt]
$p_z^{\pm} \ge 90~\mathrm{MeV}\ \vee\ p_\perp^{\pm} \ge 160~\mathrm{MeV}$\\[4pt]
$\theta_{\tilde{\gamma}} \le 15^\circ\ \vee\ \theta_{\tilde{\gamma}} \ge 165^\circ$\\[4pt]
$0.35~\mathrm{GeV}^2 \le M^2_{XX} \le 0.95~\mathrm{GeV}^2$
}
% \vspace{3mm}
\\
\bottomrule
\textbf{(c) BES 3} &
\textbf{(d) B } \\
\midrule
% \vspace{1mm}
\makecell[l]{
$\sqrt{s}=4~\mathrm{GeV}$\\[4pt]
$|\cos\theta^{\pm}| \le 0.93 \,\wedge\, p_\perp^{\pm} \ge 300~\mathrm{MeV}$\\[4pt]
$\big[|\cos\theta_\gamma| \le 0.8\,\wedge\, E_\gamma \ge 25~\mathrm{MeV}\big]$\\[4pt]
$\vee\ \big[0.86 \le |\cos\theta_\gamma| \le 0.92\,\wedge\, E_\gamma \ge 50~\mathrm{MeV}\big]$\\[4pt]
$\exists!\,\gamma:\ E_\gamma \ge 400~\mathrm{MeV}$
}
% \vspace{3mm}
&
% \vspace{1mm}
\makecell[l]{
$\sqrt{s}=10~\mathrm{GeV}$\\[4pt]
$0.65~\mathrm{rad} \le \theta^{\pm} \le 2.75~\mathrm{rad}\,\wedge\, p^{\pm} \ge 1~\mathrm{GeV}$\\[4pt]
$0.6~\mathrm{rad} \le \theta_\gamma \le 2.7~\mathrm{rad}\,\wedge\,E_\gamma \ge 3~\mathrm{GeV}$\\[4pt]
$\theta_{\tilde{\gamma},\gamma^{(h)}} \le 0.3~\mathrm{rad}$\\[4pt]
$M_{XX\gamma} \ge 8~\mathrm{GeV}$
}
% \vspace{3mm}
\\
\bottomrule
\end{tabular}
\caption{In this table, the four scenarios used for simulating the processes under study are detailed. In scenario (b), $\theta_{\tilde{\gamma}}$ is defined as the polar angle of the three-momentum given by $\vec{p}_{\tilde{\gamma}}=-\left(\vec{p}_{+}+\vec{p}_{-}\right)$. In scenario (d), $\theta_{\tilde{\gamma},\gamma^{(h)}}$ is defined as the polar angle between the three-momentum $\vec{p}_{\tilde{\gamma}}$ and the three momentum of the hardest photon passing the photon selection cuts.}
\label{tab:scenarios}
\end{table}
\subsection{Validation}
\label{sec:validation}
We validate the present formulation of radiative processes by comparing the exact NLO calculation implemented in our code with independent predictions available in the literature. 
For the $\mu^+\mu^-\gamma$ final state, our results have been benchmarked against those obtained with \textsc{McMule} and \textsc{Phokhara}, while for the $\pi^+\pi^-\gamma$ channel a dedicated comparison has been performed with \textsc{Phokhara}. 
The predictions for the other generators are taken from the \href{https://zenodo.org/records/13928140}{Zenodo} repository where all the data from the comparisons performed in~\cite{Aliberti:2024fpq} are stored.

\begin{figure}[htbp]
    \centering
    % Primo pannello
    \begin{minipage}[b]{0.48\linewidth}
        \centering
        \includegraphics[width=\linewidth]{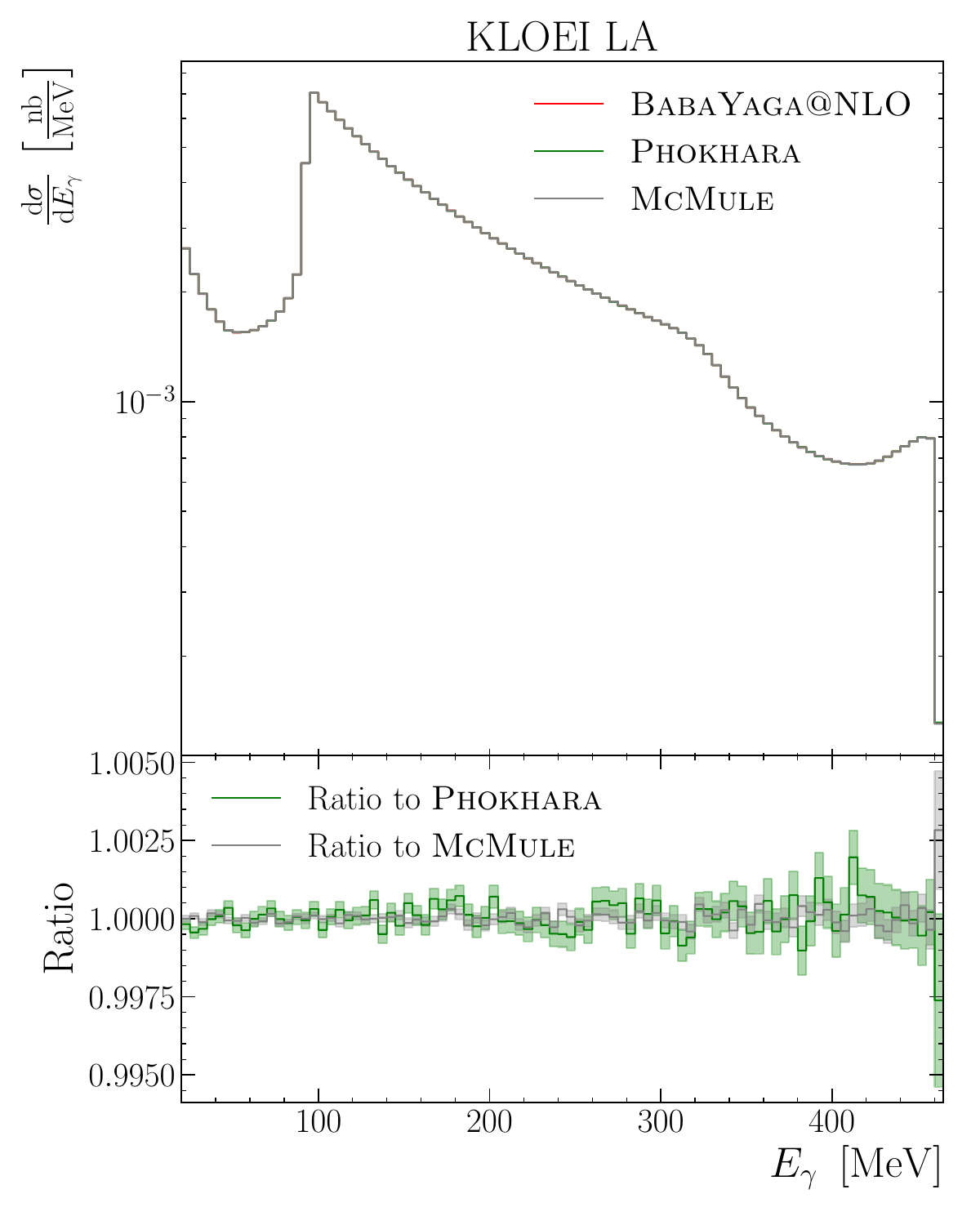} 
        
        \end{minipage}
    \hfill
    % Secondo pannello
      \begin{minipage}[b]{0.48\linewidth}
        \centering
        \includegraphics[width=\linewidth]{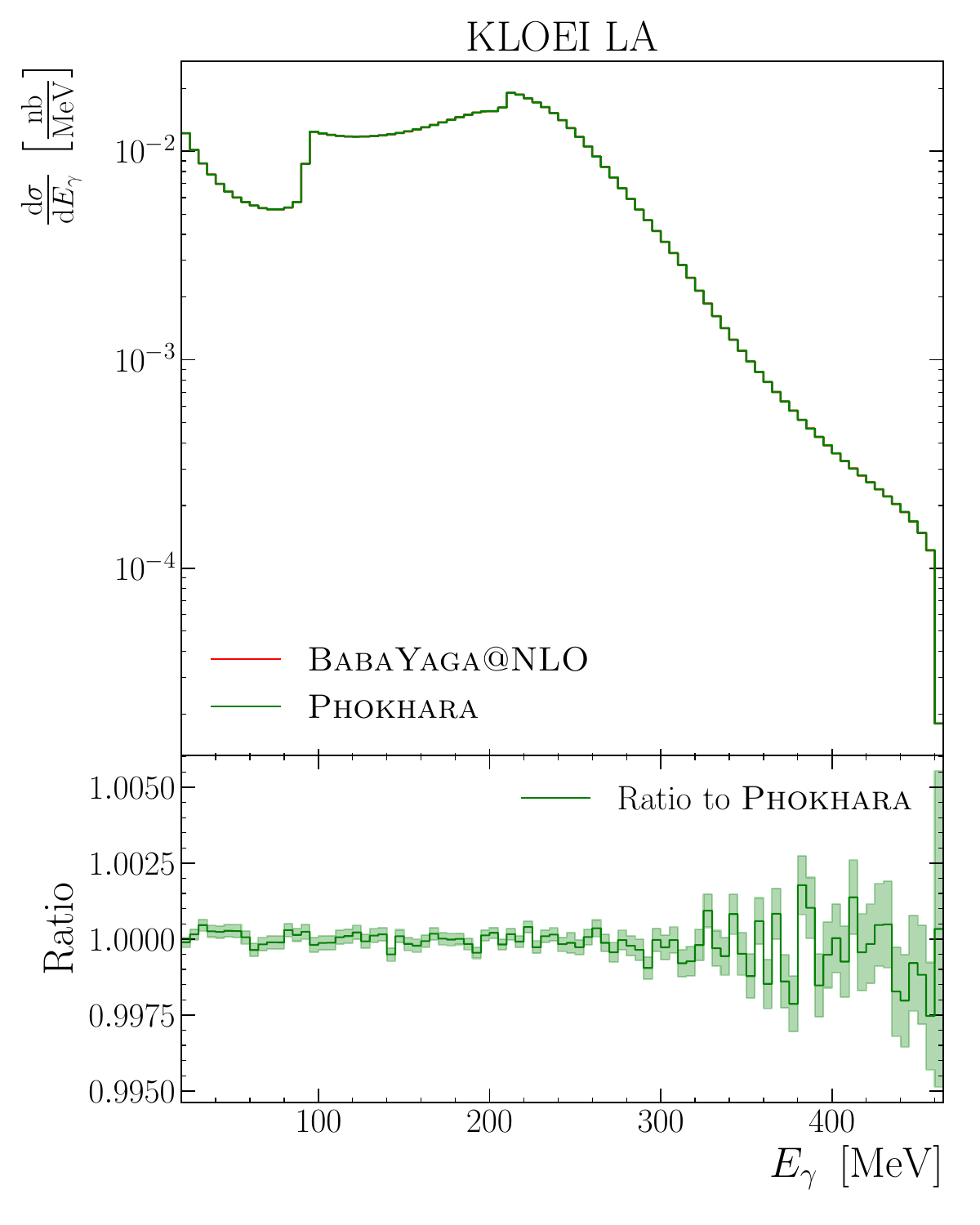}        
     \end{minipage}
     \caption{Comparison of the exact NLO calculation implemented in 
     \textsc{BabaYaga@NLO} with those of independent programs. 
Left panel: $\mu^+\mu^-\gamma$ channel, compared with \textsc{McMule} and \textsc{Phokhara}.
Right panel: $\pi^+\pi^-\gamma$ channel, compared with \textsc{Phokhara}. 
Both panels correspond to setup~(a) of Tab.~\ref{tab:scenarios}, and show the differential cross section 
as a function of the tagged photon energy.}
    \label{kloei-eg}
\end{figure}

For the scenario~(a), the differential cross section as a function of the tagged photon energy $E_\gamma$ is shown in Fig.~\ref{kloei-eg} for both final states. We observe excellent agreement, at the level of per mille accuracy, with both \textsc{Phokhara} and \textsc{McMule}. We checked that the same level of agreement holds for all the distributions in all the experimental scenarios considered in Ref.~\cite{Aliberti:2024fpq} for the $\mu^+\mu^-\gamma$ channel.

\begin{figure}[htbp]
    \centering
    % Primo pannello
    \begin{minipage}[b]{0.48\linewidth}
        \centering
        \includegraphics[width=\linewidth]{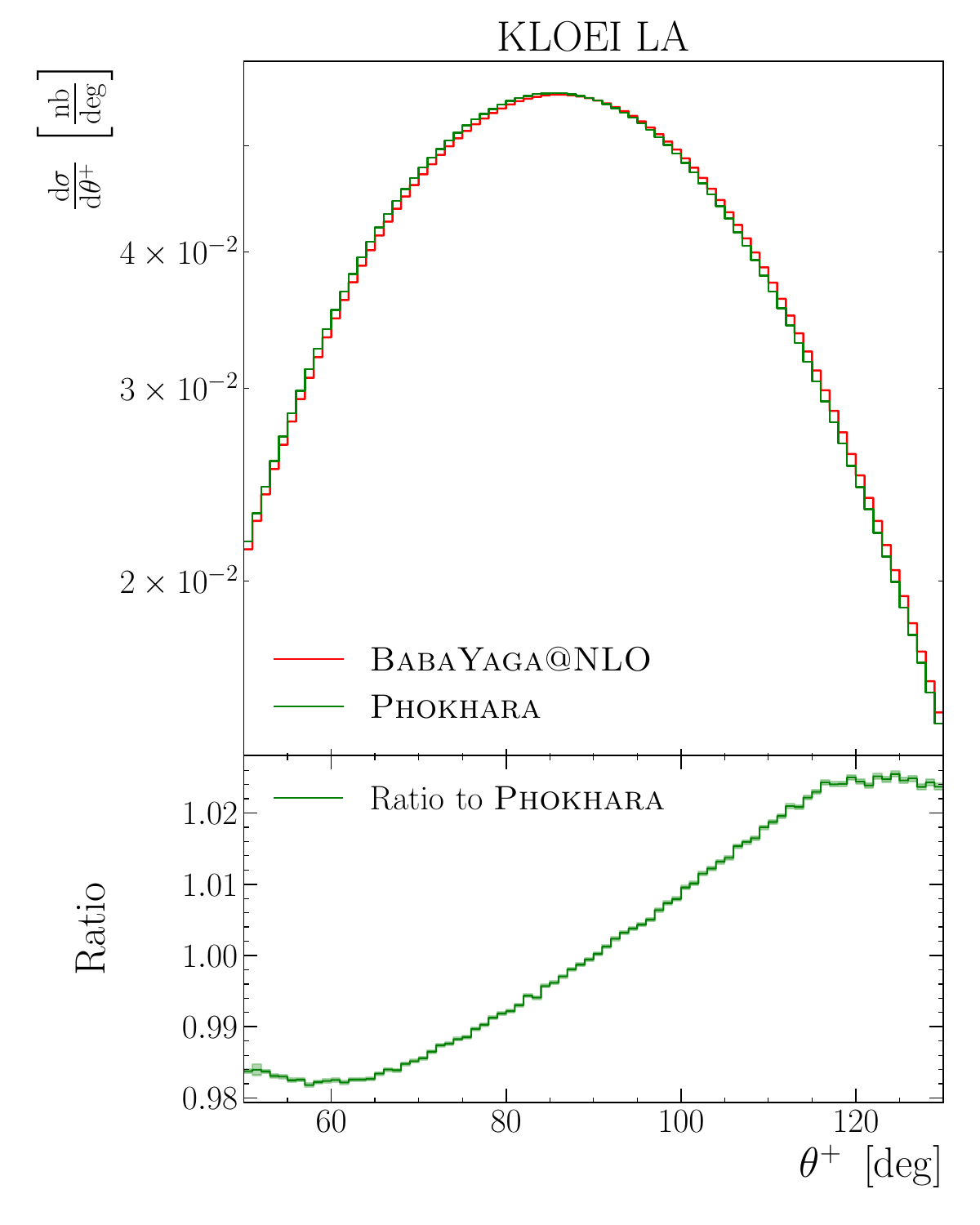} 
        
        \end{minipage}
    \hfill
    % Secondo pannello
      \begin{minipage}[b]{0.48\linewidth}
        \centering
        \includegraphics[width=\linewidth]{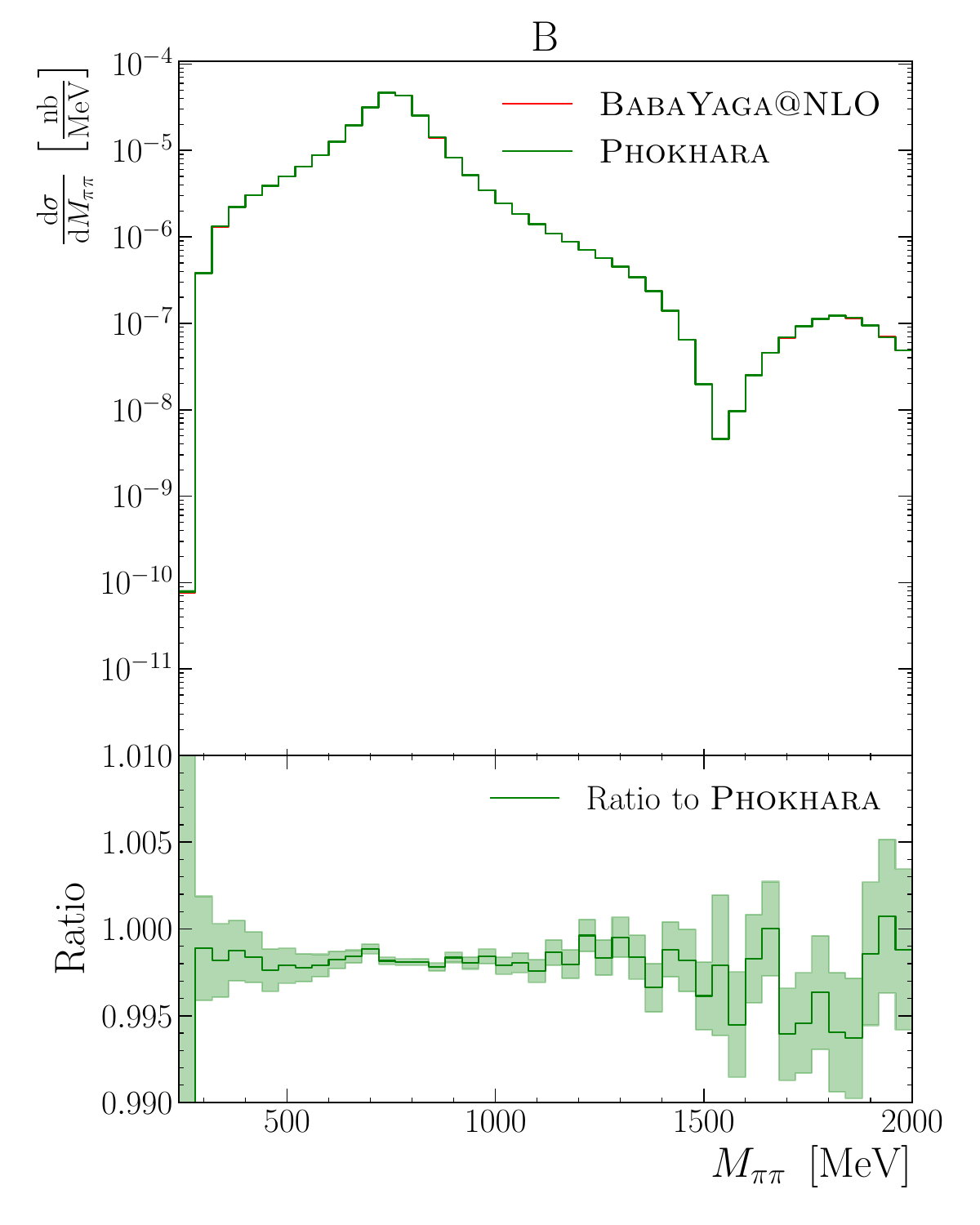}        
     \end{minipage}
     \caption{Comparison of the exact NLO calculation implemented in 
     \textsc{BabaYaga@NLO} with \textsc{Phokhara} for the $\pi^+\pi^-\gamma$ channel. 
Left panel: differential cross section as a function of $\theta^{+}$ in setup~(a). Right panel: differential cross section as a function of $M_{\pi\pi}$ in setup~(d) in the range $M_{\pi\pi}\in \left[2\,m_\pi,2 \,\, \rm GeV\right]$. Event selection provided in Tab.~\ref{tab:scenarios}.}
    \label{fig:kloei-th+_b_mxx}
\end{figure}

However, in the case of the $\pi^+\pi^-\gamma$ final state, discrepancies are observed with respect to \textsc{Phokhara}\footnote{As outlined in Sec.~\ref{sec:fix-ord}, we performed two independent calculations for the full set of NLO corrections, finding perfect agreement. Furthermore, we checked that the soft limit of the virtual corrections to the $2\to3$ processes correctly reproduces the factorised expression of Eq.~\eqref{eq:LBK}, where the $2\to2$ virtual corrections were evaluated in \cite{Budassi:2024whw}.} and concern the angular distribution of the $\pi^+$ (or, equivalently, of the $\pi^-$). The effect is only visible in the $\theta^+$ and $\theta^-$ distributions for configurations with symmetric angular cuts (left panel of Fig.~\ref{fig:kloei-th+_b_mxx}), while it impacts all the distributions in the presence of asymmetric cuts, \textit{e.g} the B scenario (right panel of Fig.~\ref{fig:kloei-th+_b_mxx}). 

\begin{figure}[htbp]
    \centering
    \includegraphics[width=0.75\linewidth]{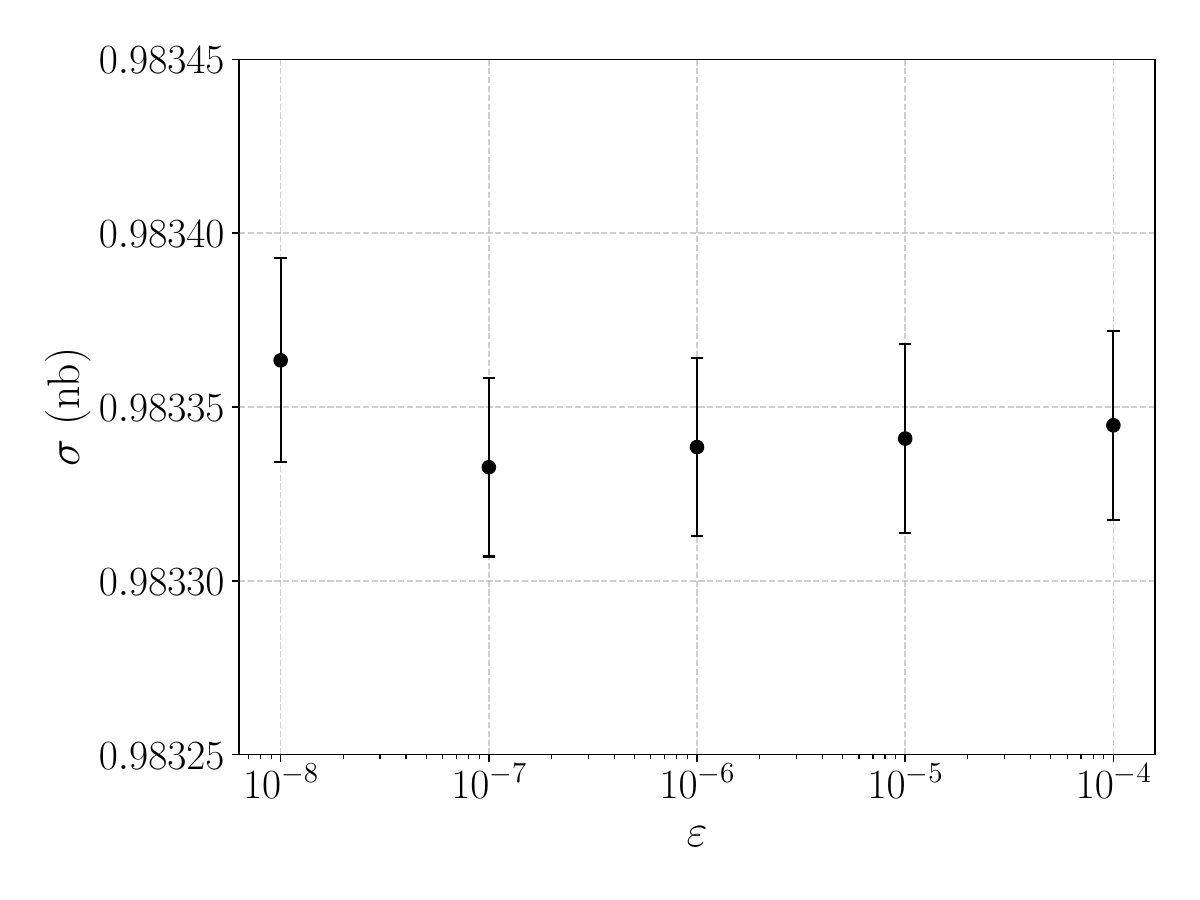}
    \caption{Integrated cross section  for the $\mu^+\mu^-\gamma$ channel at NLOPS accuracy  as a function of the infrared regulator
$\varepsilon$, according to setup~(a) of Tab.~\ref{tab:scenarios}. The error bars correspond to $1\sigma$ Monte Carlo uncertainty.}
    \label{fig:eps-independence}
\end{figure}

A further important validation of the matching between NLO corrections and the PS algorithm consists in verifying the stability of the 
NLOPS cross section under variations of the soft–hard separator $\varepsilon$. This behaviour is illustrated in Fig.~\ref{fig:eps-independence}, where the $\mu^{+}\mu^{-}\gamma$ cross section, computed as in Eq.~(\ref{eq:PSMATCH}) and under the conditions of setup~(a), is plotted as a function of $\varepsilon$ between $10^{-4}$ and $10^{-8}$. As expected, the results are compatible within  $1\sigma $ MC error intervals, essentially showing the independence of the NLOPS cross section on $\varepsilon$. This feature has also been explicitly checked for the NLO cross section, which represents a core component of the present formulation.

\begin{figure}[htbp]
    \centering

        \begin{minipage}[b]{0.48\linewidth}
        \centering
        \includegraphics[width=1\linewidth]{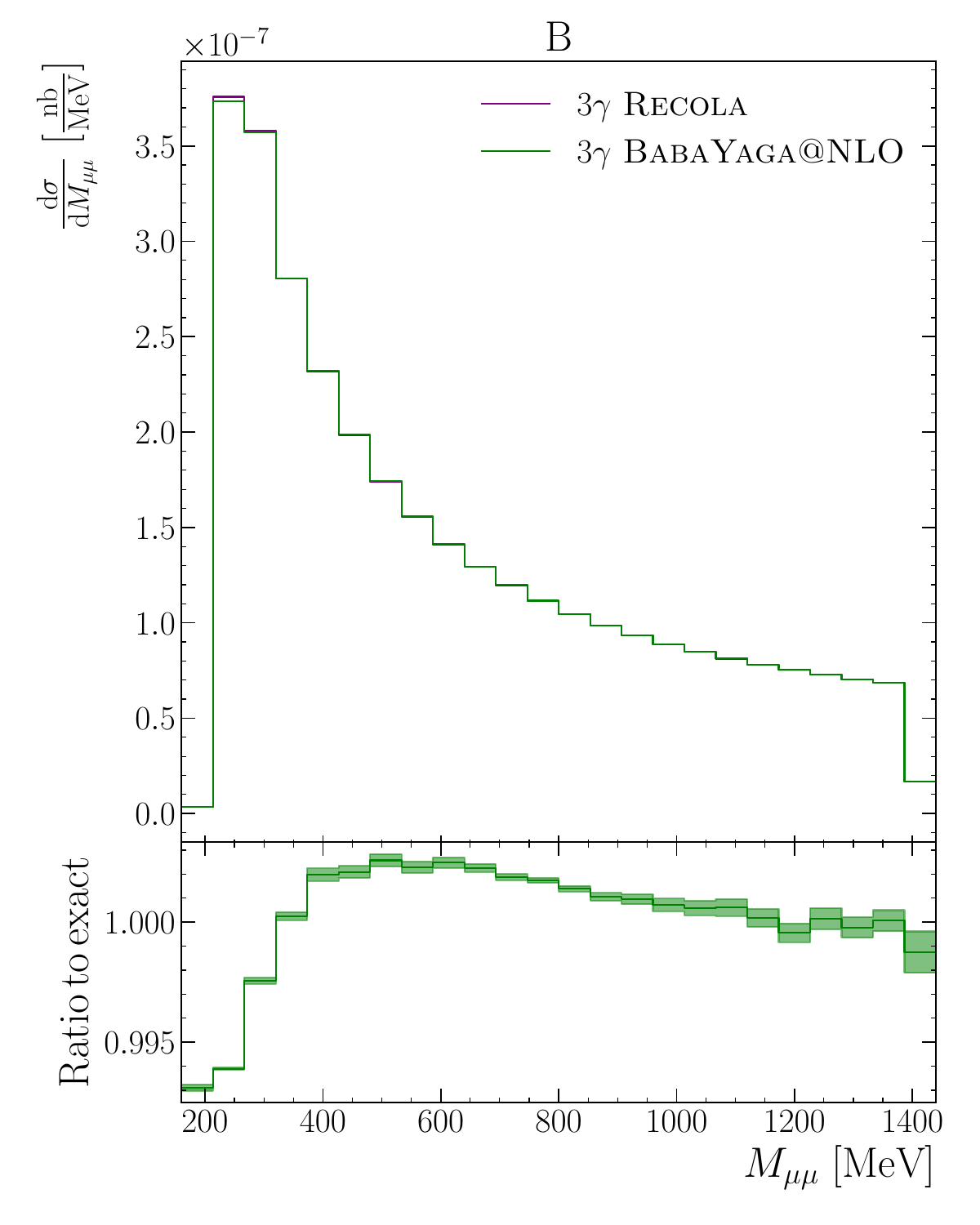}
        
        \end{minipage}
    \hfill
    % Secondo pannello
      \begin{minipage}[b]{0.48\linewidth}
        \centering
        \includegraphics[width=1\linewidth]{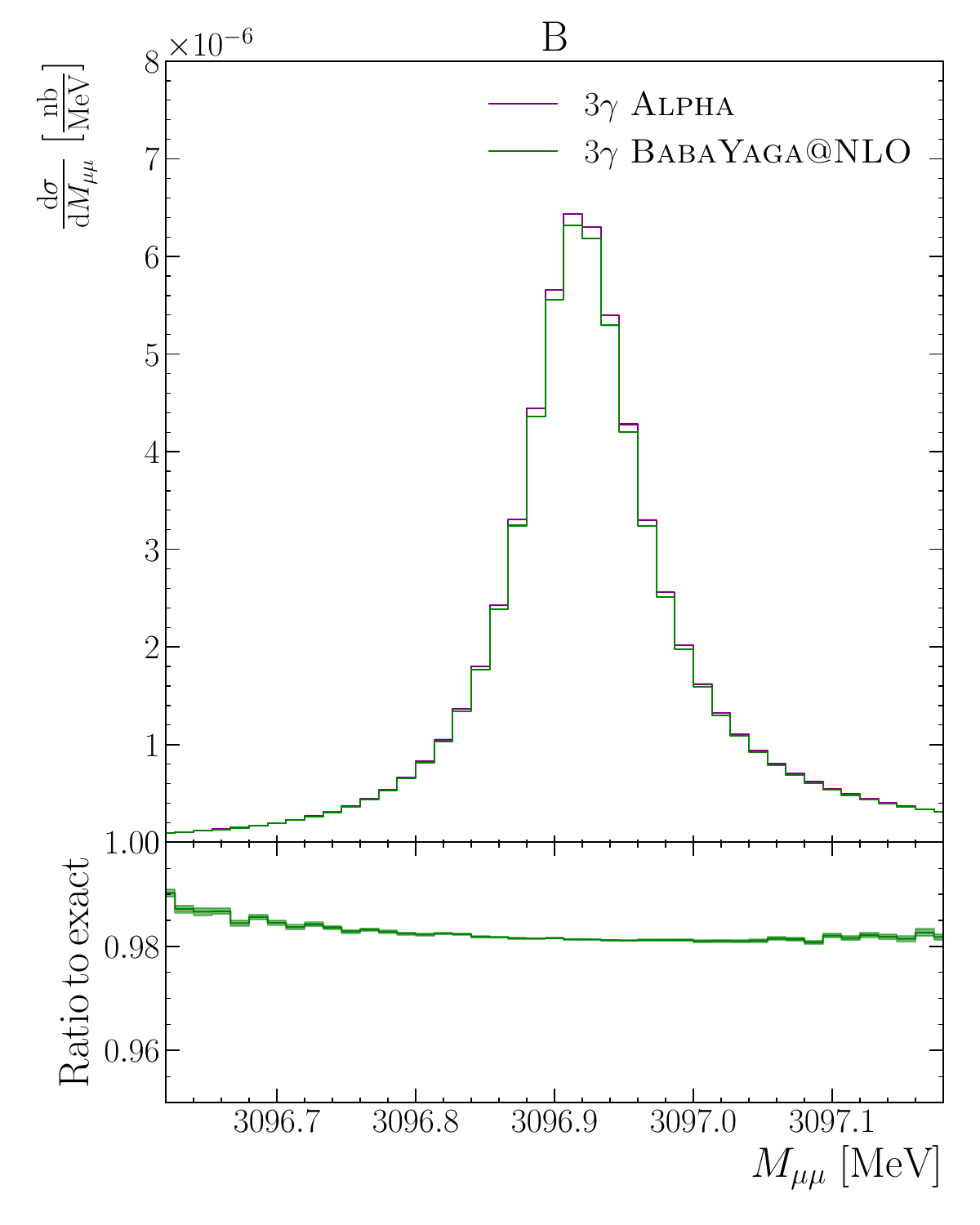}     
     \end{minipage}
    
    \caption{Left panel: comparison between the exact LO cross section of exclusive three-photon production without vacuum polarization obtained with \textsc{Recola} 
    and the LL approximation implemented 
    in \textsc{BabaYaga@NLO}, as a function of the $M_{\mu\mu}$ invariant mass in the range $\left[2 m_\mu,1.4 \,\,\rm{GeV}\right]$. Right panel: comparison between the exact LO cross section of exclusive three-photon production including vacuum polarization around the $J/\Psi$ resonance obtained with \textsc{Alpha} 
    and the LL approximation implemented 
    in \textsc{BabaYaga@NLO}, as a function of the $M_{\mu\mu}$ invariant mass in the range $M_{J/\Psi} \pm 3\, \Gamma_{J/\Psi}$. Both results refer to the B scenario of Tab.~\ref{tab:scenarios}, for the $e^+e^-\to\mu^+\mu^-3\gamma$ process, where the two additional photons are required with energy above $100$ MeV.}
    \label{fig:3gamma_excl}
\end{figure}

The consistency of the approach is also verified through the study of the approximation adopted for the calculation of the matrix elements in the presence of multiple photon emissions.  
As mentioned in Sec.~\ref{sec:PS}, the construction of the matrix elements in the LL approximation on top of the two-photon contribution allows for an accurate description of the exclusive three-photon signature.  
This feature is illustrated in Fig.~\ref{fig:3gamma_excl}, where, for 
the B scenario, two additional photons with energies above 100~$\rm{MeV}$ are required 
in the simulation of the $\mu^+\mu^-\gamma$ channel.
In the left panel, the approximate LL result is compared with the exact three-photon LO calculation obtained with \textsc{Recola}, without any vacuum polarization effect, showing very good agreement. Actually, the relative difference is of a few per mille at most, but just close to the muon pair production threshold. In the right panel, the approximate LL result is compared with the exact three-photon calculation obtained with \textsc{Alpha}, including the 
contribution of vacuum polarization around the $J/\psi$ resonance with the Novosibirsk parametrisation~\cite{NSK_VP}. In this case, the agreement is at the percent level, because of the relevance of narrow resonance effects, that are exactly included in the computation with \textsc{Alpha} and only approximately captured in the LL calculation.

\subsection{Size of LO and NLO gauge-invariant 
subsets}
\label{sec:subsec-giss}

Before discussing the impact of NLOPS corrections, which is one of the main original results of our work, we show in this Section to which extent the gauge-invariant subsets associated to ISR, FSR and IFI contribute to the full predictions at LO and NLO. We focus on the large-angle scenario (a) and the small-angle setup~(b) of the KLOE experiment, as we are interested to point out how the different classes of contributions depend on the event selection criteria, when increasing the theoretical accuracy from LO to NLO. For this purpose, we consider the pion invariant mass distribution and we define the relative size of the gauge-invariant subsets of contributions due to ISR, FSR and IFI as follows 
\begin{equation}
\tilde{\rm K}_i^{\rm LO}=\frac{\rmd \sigma_i^{\rm LO}}{\rmd \sigma^{\rm LO}} \,, \, \quad \quad \tilde{\rm K}_i^{\rm NLO}=\frac{\rmd \sigma_i^{\rm NLO}}{\rmd \sigma^{\rm NLO}} \, , \,
\label{eq:ratios}
\end{equation}
for the LO and NLO approximation, respectively, and where the pedex $i$ stands for ISR, FSR and IFI. In the denominators of Eq.~(\ref{eq:ratios}), $\rmd \sigma_{\rm LO}$ and $\rmd \sigma_{\rm NLO}$ are the full results at LO and NLO, respectively. At NLO, the ISR and FSR gauge-invariant cross sections are defined as
\begin{equation}
   \rmd\sigma_{i}^{\rm NLO} \, = \, \left[1+C_{\alpha, i}\right]\left|{\cal M}_{1, i}^{\rm ex}\right|^2\rmd\Phi_{1}+\left|{\cal M}_{2, i}^{\rm ex}\right|^2\rmd\Phi_{2}\,\, ,
\end{equation}
for both $i$ = ISR and $i$ = FSR. Then, the gauge-invariant IFI contribution is computed by subtraction as
\begin{equation}
\rmd\sigma_{\rm IFI}^{\rm NLO} \, = \, \rmd\sigma^{\rm NLO} - \rmd\sigma^{\rm NLO}_{\rm ISR} - \rmd\sigma^{\rm NLO}_{\rm FSR} \, , 
\end{equation}
and, for convenience, we split the latter as the sum of an even and odd contribution under the exchange of $\pi^+$ and $\pi^-$ momenta as
\begin{equation}
    \rmd\sigma_{\rm IFI}^{\rm NLO}=\left(\rmd\sigma_{\rm IFI}^{\rm NLO}\right)_{\rm odd}+\left(\rmd\sigma_{\rm IFI}^{\rm NLO}\right)_{\rm even}\,.
\end{equation}
\begin{figure}[htbp]
        \centering
        \includegraphics[width=\linewidth]{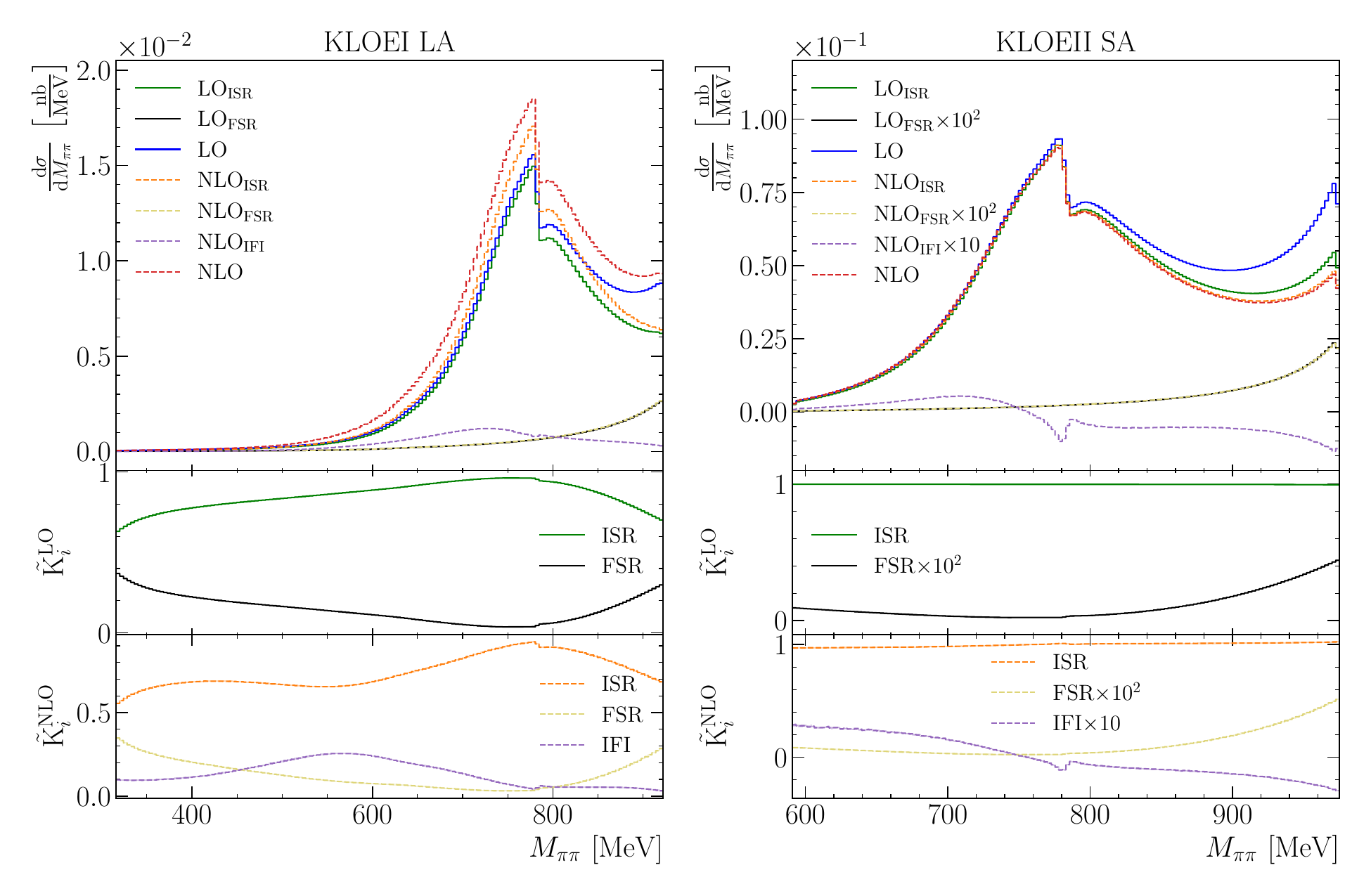}        
       \caption{Left panels: the LO and NLO ISR, FSR and IFI gauge-invariant contributions to $\rmd \sigma / M_{\pi\pi}$ in the KLOEI LA scenario of Tab.~\ref{tab:scenarios}. Right panels: the same as in the left panels for the KLOEII SA setup of Tab.~\ref{tab:scenarios}.}
\label{gaugeinvsubsets}
\end{figure}
In our numerical analysis, we also consider the forward-backward asymmetry $A_{\rm FB}$ as a function of the $M_{\pi\pi}$ invariant mass. It is defined as follows~\cite{Binner:1999bt,WorkingGrouponRadiativeCorrections:2010bjp}
\begin{equation}
    A_{\rm FB}
    \left(M_{\pi\pi}\right)=\frac{\rmd \sigma_{\rm F}-\rmd \sigma_{\rm B}}{\rmd \sigma_{\rm F}+\rmd \sigma_{\rm B}}
    \label{eq:asymmetry}
\end{equation}
where
\begin{equation}
    \rmd \sigma_{\rm F}=\int_{0}^{1}\frac{\rmd \sigma}{\rmd M_{\pi\pi}\,\,\rmd \cos{\theta_{+}}} \rmd \cos{\theta_+}\;\;\;\;\;\;\;\;\;\rmd \sigma_{\rm B}=\int_{-1}^{0}\frac{\rmd \sigma}{\rmd M_{\pi\pi}\,\,\rmd \cos{\theta_{+}}} \rmd \cos{\theta_+}
\end{equation}
and $\cos{\theta_{+}}$ is the cosine of the $\pi^+$ scattering angle defined with respect to the direction of the incoming electron. Note that, for symmetric cuts on the photon angular variables, as in scenario (a) and (b) of our simulations, the numerator of Eq.~(\ref{eq:asymmetry}) is non zero only due to the IFI term, both at LO and NLO. At LO, this holds for the complete IFI contribution, since it is an odd function under the exchange of $\pi^+$ and $\pi^-$ momenta (or electron and positron momenta), whereas at NLO this applies to $\left(\rmd\sigma_{\rm IFI}^{\rm NLO}\right)_{\rm odd}$. Moreover, the forward-backward asymmetry is an interesting observables because it is, at least to a first approximation, a test for the description of FSR~~\cite{Binner:1999bt,WorkingGrouponRadiativeCorrections:2010bjp} and can therefore be used to check the 
model adopted to describe the interaction
of the pions to photons.

\begin{figure}[htbp]
    \centering        
    \includegraphics[width=\linewidth]{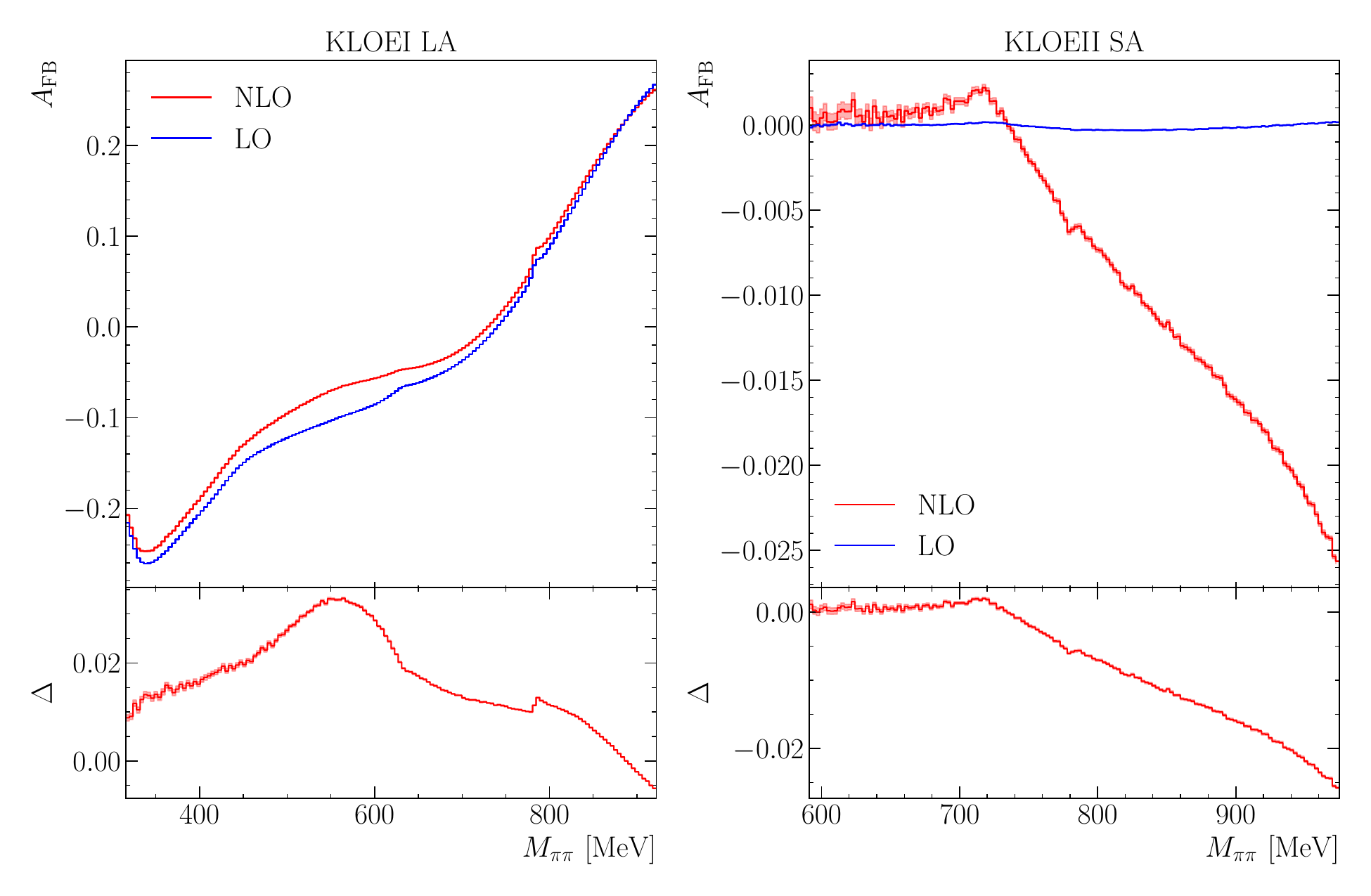}            
    \caption{The forward-backward asymmetry $A_{\rm FB}$ as a function of $M_{\pi\pi}$ at LO and NLO in the KLOEI LA (left plot) and KLOEII SA (right plot) scenarios of Tab.~\ref{tab:scenarios}. In the lower panels, $\Delta$ is defined as the absolute difference between the NLO and LO asymmetry, namely $\Delta= A_{\rm FB}^{\rm NLO} - A_{\rm FB}^{\rm LO}$.}
\label{asymm}
\end{figure}

We show the results of our study in Fig.~\ref{gaugeinvsubsets} for $\rmd\sigma /M_{\pi\pi}$ and in Fig.~\ref{asymm} 
for $A_{\rm FB} (M_{\pi\pi})$. 

Concerning the pion invariant mass cross section distribution, the IFI contribution at LO is zero since it  
cancels out when integrated over the symmetric pion angular setups used in our simulations, because of its odd nature. As can be seen from Fig.~\ref{gaugeinvsubsets}, the LO FSR contribution can amount to a few tens of percent 
in the KLOEI LA setup (left panels), whereas it is practically negligible in the KLOEII SA (right panels), where it can reach at most the few per mille level in the right tail of the distribution. At NLO, the relative impact of FSR is essentially the same as at LO, both in scenario (a) and (b). This can be simply understood as the FSR at NLO is driven by relatively small logarithms of the kind 
$\log (s / m_\pi^2)$ or 
$\log (M_{\pi\pi} / m_\pi)$. The  contribution $\left(\rmd\sigma_{\rm IFI}^{\rm NLO}\right)_{\rm even}$ is of the same order as the FSR in the large-angle setup, while it amounts to a few percent in the small-angle scenario. In a nutshell, the IFI and FSR contributions to $\rmd\sigma / M_{\pi\pi}$ are significant in a large-angle setup but they are quite suppressed, as expected, at small photon angles.

In Fig.~\ref{asymm}, we show our predictions for $A_{\rm FB}$ as a function of $M_{\pi\pi}$ at LO and NLO, in the KLOEI LA scenario (left plot) and KLOEII SA one (right plot). In the lower panels, we also show the absolute difference between the NLO and LO calculations, $i.e.$ 
$\Delta= A_{\rm FB}^{\rm NLO}- A_{\rm FB}^{\rm LO}$. As can be seen, the LO asymmetry is quite large in the large-angle setup as it is given in the numerator by the full (odd) IFI contribution at LO, which is sizeable at large photon angles. However, the NLO corrections to $A_{\rm FB}$, that receive contributions from $\left(\rmd\sigma_{\rm IFI}^{\rm NLO}\right)_{\rm odd}$ only, are moderate and at a few per cent level. On the other hand, the LO $A_{\rm FB}$ is extremely small at small angles, being of the order of $10^{-4}$, because of the overwhelming role of ISR and strong suppression of FSR. At NLO, the asymmetry in the small-angle scenario remains small, varying between one per mille and two percent.

To summarize, in a setup with photons detected at large angles and with symmetric acceptance cuts on the pion angular variables, the $\rmd\sigma / M_{\pi\pi}$ distribution and the forward-backward asymmetry provide different and complementary information about the role of IFI. Indeed, $\rmd\sigma / \rmd M_{\pi \pi}$ receives a contribution from $\left(\rmd\sigma_{\rm IFI}^{\rm NLO}\right)_{\rm even}$ only, whereas $\left(\rmd\sigma_{\rm IFI}^{\rm NLO}\right)_{\rm odd}$ affects the shape of $A_{\rm FB}$. The contribution $\left(\rmd\sigma_{\rm IFI}^{\rm NLO}\right)_{\rm even}$ to the invariant mass distribution is relevant, while the NLO correction to 
$A_{\rm  FB}$, that is driven in the numerator by the odd NLO IFI contributions, is moderate and at the percent level, which is the same order of magnitude as the one-loop charge asymmetry of the two-pion production process 
$e^+ e^- \to \pi^+\pi^-$~\cite{Ignatov:2022iou,Colangelo:2022lzg,Budassi:2024whw}. As discussed in~\cite{Ignatov:2022iou,Colangelo:2022lzg,Budassi:2024whw}, the latter receives relevant distortions when inserting the composite nature of the pion in the loop calculation. Note, however, that in our evaluation of the NLO IFI corrections to both  $\rmd\sigma / \rmd M_{\pi \pi}$ and $A_{\rm FB}$
there are contributions that will be affected by these structure-dependent corrections and other ones that are insensitive to the modelling of the pion-photon interaction.

In this respect, the results of our analysis can serve as a guideline for the study of the phenomenological implications of a calculation of the structure-dependent corrections, that will be addressed in a future publication. Possibly, they can also stimulate novel experimental investigations on the accuracy of the FSR model as
implemented in MC generators used in radiative return measurements.

\subsection{NLOPS predictions for $e^+e^-\to X^+X^-\gamma $}
\label{sec:subsec-numerical}

In this Section, we present the numerical results at NLOPS accuracy obtained with the new version of \textsc{BabaYaga@NLO} according to realistic event selection criteria, as given by the four experimental scenarios described in Tab.~\ref{tab:scenarios}. 

\begin{table}[htbp]
\centering
\begin{tabular}{llllll}
\toprule
$XX\gamma$&\textbf{Scenario} & $\sigma_{\rm LO} \,(\rm nb)$& $\sigma_{\rm NLO} \,(\rm nb)$ & $\sigma_{\rm NLOPS} \,(\rm nb)$\\
\midrule
\multirow{4}{*}{$\mu^+\mu^-\gamma$}&KLOEI LA  &$8.792\cdot 10^{-1} $ &$ 9.925\cdot 10^{-1}$ & $9.833\cdot10^{-1}$  \\
&KLOEII SA&   3.732 & 3.487 &  3.485 \\
&BES3  & $2.680\cdot 10^{-1}$   & $2.643\cdot 10^{-1}$  & $2.640\cdot 10^{-1}$ \\
&B     &  $1.750\cdot10^{-2}$   & $1.618\cdot10^{-2}$ & $1.618\cdot10^{-2}$ \\
\midrule
\multirow{4}{*}{$\pi^+\pi^-\gamma$}&KLOEI LA  & 2.652 &3.227 & 3.204\\
&KLOEII SA& 16.46 & 16.06 & 16.01 \\
&BES3   &  $8.009\cdot10^{-2}$ & $8.381\cdot10^{-2}$ &  $8.374\cdot10^{-2}$  \\
&B        & $1.041\cdot10^{-2}$ &  $1.008\cdot10^{-2}$ & $1.005\cdot10^{-2}$\\
\bottomrule
\end{tabular}
\caption{Integrated cross sections in $\rm nb$ at LO, NLO and NLOPS accuracy for the 
processes $e^+ e^- \to X^+ X^-\gamma$, $X = 
\{\mu, \pi\}$, according to the four scenarios of Tab.~\ref{tab:scenarios}. The MC statistical uncertainty is not shown as it is below the 
quoted digits.}
\label{tab:isigma}
\end{table}

First, we present in Tab.~\ref{tab:isigma} the values of the integrated cross sections, without vacuum polarization effects, of the processes $e^+ e^- \to X^+ X^-\gamma$, $X = 
\{\mu, \pi\}$, in order to compare the cross sections of the hadronic and leptonic channels and to show their scaling with the c.m. energy.

From Tab.~\ref{tab:isigma}, it can be noticed that, for both the muon and pion final states, the cross section in the KLOEII SA setup is considerably larger than the one in the KLOEI LA scenario at the same c.m. energy, as a consequence of the strong enhancements given by the large collinear logarithms $\log (s / m_e^2)$ associated to ISR. For the $\mu^+\mu^-\gamma$ channel, the large-angle QED cross section quickly decreases from $\phi$ to $B$ factories because of the increasing c.m. energy, but with a not trivial scaling factor due to the different event selection criteria. A more sizeable scaling occurs for the $\pi^+\pi^-\gamma$ final state, as the contribution of the pion form factor falls down from the c.m. energy of KLOE close to the $\rho$ resonance to the higher energies of the BES3 and B event selections. Moreover, whereas at KLOE the pion channel cross section is significantly larger than that of the muon channel one, in the BES3 and B scenarios $\sigma (e^+ e^- \to \mu^+ \mu^- \gamma)$ exceeds $\sigma (e^+ e^- \to \pi^+ \pi^- \gamma)$.

The NLO corrections to the integrated cross sections are particularly visible in the KLOEI LA setup and, to a less extent, at KLOE small angle, as well as at BES3 and $B$ factories. The higher-order contributions to the integrated cross sections are about one percent in the KLOEI LA scenario and at the per mille level elsewhere.

\begin{figure}[htbp]
    \centering
    \includegraphics[width=\imagewidth]{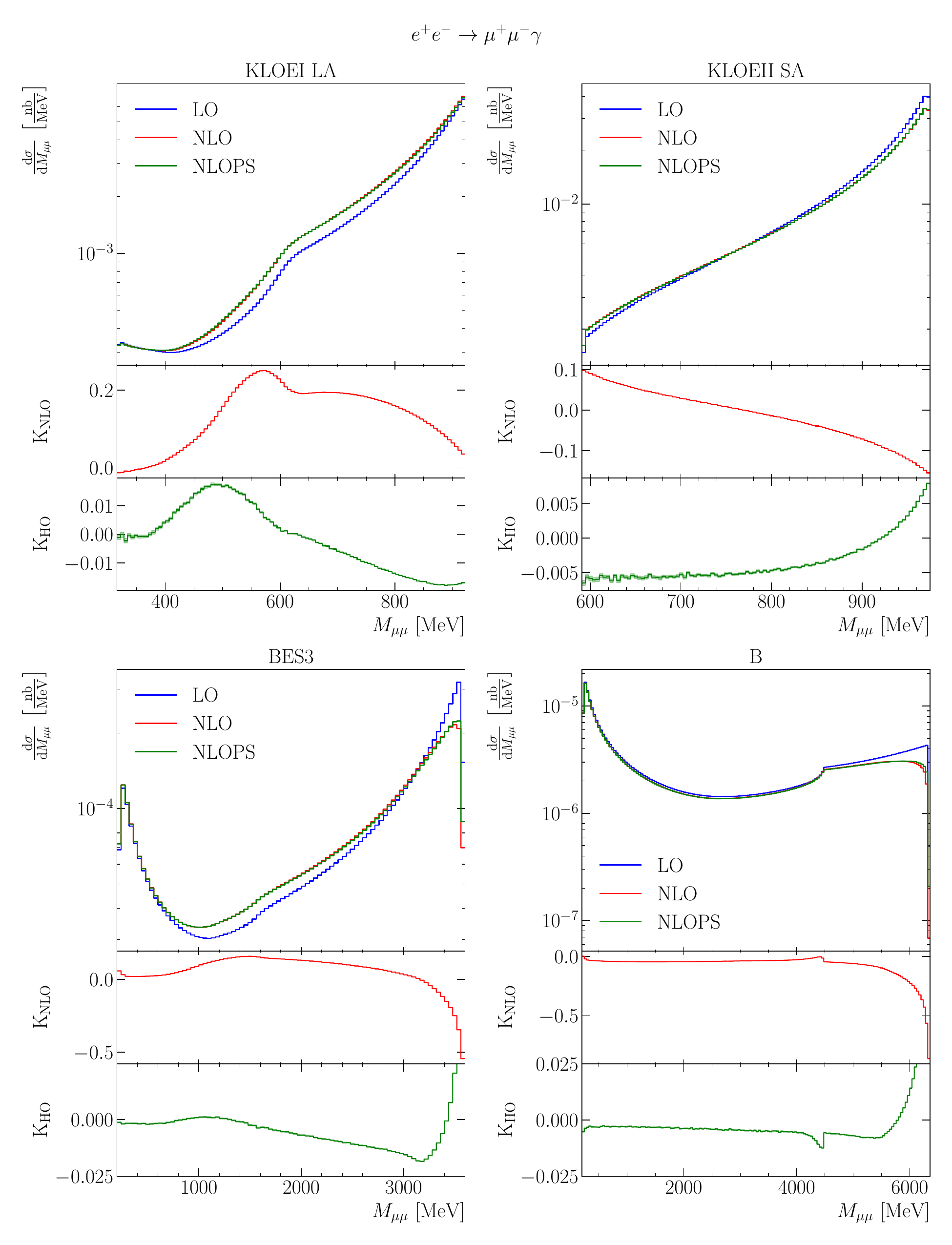} 
     \caption{The differential cross section of the $e^+e^-\to\mu^+\mu^-\gamma$ process as a 
     function of the invariant mass $M_{\mu\mu}$ with different approximations for radiative corrections in the four scenarios 
     of Tab.~\ref{tab:scenarios}. The quantities $K_{\rm HO}$ and $K_{\rm NLO}$ are defined in Eqs.~(\ref{eq:nlo}, \ref{eq:ho}).}
    \label{m_mumu}
\end{figure}

\begin{figure}[htbp]
    \centering
    \includegraphics[width=\imagewidth]{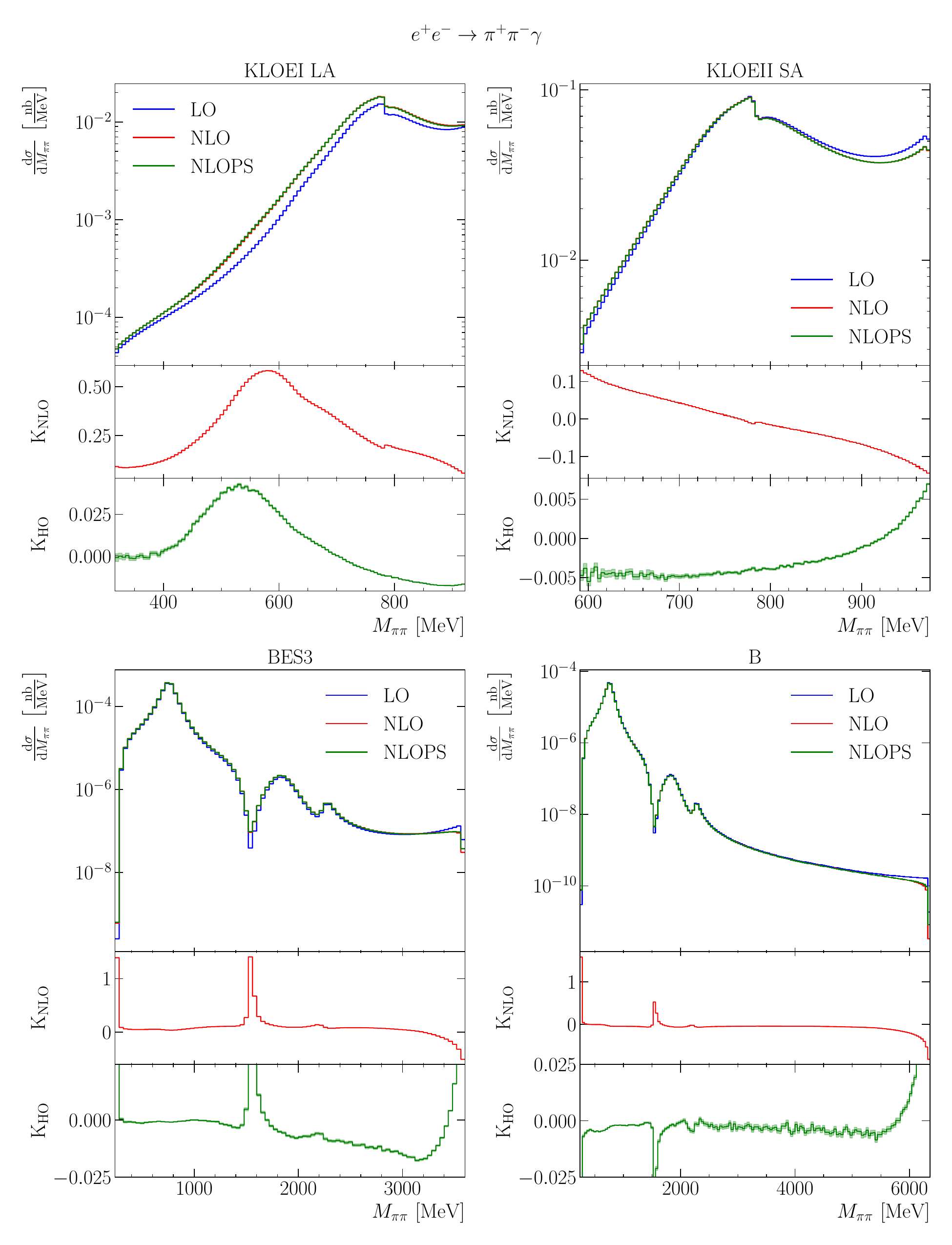} 
     \caption{The same as in Fig.~\ref{m_mumu} for the $\pi^+\pi^-\gamma$ channel.}
    \label{m_pipi}
\end{figure}

For the extraction of the pion form factor from the data, the observables of interest are the 
differential cross sections as a function of the invariant mass $M_{XX}$. We show our results
for these quantities in Fig.~\ref{m_mumu}
for the leptonic channel $e^+ e^- \to \mu^+ \mu^-\gamma$ and
in Fig.~\ref{m_pipi} for the hadronic channel $e^+ e^- \to \pi^+ \pi^-\gamma$. For the muon final state, we do not include in our predictions the effect of $\alpha$ running.

The relative contributions shown in the plots and due to NLO and higher-order multi-photon corrections are defined as follows:
\begin{eqnarray}
    &&{\rm K}_{\rm NLO} = \frac{\rmd\sigma_{\rm 
    NLO} - \rmd\sigma_{\rm LO}}{\rmd\sigma_{\rm LO}} \, ,
    \label{eq:nlo}
     \\
    &&{\rm K}_{\rm HO} = \frac{\rmd\sigma_{\rm 
    NLOPS} - \rmd\sigma_{\rm NLO}}{\rmd\sigma_{\rm LO}} \, ,
    \label{eq:ho}
\end{eqnarray}
where $\rmd\sigma$ is a short-hand notation that stands for the differential cross section as a function of the invariant mass. The same convention holds for the NLO and higher-order relative contributions shown for other distributions in Appendix~\ref{sec:app1}.

Concerning the muon pair invariant mass cross section distribution shown in 
Fig.~\ref{m_mumu}, it can be seen that the NLO corrections are particularly large in the high 
$M_{\mu\mu}$ region, where they can reach the level of tens of percent because of the relevance of one soft photon emission. Around the 
$\rho$ resonance, their contribution is of the order of some percent, from $\phi$ to $B$ factories. Broadly speaking, the impact of higher-order corrections is clearly visible at high values of 
$M_{\mu\mu}$, where multiple soft photon radiation dominates. However, their contribution 
depends non trivially on the considered scenario, both in shape and size. In setups~(a) and (b) of the KLOE experiment, the higher-order corrections modify the high invariant mass region at the percent level, whereas they amount to one percent for KLOEI LA scenario and 
$0.5\%$ for KLOEII SA scenario around the $\rho$. In particular, the shape of multi-photon contributions is strongly varying according to event selection (a), analogously to the behaviour observed for NLO corrections. For BES3 and B scenarios, the main effect of higher-order corrections occurs in the high $M_{\mu\mu}$ tail, where they partially compensate the large NLO corrections and can reach the level of some percent. At low-intermediate invariant mass, the multiple photon contributions are flat and are below the $0.1\%$ and of a few per mille around the $\rho$ at BES3 and B scenario, respectively.

The results for the hadronic invariant mass cross section distribution shown in Fig.~\ref{m_pipi} display similarities with those for $M_{\mu\mu}$, for both NLO and higher-order corrections. Again, the NLO contributions are particularly enhanced in the high $M_{\pi\pi}$ region for all the setups. However, they are also quite relevant below the $\rho$ resonance in KLOEI LA conditions and around the $\rho'$ in BES3 and B scenarios. As for $M_{\mu\mu}$, the multiple photon corrections to $M_{\pi\pi}$ strongly depend on the event selection criteria. For the KLOEI LA scenario, the higher-order contributions are about 3\% and 1\% below and around the $\rho$ resonance, respectively. They are at the level of 0.5\% around the $\rho$ for the KLOEII SA setup. For the BES3 and B scenario, the corrections beyond NLO have a relevant impact on $\rmd \sigma / M_{\pi\pi}$ at high values of the invariant mass, where they reach the some percent level 
and mitigate the effect of NLO corrections. Around the $\rho$, they are flat and negligible at BES3 and of 0.5\% at B factories. Around the $\rho'$, the higher-order corrections amount to about 3\% both in BES3 and B scenario.

\begin{figure}[htbp]
    \centering
     \includegraphics[width=\imagewidth]{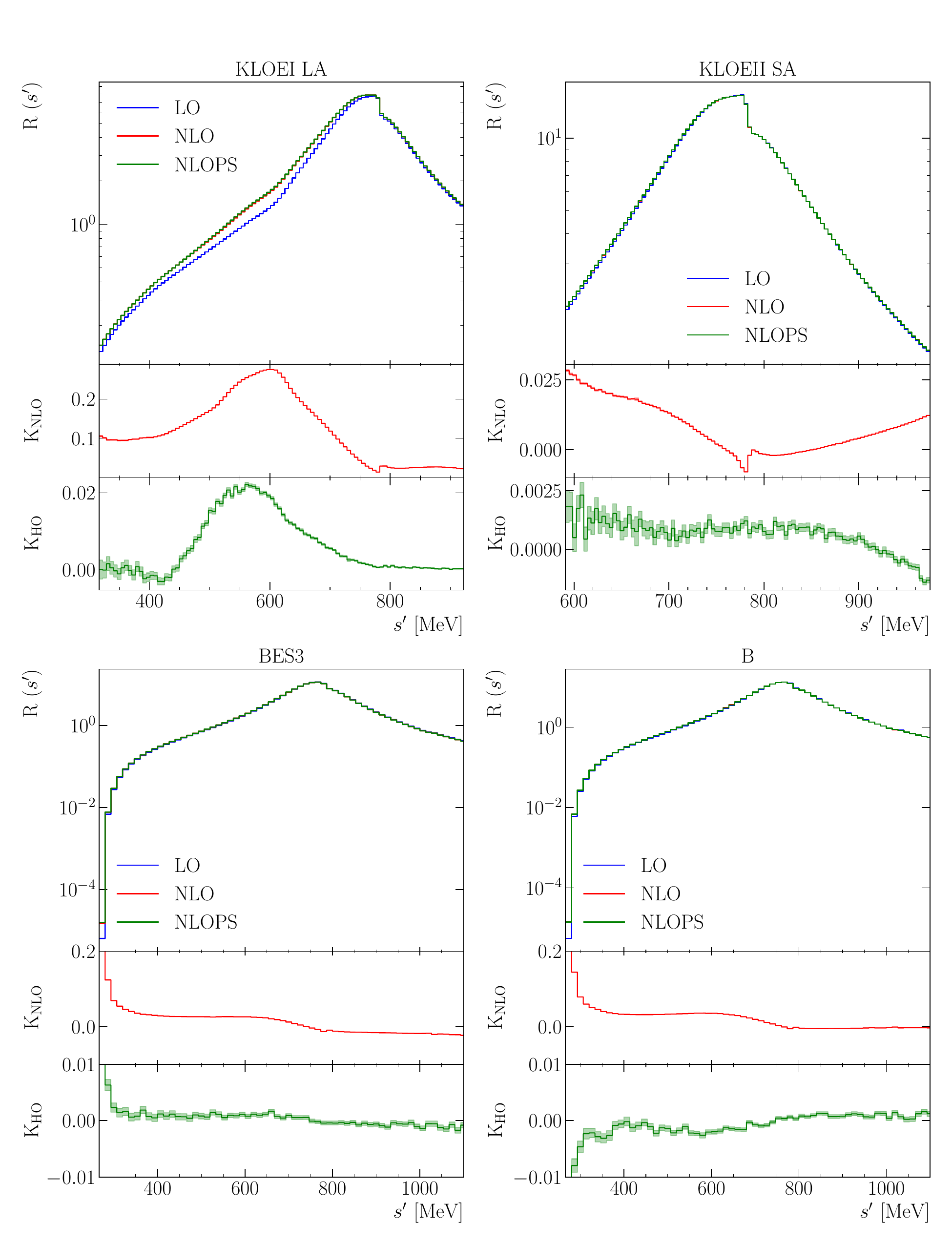}        
    \caption{The ratio ${\rm R}\left(s'\right)$, as defined in Eq.~\eqref{eq:hadr_ratio}, of the differential cross sections $\rmd \sigma(e^+e^-\to\pi^+\pi^-\gamma)$ and $\rmd\sigma(e^+e^-\to\mu^+\mu^-\gamma)$ as a function of the invariant mass $s'$ of the charged pairs with different approximations for radiative corrections in the four scenarios of Tab.~\ref{tab:scenarios}. The quantities $K_{\rm HO}$ and $K_{\rm NLO}$ are defined in Eqs.~(\ref{eq:nlo}, \ref{eq:ho}).}
\label{ratio}
\end{figure}

To complete the discussion on the effects due to 
multi-photon radiation, we show in Fig.~\ref{ratio} our predictions for the ratio of the differential cross section in the radiative pion channel to the $\mu^+\mu^-\gamma$
\begin{equation}
    {\rm R}\left(s'\right)=\frac{\rmd \sigma (e^+ e^- \to \pi^+ \pi^-\gamma) / \rmd s'}{\rmd \sigma (e^+ e^- \to \mu^+\mu^-\gamma) / \rmd s'} \,,
    \label{eq:hadr_ratio}
\end{equation}
including NLO and higher-order corrections, where $s'$ is the invariant mass of the charged pairs. This ratio is relevant for those experiments that normalise the yield of hadronic data to the number of QED events. Again, we present numerical results for all the 
four scenarios of Tab.~\ref{tab:scenarios}. As can be seen 
in Fig.~\ref{ratio}, in the KLOEI LA setup the LL corrections beyond NLO have a varying shape and can reach the two percent level, whereas they vary slightly and are of the order of 0.1\% in the KLOEII SA scenario. For the BES3 and B scenarios, the higher-order contributions to the ratio are almost flat 
and quite small, reaching the one percent only at the lower edge, where the ratio is 
largely suppressed.

Other experimentally relevant observables at NLOPS accuracy are shown in Appendix~\ref{sec:app1}, along with the contribution of NLO and multi-photon corrections.

\section{Summary and outlook}
\label{sec:conc}
We have computed for the first time the full set of NLO corrections matched to PS to the radiative processes 
$e^+ e^- \to \pi^+ \pi^- \gamma$ and 
$e^+ e^- \to \mu^+ \mu^- \gamma$,
that are relevant for radiative return experiments at flavour factories. All the radiative corrections due to ISR, FSR and IFI are included in our calculation according to QED for $e^+e^-\to\mu^+\mu^-\gamma$ and QED$\oplus$F$\times$sQED for $e^+e^-\to\pi^+\pi^-\gamma$. The formulation here presented contains some novel features in comparison with the NLOPS approach described in~\cite{Budassi:2024whw,Balossini:2006wc,Balossini:2008xr} for $2 \to 2$ processes. The main improvements concern both the description of the dynamics associated with exclusive hard photon events, which are a crucial ingredient in modelling photonic corrections to radiative return, and the phase-space treatment, which enables efficient MC integration and event generation even when non-perturbative effects are included. Our computation is implemented in a new release of the event generator \textsc{BabaYaga@NLO}, that can be used for data analysis in radiative return experiments. The code is publicly available on \href{https://github.com/cm-cc/BabaYagaNLO}{GitHub \faGithub}.

In the paper, we have shown a number of validation tests to cross-check the main features of our MC implementation. We have performed tuned comparisons of our NLO predictions with the fixed-order results available in the literature. We have also shown that our simulations at NLOPS accuracy are 
independent of the soft-hard separator used in the PS algorithm and that our LL predictions for the exclusive signature with three photons in the final state are in good agreement with those of an exact matrix element calculation.

Furthermore, we have quantified the impact of the gauge-invariant subsets associated to ISR, FSR and IFI at LO and NLO in the presence of different experimental cuts. We have pointed out that, in a setup with large-angle photons and symmetric angular cuts on the pions, the two-pion invariant mass cross section distribution and the forward-backward asymmetry provide useful and complementary information on the role played by IFI and can therefore be jointly used to test the model describing the interaction of the pions to photons.

To highlight the contribution of exclusive multiple photon emission, that is missing in the standard tool \textsc{Phokhara} used by the experiments, we have presented a sample of phenomenological results at NLOPS accuracy according to realistic event selection criteria for precision measurements
at flavour factories. We have shown that higher-order corrections beyond NLO are relevant for accurate simulations of radiative processes, from $\phi$ to B factories. The contribution of soft and collinear logarithms due to multi-photon radiation has a non-trivial effect on the shape of the main differential cross sections, depending on the considered experimental scenario. Both for muon and pion final states, the higher-order corrections amount to about one percent in the high invariant mass region at KLOEI LA and KLOEII SA and can reach the level of some percent in the same region in BES3 and B scenarios. This is due to the relevance of multiple soft-photon emission at high values of the invariant mass. Around the 
$\rho$ resonance, the higher-order corrections to the pion invariant mass cross section distribution amount to about one per cent and to 0.5\% for KLOEI LA and KLOEII SA scenario, respectively. In the same region, the multiple photon contributions are 
negligible at BES3 and of 0.5\% in B scenario. 
As detailed in Appendix~\ref{sec:app1}, 
the exponentiation of soft and collinear corrections plays a role at the level of some percent also for other experimentally interesting distributions, reinforcing the importance of 
precise theoretical predictions beyond fixed order for sub-percent measurements of radiative processes at low-energy 
$e^+ e^-$ colliders.

Broadly speaking, the observed higher-order effects are somewhat larger than or comparable to the systematic uncertainty in radiative return measurements at meson factories. In this respect, new data analyses based on our code 
\textsc{BabaYaga@NLO} will be likely to have an impact on the extraction of the pion form factor via radiative return and will benefit from a reduction of the theoretical uncertainty affecting this kind of measurements.

Concerning the perspectives of our work, we are interested to address other hadronic 
and leptonic channels of experimental concern, such as $e^+ e^- \to K^+ K^-\gamma$ and $e^+ e^- \to e^+ e^- \gamma$. More importantly, we plan to calculate the radiative corrections to $e^+ e^- \to \pi^+ \pi^-\gamma$ going beyond the F$\times$sQED approximation 
and computing the structure-dependent contributions due to the inclusion of the pion composition in the loops, as recently done by us for the two pion production in~\cite{Budassi:2024whw}. These developments are left to future work.

\acknowledgments
We are grateful to Riccardo Aliberti, Achim Denig, Fedor Ignatov, Bogdan Malaescu and Graziano Venanzoni for useful discussions and interest in our work. We thank Mauro Chiesa for helpful discussions on automatic loop calculations.
We also wish to thank Pau Petit Ros{\`a}s and William J. Torres Bobadilla for their assistance in the comparison with the \textsc{Phokhara} code. We acknowledge financial support by the Italian Ministero dell'Universit\`a e Ricerca (MUR) and European Union - Next Generation EU through the research grant number 20225X52RA ``MUS4GM2: Muon Scattering for $g\!-\!2$'' under the program Prin~2022. AG is supported by the Italian Ministero dell'Universit\`a e Ricerca (MUR) and European Union - Next Generation EU through the research grant number 2022ENJMRS ``INSIGHT:  INdirect SIGnature of new Higher TeV energy scale physics'' under the program Prin~2022. MG and FPU acknowledge the Galileo Galilei Institute for Theoretical Physics for hospitality during the completion of this work. FPU is grateful to the Instituto de Fisica Teorica UAM/CSIC for hospitality and to the Erasmus$+$Traineeship program for partial support. 
\appendix
\section{Other distributions at NLOPS accuracy}
\label{sec:app1}
\begin{figure}[htbp]
    \centering        \includegraphics[width=\linewidth]{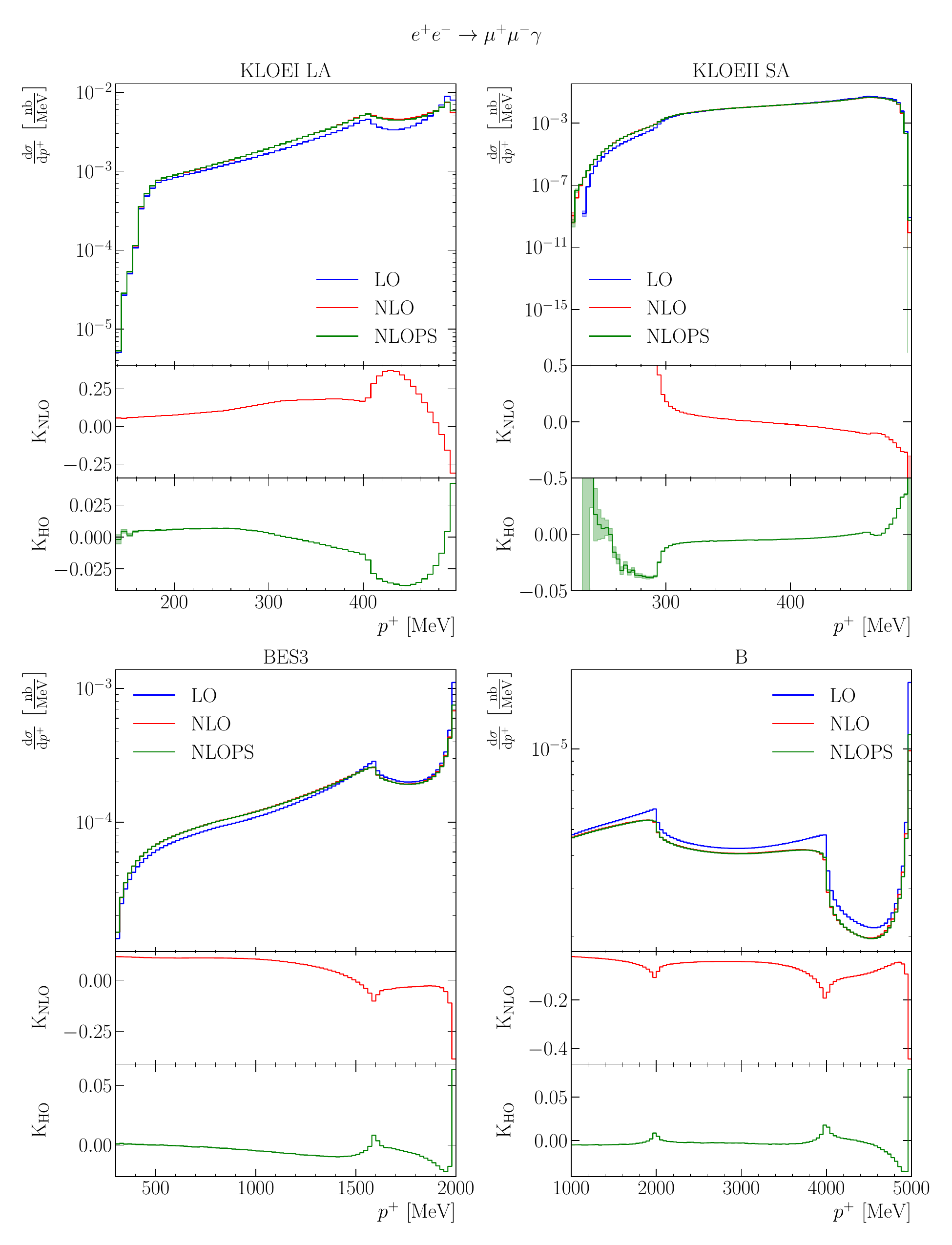} 

    \caption{The differential cross section of the $e^+e^-\to\mu^+\mu^-\gamma$ process as a 
     function of the modulus of the $\mu^+$ momentum with different approximations for radiative corrections in the four scenarios 
     of Tab.~\ref{tab:scenarios}.}
    \label{ap_mumu}
\end{figure}

\begin{figure}[htbp]
    \centering        \includegraphics[width=\linewidth]{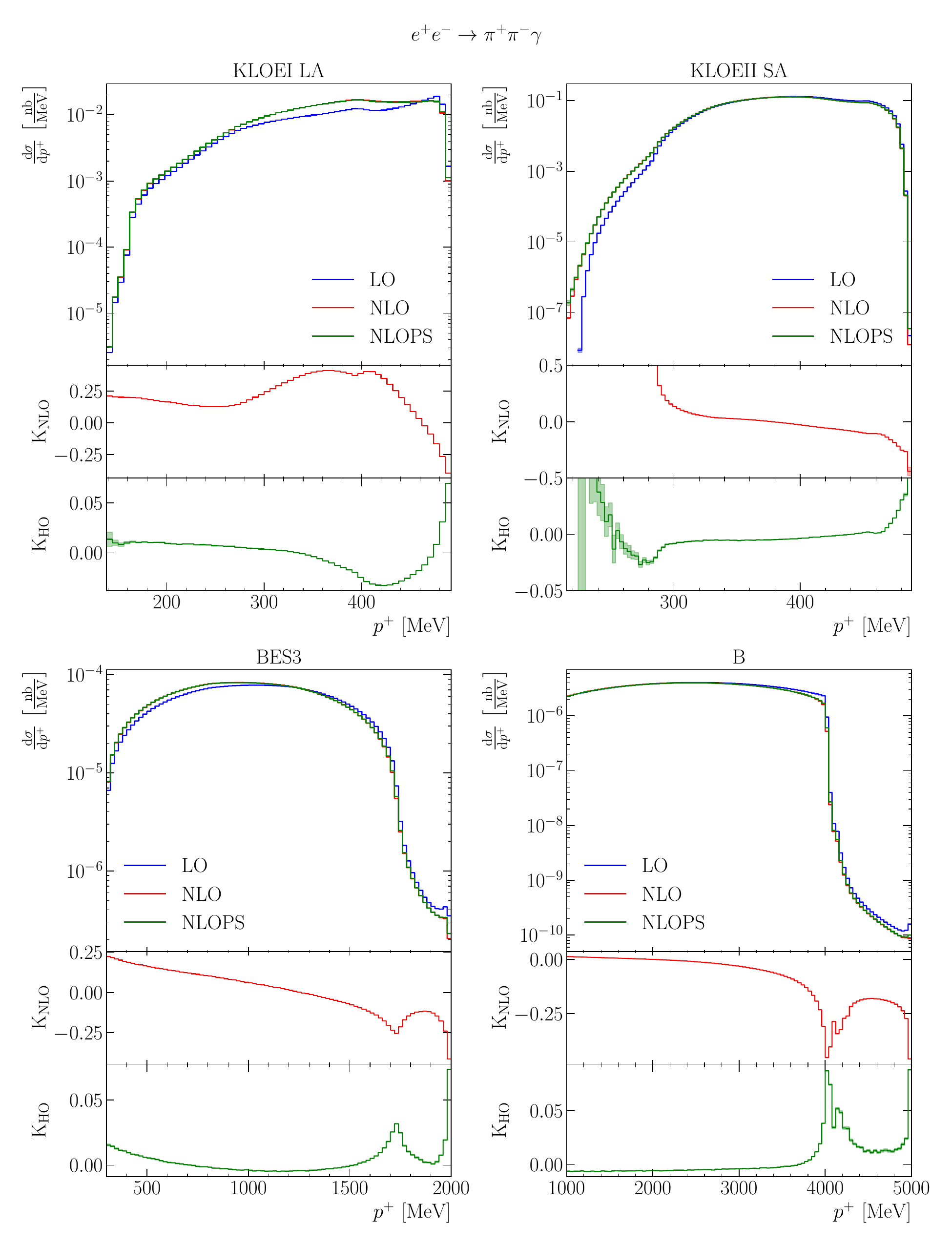} 

     \caption{The same as in Fig.~\ref{ap_mumu} for the $\pi^+\pi^-\gamma$ channel.}
    \label{ap_pipi}
\end{figure}

\begin{figure}[htbp]
    \centering
    \includegraphics[width=\linewidth]{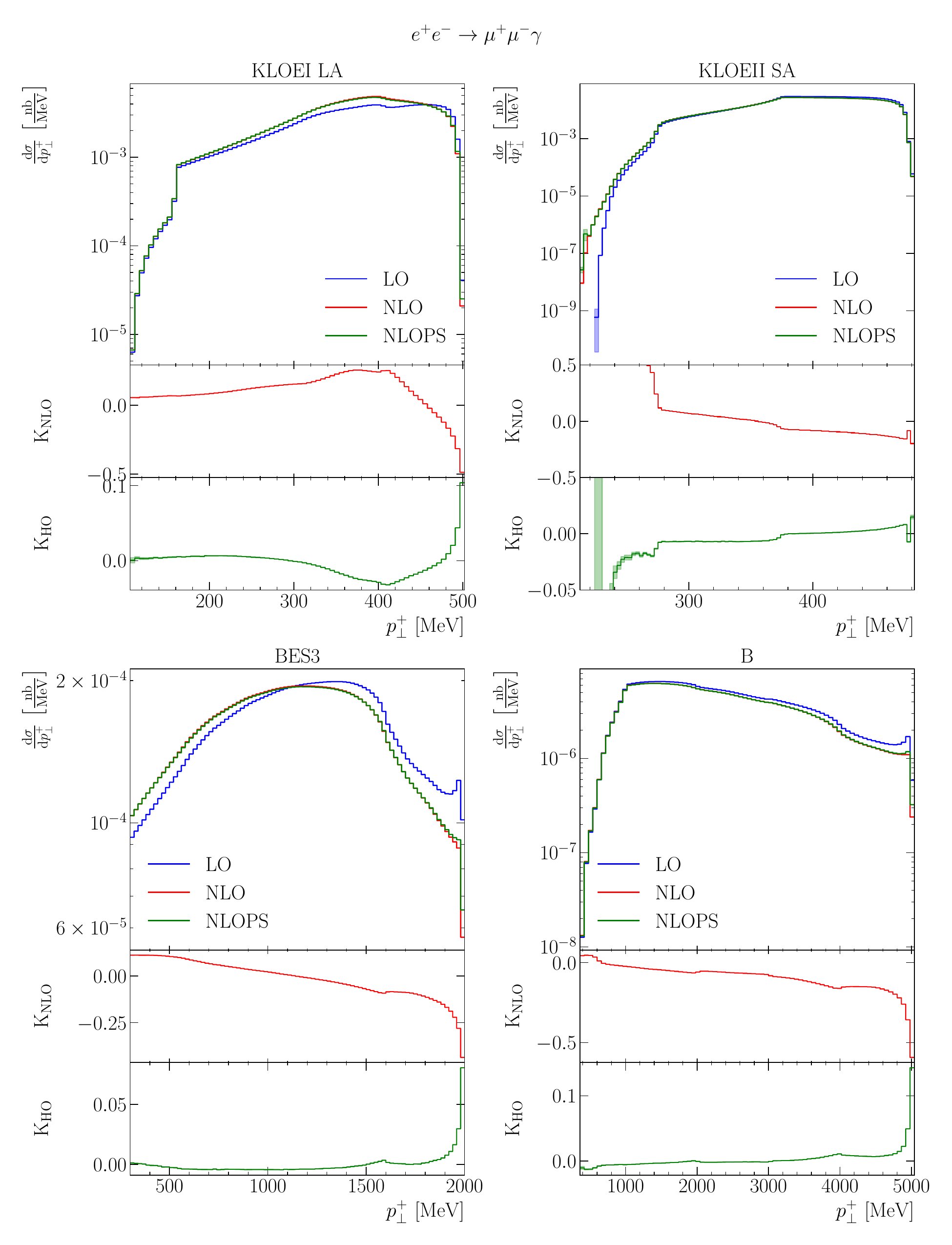}

     \caption{The differential cross section of the $e^+e^-\to\mu^+\mu^-\gamma$ process as a 
     function of the $\mu^+$ transverse momentum with different approximations for radiative corrections in the four scenarios 
     of Tab.~\ref{tab:scenarios}.}
    \label{pt_mumu}
\end{figure}

\begin{figure}[htbp]
    \centering
    \includegraphics[width=\linewidth]{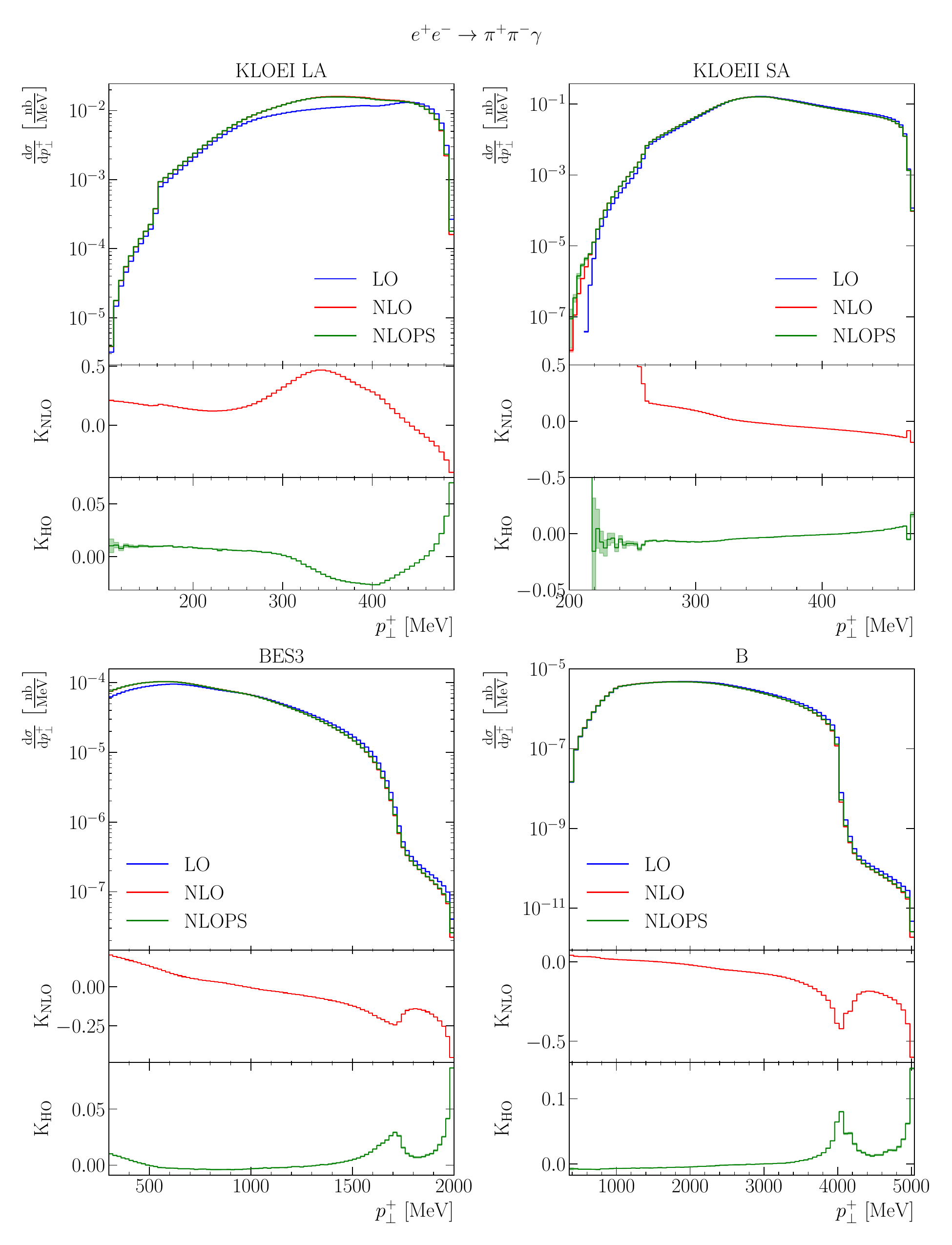} 
        
     \caption{The same as in Fig.~\ref{pt_mumu} for the $\pi^+\pi^-\gamma$ channel.}
    \label{pt_pipi}
\end{figure}

In this Appendix, we present further numerical results at NLOPS accuracy, by considering other observables that are experimentally relevant for radiative return measurements. In particular, we show predictions for higher-order corrections to the cross section distribution w.r.t. the modulus of the momentum of the charged tracks in Fig.~\ref{ap_mumu} (muon channel) and in Fig.~\ref{ap_pipi} 
(pion channel), as well as w.r.t. their transverse momentum in  Fig.~\ref{pt_mumu} (muon channel) and in Fig.~\ref{pt_pipi} 
(pion channel). It can be noticed that multiple photon corrections can reach the some percent level in specific phase-space regions and generally tend to compensate the NLO predictions in all the experimental scenarios, for both 
$e^+ e^- \to \pi^+\pi^-\gamma$ and 
$e^+ e^- \to \mu^+\mu^-\gamma$.

\FloatBarrier

\section{NLOPS predictions for 
$e^+ e^- \to X^+ X^-$}
\label{sec:app2}

As mentioned in 
Section~\ref{subsec:PS-match}, the treatment of 
hard exclusive emission here presented, as well as its matching to the PS algorithm, has been implemented in \textsc{BabaYaga@NLO} also for the simulation of the $2 \to 2$ processes
$e^+ e^- \to \pi^+ \pi^-$
and $e^+ e^- \to \mu^+ \mu^-$. 
As discussed in Section~\ref{subsec:PS-match},
this refinement improves the formulation developed in~\cite{Budassi:2024whw,Balossini:2006wc,Balossini:2008xr}, when 
considering observables that are sensitive to 
hard real radiation with associated resonant enhancements. 

In this Appendix we discuss the accuracy of the new  NLOPS formulation for $2\to2$ processes, both in energy scan scenarios (Sec.~\ref{app:sub_es}) and in radiative return setups (Sec.~\ref{app:sub_rr})

\subsection{Energy scan}
\label{app:sub_es}

\begin{figure}[htb]
    \centering
    \includegraphics[width=\imagewidth]{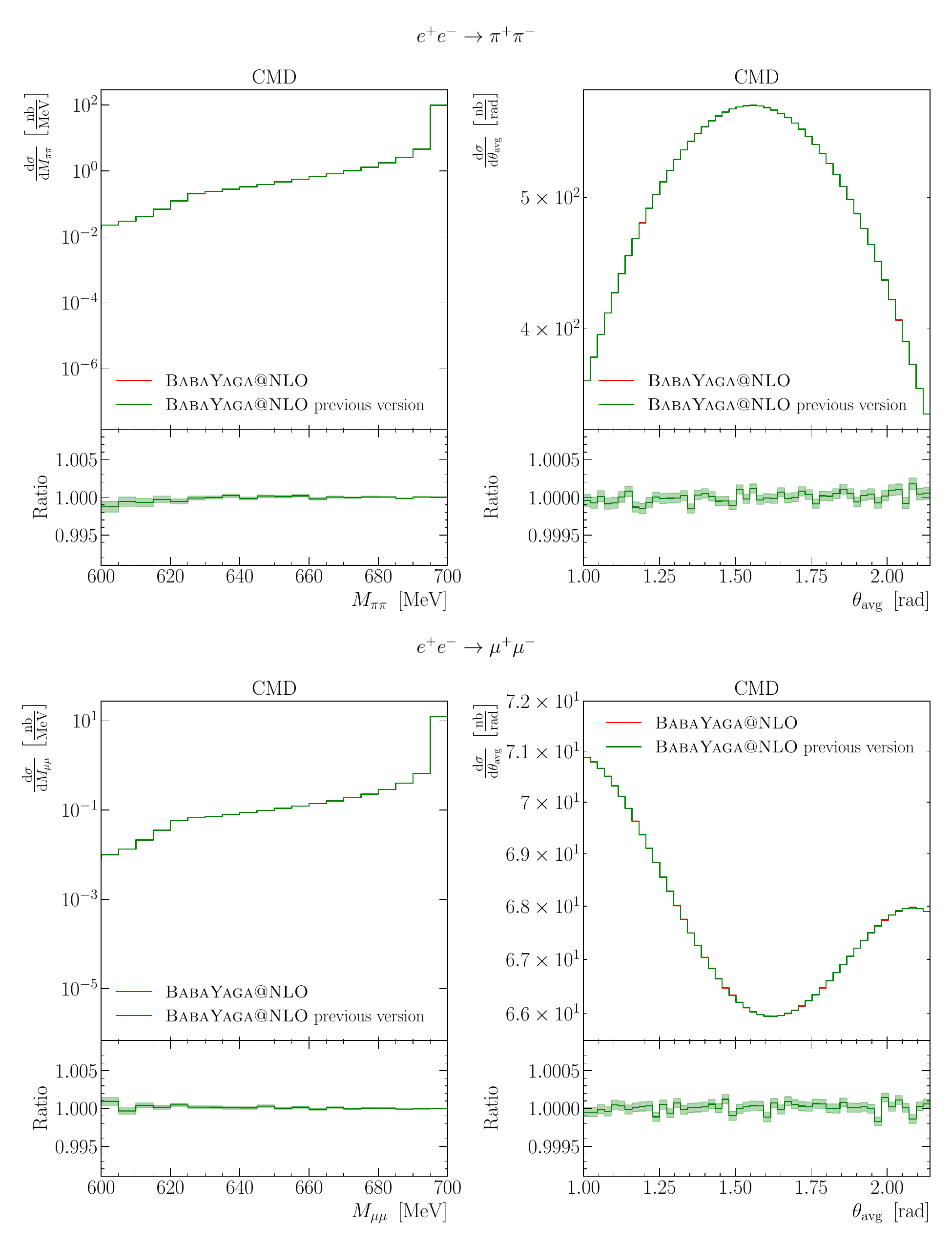}        \caption{The differential cross sections 
    of the $e^+ e^- \to \pi^+\pi^-$ process (top 
    panels) and of the $e^+ e^- \to \mu^+\mu^-$ 
    process (lower panels) 
    as 
    functions of the invariant mass of the charged pair (left plots) and of the average scattering angle $\theta_{\rm avg}$ (right plots), according to two different NLOPS formulations implemented in the code \textsc{BabaYaga@NLO}.
    The figure shows a comparison for 
    $2 \to 2$ observables between the NLOPS 
    formulation of this work ($\textsc{BabaYaga@NLO}$)
    and the approach of~\cite{Budassi:2024whw,Balossini:2006wc,Balossini:2008xr} ($\textsc{BabaYaga@NLO}$ previous version) in the CMD-like scenario of~\cite{Aliberti:2024fpq}. The ratio between the predictions of the two formulations is shown in the lower panels.}
\label{fig:nlops22_cmd3}
\end{figure}

First, it is crucial to check that the formulation of ~\cite{Budassi:2024whw,Balossini:2006wc,Balossini:2008xr} and the present approach provide the same NLOPS predictions, within the quoted theoretical accuracy (namely $0.1\%$), for $2 \to 2$ observables
in the presence of typical selection criteria of energy scan experiments, that are inclusive over photon radiation. For this purpose, we considered the CMD-like setup of~\cite{Aliberti:2024fpq} both for muon pair and two pion final state. We checked that the results of the two formulations for the integrated cross sections agree at the 
$10^{-5}$ level. As shown in Fig.~\ref{fig:nlops22_cmd3}, we also compared the differential cross sections of $e^+ e^- \to 
X^+ X^-$, $X = \{\pi, \mu\}$, as a function of the invariant mass and of the average scattering angle $\theta_{\rm avg}$.\footnote{The definition of $\theta_{\rm avg}$ can be found in \cite{Aliberti:2024fpq}.} As can be seen from 
Fig.~\ref{fig:nlops22_cmd3}, the predictions of the two NLOPS formulations for $2\to 2$ processes are in excellent agreement, below $0.1\%$.

The results of this investigation confirm that the NLOPS approach to $2\to 2$ processes implemented in the previous version of \textsc{BabaYaga@NLO} is well under control. Moreover, the present extension significantly improves the description of phase-space regions dominated by hard radiation with associated resonant effects.

\subsection{Radiative return}
\label{app:sub_rr}

Now we focus on event selection criteria of radiative return experiments and we consider our NLOPS predictions for $2\to 3$ processes in comparisons with those of the new formulation for
$2 \to 2$ processes. Note that, by construction, the application of the method to 
$2 \to 2$ processes differs from the formulation of $2 \to 3$ processes only at the level of NLO SV corrections.

\begin{figure}[!htbp]
    \centering 
    \includegraphics[width=\imagewidth]{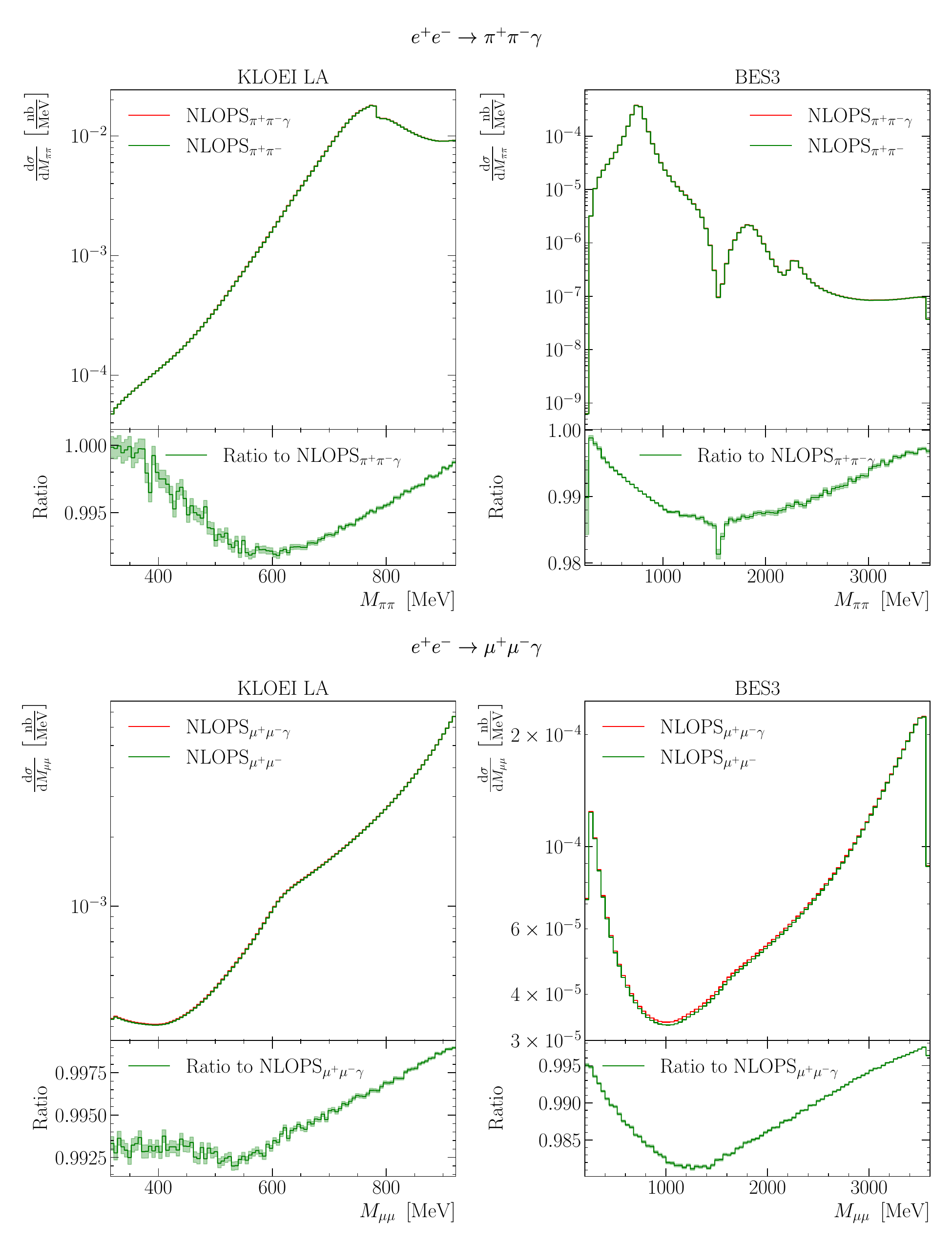}        
    \caption{The differential cross section as a 
         function of the invariant mass of the charged pair 
         at NLOPS accuracy according to the formulation presented in this work as applied to the processes $e^+ e^- \to \pi^+\pi^-\gamma$ vs. $e^+ e^- \to \pi^+\pi^-$ (top panels) and to
         $e^+ e^- \to \mu^+\mu^-\gamma$ vs. $e^+ e^- \to \mu^+\mu^-$ (bottom panels). The ratio between the predictions of the two formulations is shown in the lower panels, for the KLOEI LA (left) and BES3 (right) scenarios of Tab.~\ref{tab:scenarios}.}
\label{nlops2to2}
\end{figure}

The results of this investigation are shown in Fig.~\ref{nlops2to2} 
for the differential cross sections as a function of the invariant mass of the charged pair at NLOPS accuracy for the $2 \to 3$ processes (${\rm NLOPS}_{X^+X^-\gamma}$) in comparison with the same prediction for the 
$2 \to 2$ processes  (${\rm NLOPS}_{X^+X^-}$), with 
$X = \pi$ (left panel) and $X = \mu$ (right panel). For definiteness, we present results for the KLOEI LA and BES3 scenarios. As can be seen from 
Fig.~\ref{nlops2to2}, the NLOPS formulation of this work applied to $2 \to 2$ processes provides results in reasonable agreement with the predictions for the $2 \to 3$ processes. The deviations between the two descriptions are visible in the lower panels of Fig.~\ref{nlops2to2} and come from the different definition of the SV matching factor $F_{\rm SV}$, which is evaluated on the $2\to2$ kinematics for ${\rm NLOPS}_{X^+X^-}$ 
and on the $2\to3$ kinematics for ${\rm NLOPS}_{X^+X^-\gamma}$. In particular, it can be noticed that the differences between the two matching procedures 
are at the percent level in the KLOEI LA setup, where the tagged photon is
relatively soft, but are more pronounced in the BES3 scenario, due to the impact of non-factorisable SV contributions associated to hard tagged photons.

On the other hand, this comparison clearly shows that accurate NLOPS predictions for radiative return signatures, with a hard photon in the final state, necessarily require the inclusion of multiple photon contributions on top of the exact NLO corrections to $2 \to 3$ processes. Indeed, both for KLOEI LA and BES3 scenario, 
the $2\to2$ NLOPS approach can not model the invariant mass distributions in most part of the spectrum with sufficiently good accuracy and is therefore inadequate for sub-percent measurements of the pion form factor in radiative return experiments. On the grounds of these considerations and the evaluation of higher-order corrections addressed in Section~\ref{sec:subsec-numerical}, one can conclude that both NLO and  
multi-photon corrections to $e^+ e^- \to X^+ X^-\gamma$ processes are needed for radiative return simulations with per mille level accuracy.

\FloatBarrier

\FloatBarrier

\section{Phase-space parametrisation and MC importance sampling}
\label{sec:app_ps_and_is}
In this Appendix, we detail the multi-photon phase-space
parametrisation and the importance sampling of the independent variables.

The peaks of the multi-differential cross section
originate from the emission of soft and/or collinear (to the charged
particles) photons or from the behaviour of the effective internal photon
propagator, which can have enhancements at low $Q^2$, when hadronic
vacuum polarisation effects
are included or when $F_\pi(Q^2)\ne 1$. When sampling the independent
variables from which the generated momenta are derived, without loss
of generality we try to follow at our best the absolute squared value
of all possible matrix elements, neglecting any interference among
them. Needless to say, this is a general strategy and any other choice
is equivalent from the MC point of view: different alternatives impact
only on the rate of convergence of
the MC integration or the generation efficiency of unweighted events.

In the following, we indicate as $m_e = m_1 = m_2$ and
$m_F = m_3 = m_4$ the masses of the initial-state ($p_1$ and $p_2$)
and final-state momenta ($p_3$ and $p_4$) respectively.

In view of a multi-channel importance sampling~\cite{Jadach:1999bv},
once the photons' multiplicity $n$ is selected (the sampling of $n$ is
described at the end of this Appendix), we indicate with
$\mathcal{P} =\{k_1,\cdots,k_n\}$ the set of their four-momenta $k$ and we
consider all the $2^n$ subsets $S_{I}$ of $\mathcal{P}$ and their
complement $S_{F} = \mathcal{P} \setminus S_{I}$. We interpret the momenta
in the sets $S_{I}$ ($S_{F}$) as attached to initial(final)-state charged
currents. We then multiply
Eq.~\eqref{eq:phsp} by $1$, in a form where the MC channels are
readily read:
\begin{align}
1 & = \frac{\mathcal{N}}{\mathcal{D}}
= \frac{\sum_\lambda\mathcal{N}_\lambda}{\mathcal{D}} = \frac{\sum_{S_{I} \subseteq \mathcal{P}} \mathcal{C}_{I}(S_{I})\; \mathcal{C}_{F}(S_{F})}{\sum_{S_{I} \subseteq \mathcal{P}} \mathcal{C}_{I}(S_{I})\; \mathcal{C}_{F}(S_{F})}\,,\label{eq:much1}\\
\mathcal{C}_{I}(S_{I}) & = \frac{1}{n_I}\sum_{k\in S_{I}} \mathcal{F}(k,Q_{S_{I}},m_F)
\prod_{k\in S_{I}}\left[\mathcal{R}\left(p_1,k\right)
+ \mathcal{R}\left(p_2,k\right)\right]\,, \label{eq:much2}\\
\mathcal{C}_{F}(S_{F}) & = \prod_{k\in S_{F}}\left[\mathcal{R}\left(p_3,k\right) + \mathcal{R}\left(p_4,k\right)\right]\,, \label{eq:much3}\\
\mathcal{R}\left(p_i,k\right) & = \frac{E_i}{p_i\cdot k}
= \frac{1}{k^0}\;\frac{1}{1-\beta_i\cos{\measuredangle(\vec{p}_i,\vec{k})}}\qquad\beta_i
= \sqrt{1-\frac{m_i^2}{E_i^2}}\,.\label{eq:much4}
\end{align}
In the previous definitions, $E_i$ are the energies of $p_i$, $n_I$ is
the number of photons in the set
$S_{I}$, $Q_{S_{I}} = p_1 + p_2 - \sum_{k\in S_{I}} k = p_3 + p_4 + \sum_{k\in S_{F}} k$ and the
function $\mathcal{F}(k,Q,m)$ reads:
% ro = (q+k)^2
\begin{align}
\mathcal{F}(k,Q,m) = & \left(c_0\frac{\sqrt{1-\frac{4m^2}{Q^2}}}{Q^2}
+ \sum_{i=1}^{n_\text{res}}c_i\;\frac{{\bar{g}}^2_iM_i^2}{\left(Q^2-M^2_i\right)^2 + M_i^2\Gamma^2_i}\right)\;\mathcal{H}(k,Q)\,,\nonumber\\
\mathcal{H}(k,Q) = & \frac{k^0}{\left(Q+k\right)^2 - Q^2}\;\left[\left(Q+k\right)^2\right]^{\frac{3}{2}}\,.
\label{eq:Freg}
\end{align}
In $\mathcal{F}(k,Q,m)$ the coefficients $c_i$ are set to $0$ or
$1$ according to the running conditions ({\it i.e.} if
$e^+e^-\to\mu^+\mu^-\gamma(+n\gamma)$ or
$e^+e^-\to\pi^+\pi^-\gamma(+n\gamma)$ is simulated, if hadronic
vacuum polarisation effects are switched on or if $F_\pi(Q^2) \ne 1$)
and the sum over the Breit-Wigner functions (with appropriate
``couplings'' ${\bar{g}}^2_i$) is inserted to flatten
the spikes in the differential cross sections due to hadronic resonances.

We would like to
stress that $\mathcal{F}$ takes care of the flattening of
enhancements due to the effective internal
photon propagator and the infrared singularity of one photon from the
initial-state (in Eq.~\eqref{eq:much2} the sum over initial-state
photons is then taken). We notice that, if $\mathcal{F} = 1$, the
numerator $\mathcal{N}$ and the
denominator $\mathcal{D}$  of Eq.~\eqref{eq:much1} reduce to the much simpler form
\begin{align}
\mathcal{N}_{\mathcal{F} = 1} & = \mathcal{D}_{\mathcal{F} = 1}
= \prod_{k\in\mathcal{P}}\left(\frac{E_1}{p_1\cdot k}
+ \frac{E_2}{p_2\cdot k} + \frac{E_3}{p_3\cdot k}
+ \frac{E_4}{p_4\cdot k} \right)\nonumber \\
& = \prod_{i=1}^n\left(\frac{E_1}{p_1\cdot k_i}
+ \frac{E_2}{p_2\cdot k_i} + \frac{E_3}{p_3\cdot k_i} + \frac{E_4}{p_4\cdot
k_i} \right) = \prod_{i=1}^n\left(\frac{1}{k^0_i}\sum_{j=1}^4 \frac{1}{1-\beta_j\cos{\measuredangle(\vec{p}_j,\vec{k}_i)}}\right)\,,
\label{eq:simplerform}
\end{align}
in which the soft and collinear singularities for all the photons
emitted from all the charged particles are manifest at a glance.

Every single term in the numerator of Eq.~\eqref{eq:much1}, identified
by the index $\lambda$, defines a single channel for the importance
sampling, to which we associate the MC weight
$\mathcal{N}_\lambda/\mathcal{D}$. The sum over $\lambda$ is done as a
MC sum. In order to perform it and to choose randomly a channel, we attach with
equal probability each of the $n$ photons to $p_1$, $p_2$, $p_3$ or
$p_4$ (this associates $n_I$ photons to the initial state, {\it i.e.}
emitted by $p_1$ or $p_2$, and $n_F$ to the final state, {\it i.e.}
emitted by $p_3$ or $p_4$, with $n_I + n_F = n$). We then pick with
probability $1/n_I$ one of the initial state photons, labelling it
with the index $\bar{l}$. This procedure ensures that we select with
equal probability a single term resulting from unfolding products and
sums implied in
$\sum_{S_{I} \subseteq \mathcal{P}} \mathcal{C}_{I}(S_{I})\; \mathcal{C}_{F}(S_{F})$
of Eq.~\eqref{eq:much1}.

Once one channel is selected, we split the phase-space for the
reaction $p_1 + p_2 \to p_3 + p_4 + \sum_{i=1}^n k_i$ in
two subsequent decays, $p_1+p_2 \to Q + \sum_{i=1}^{n_I}k_i$ and
$Q\to p_3 + p_4 + \sum_{j=1}^{n_F}k_j$, by inserting the identity
\begin{equation}
1 = \dd^4 Q\; \delta^{(4)}\left(Q - p_3-p_4 - \sum_{j=1}^{n_F} k_j\right)
= \dd Q^2\; \frac{\dd^3 \vec{Q}}{2Q^0}\; \delta^{(4)}\left(Q - p_3-p_4 - \sum_{j=1}^{n_F}
k_j\right)\,,
\label{eq:phspinsertion}
\end{equation}
into Eq.~\eqref{eq:phsp}. The latter (omitting the flux factor and all
the $2\pi$ factors for the sake of clarity) can therefore be cast in
the form
\begin{align}
\dd\Phi_n(\{p\},\{k\}) \;
=\; & \frac{\dd^3 \vec{p}_3}{2E_3}\; \frac{\dd^3 \vec{p}_4}{2E_4}\;
\prod_{i=1}^{n_I}\frac{\dd^3\vec{k}_i}{2\omega_i}\;
\prod_{j=1}^{n_F}\frac{\dd^3\vec{k}_j}{2\omega_j}\;
\delta^{(4)}\left(P - p_3 - p_4 - \sum_{l=1}^n k_l\right) = \nonumber \\
& \dd Q^2\;  \frac{\dd^3 \vec{Q}}{2Q^0}\;
\prod_{i=1}^{n_I}\frac{\dd^3\vec{k}_i}{2\omega_i}\;
\delta^{(4)}\left(P - Q - \sum_{i=1}^{n_I} k_i\right)
\label{eq:phspwQ1}\\
&\times\;\frac{\dd^3 \vec{p}_3}{2E_3}\;\frac{\dd^3 \vec{p}_4}{2E_4}\;
\prod_{j=1}^{n_F}\;\frac{\dd^3\vec{k}_j}{2\omega_j}\;
\delta^{(4)}\left(Q - p_3-p_4 - \sum_{j=1}^{n_F} k_j\right)\,.
\label{eq:phspwQ2}
\end{align}
In the previous equations, we defined $P = p_1 + p_2$ and
$Q^0 = \sqrt{Q^2 + |\vec{Q}|^2}$. The idea underlying this
phase-space factorization is to keep under control with the importance sampling of $Q^2$ also the propagator
of the internal photon\footnote{This feature is an improvement over
the version of \textsc{BabaYaga@NLO} of Ref.~\cite{Balossini:2006wc},
where it was not implemented.}, whose momentum is
$Q=p_1+p_2-\sum_{i=1}^{n_I} k_i =  p_3+p_4+\sum_{j=1}^{n_F} k_j$.
The last two lines Eqs.~\eqref{eq:phspwQ1}
and~\eqref{eq:phspwQ2} are
separately and manifestly Lorentz invariant, a property that we exploit.

In order to perform the multi-dimensional phase-space integral, we
choose as independent integration
variables $Q^2$, $\omega_i$ for $i = 1,\dots,\bar{l}-1,\bar{l}+1,\dots,n_I$,
two angles for each of the $n_I$ photons, $\omega_j$ for
$j=1,\dots,n_F$, two angles for each of the $n_F$ photons and the
angles of $\vec{p}_3$. The rest of the differentials are eliminated by
using the momentum conserving 4-dimensional Dirac deltas. With this
choice, we resolve the part~\eqref{eq:phspwQ1} of
$\dd\Phi(\{p\},\{k\})$ in the c.m. reference frame (where
$P = p_1 + p_2$ is at rest), where it becomes
\begin{equation}
\eqref{eq:phspwQ1}
= \frac{1}{2^{1+n_I}}\frac{\omega_{\bar{l}}}{Q^0
+ \omega_{\bar{l}}+\vec{\kappa}\cdot \vec{v}_{\bar{l}}} \dd
Q^2\dd\Omega_{\bar{l}}\prod_{i=1, i\ne\bar{l}}^{n_I}\omega_i\dd\omega_i \dd\Omega_{i}\,.
\label{eq:phspCMSbit}
\end{equation}
In the above expression, in terms of the independent
integration variables we define $\dd\Omega_i = \dd\cos\theta_i\dd\phi_i$, $\vec{v}_{\bar{l}}$ is a
unit vector directed along the $\bar{l}^\text{th}$ photon (such that
$\vec{k}_{\bar{l}} = \omega_{\bar{l}}\;\vec{v}_{\bar{l}}$),
$\kappa = \sum_{i=1,i\ne\bar{l}}^{n_I} k_i$, $\vec{Q}=-\sum_{i=1}^{n_I} \vec{k}_i$ and
the energy $\omega_{\bar{l}}$ is given by
\begin{equation}
\omega_{\bar{l}} = \frac{1}{2}\frac{\left(P-\kappa\right)^2 -
Q^2}{\sqrt{P^2}-\kappa^0 + \vec{\kappa}\cdot\vec{v}_{\bar{l}}}\,.
\label{eq:omegalbar}
\end{equation}

Concerning the part~\eqref{eq:phspwQ2} of $\dd\Phi(\{p\},\{k\})$ we
decide to express it in the so-called {\it comoving reference frame}
(CRF)~\cite{StaszekTorinoLectures,Jadach:1999vf,Schonherr:2008av,Boselli:2015aha},
which is the reference frame where $p_3+p_4$ is at rest, {\it i.e.}
where $p_3$ and $p_4$ are back-to-back. The choice is dictated by the
ease of dealing with the sampling of collinear enhancements for multi-photon
emissions from $p_3$ and $p_4$~\footnote{In the original version
of \textsc{BabaYaga@NLO} of Ref.~\cite{Balossini:2006wc} we adopted a different
solution to deal with collinear singularities for emissions from the
final state charged particles, based on a properly chosen
multi-channel strategy.}. We follow our implementation of
the phase-space in the CRF described in Appendix A of Ref.~\cite{Boselli:2015aha}.

In the CRF, we can write
\begin{equation}
\eqref{eq:phspwQ2}
= \frac{\beta_3}{2^{1+n_F}}\;\frac{\left(p_3+p_4\right)^2}{Q^2}\;\frac{1}{1 + \frac{K_F^0}{\sqrt{\left(p_3+p_4\right)^2}}}\;\dd\Omega_3\;\prod_{j=1}^{n_F}\omega_j\dd\omega_j\dd\Omega_j\,,
\label{eq:phspCRFbit}
\end{equation}
where $K_F=\sum_{j=1}^{n_F}k_j$, $\beta_3 = \sqrt{1-m_F^2/E_3^2}$ with
$E_3=\frac{1}{2}\left(\sqrt{Q^2 + |\vec{K}_F|^2} - K_F^0\right)$,
$\dd\Omega_3=\dd\cos\theta_3\dd\phi_3$ and $\dd\Omega_j
= \dd\cos\theta_j\dd\phi_j$. We finally re-express the last equation by performing a change of the
integration variables from $\omega_j$ to a new set of variables 
$y_j =\frac{x_j}{1 + \sum_l x_l}$. With this set of variables, we
define $V = \sum_{j=1}^{n_F} x_j\;v_j$ where $v_j$ are light-like
vectors with $v_j^0=1$ directed along the direction of the
$j^\text{th}$ photon and $x_j
= \frac{y_j}{1-\sum_ly_l}$. Equation~\eqref{eq:phspCRFbit} is then equivalent to
\begin{equation}
%\frac{\beta_3}{2^{1+2n_F}}\;
8\beta_3
\frac{\left(Q^{2}\right)^{n_F}}{\left(4+ V^2 +
4V^0\right)^{2+n_F}}\;\frac{1+V^0}{\left(1-\sum_ly_l\right)^{n_F}}\;\dd\Omega_3\prod_{j=1}^{n_F} \frac{y_j}{1-\sum_ly_l}\dd y_j\dd\Omega_j\,.
\label{eq:phspCRFbit2}
\end{equation}
With the above definitions, $E_3$ entering $\beta_3$ can be written as
$E_3 = \sqrt{\frac{Q^2}{4 + V^2+4V^0}} $. The advantage of
Eq.~\eqref{eq:phspCRFbit2} is that the variables
$y_j$ lie in the range $[0,1]$ and the maximum value for $\omega_j$ in
the CRF does not need to be known a priori, at the cost of a slightly more
involved expression. We stress that in Eqs.~\eqref{eq:phspCRFbit} and
~\eqref{eq:phspCRFbit2} all variables must be intended as expressed in
the CRF and that, once generated, all the momenta must be boosted back in
the c.m. reference frame.

We finally describe how the independent variables are sampled to
integrate over the phase-space. We write for short 
``$x \propto f(x)$'' by meaning that we distribute $x$ according to the
function $f(x)$; more explicitly, when we want to integrate over the
variable $x$ by importance sampling we transform the differential
$\dd x$ as $\dd x = \frac{1}{f(x)}\;f(x)\;\dd x = \frac{\dd y}{f(x)}$, {\it i.e.} we change the
integration variable from $x$ to $y$ and we multiply the differential
cross-section by the weight
$1/f(x)$. In doing so, we use standard MC techniques
(inversion of cumulative distribution function, {\it hit or miss},
multi-channel or combinations of them).

For definiteness, we set $p_1$ directed along the positive $z$ axis in the c.m.
reference frame and $p_2$ along the negative $z$ axis. In the
part~\eqref{eq:phspCMSbit} of $\dd\Phi_n\left(\{p\},\{k\}\right)$ that we
resolve in the c.m. reference frame, we first generate the angles of
the $n_I$ photons as ($\beta_e = \sqrt{1-m_e^2/E_1^2}$)
\begin{align}
\phi_i & \propto 1\qquad\phi_i\in [0,2\pi]\nonumber\\
\cos\theta_i & \propto \frac{1}{1\pm\beta_e\cos\theta_i}\qquad\text{[\;$-$
($+$) if $i$ is attached to $p_1$ ($p_2$)\;]}\qquad \cos\theta_i\in [-1,1]\,.
\label{eq:kisangles}
\end{align}
We generate the energies of the $n_I-1$ photons (all the
initial-state photons except the $\bar{l}^\text{th}$ one) as ($s =(p_1 + p_2)^2$)
\begin{align}
\omega_i \propto \frac{1}{\omega_i}\qquad\omega_i\in
[\varepsilon\;\sqrt{s}/2, \sqrt{s}/2] = [\varepsilon\; E_1, E_1]\,.
\label{eq:kisenergies}
\end{align}
We can therefore calculate $\kappa$ and $\vec{v}_{\bar{l}}$ as defined
above Eq.~\eqref{eq:omegalbar} and we notice that, in order to catch
the $1/\omega_{\bar{l}}$ infrared singularity, we should sample
$Q^2\propto\frac{1}{(P-\kappa)^2 - Q^2}$ (see the relation between
$Q^2$ and $\omega_{\bar{l}}$ in Eq.~\eqref{eq:omegalbar}). With this
in mind and considering that $Q^2$ must sample also the internal
photon propagator, we choose to distribute $Q^2$ as 
\begin{align}
& Q^2  \propto \frac{1}{\rho - Q^2}\left( c_0\;\frac{\sqrt{1-\frac{4m_F^2}{Q^2}}}{Q^2}+  \sum_{i=1}^{n_\text{res}}c_i\;\frac{{\bar{g}}_i^2M_i^2}{\left(Q^2-M_i^2\right)^2 +
M_i^2\Gamma_i^2}\right)\qquad Q^2\in [Q^2_\text{min},Q^2_\text{max}]\nonumber\\
& Q^2_\text{min} = \min(4m_F^2,\rho-\xi\sqrt{s},Q^2_\text{cut})\qquad Q^2_\text{max}
= \rho - \varepsilon\xi\sqrt{s}\,.
\label{eq:Q2sampling}
\end{align}
where $\rho = (P-\kappa)^2$,
$\xi = \sqrt{s}-\kappa^0+\vec{\kappa}\cdot\vec{v}_{\bar{l}}$ and
$Q^2_\text{cut}$ is a user-defined cut directly on $(p_3+p_4)^2$ 
or derived from other user-defined cuts. We stress that the sampling
of $Q^2$ in Eq.~\eqref{eq:Q2sampling} factorises also the IR singularity
for the $\bar{l}^\text{th}$ photon.

Moving to the part~\eqref{eq:phspCRFbit} of
$\dd\Phi_n\left(\{p\},\{k\}\right)$ that we make explicit in the CRF,
we firstly assume that $p_3$ ($p_4$) is directed along the positive
(negative) $z$ axis. We generate for the $n_F$ photons attached
to $p_3$ or $p_4$ ($\bar{\beta} = \sqrt{1-4m_F^2/Q^2}$)
\begin{align}
\phi_j & \propto 1\qquad\phi_j\in [0,2\pi]\nonumber\\
\cos\theta_j & \propto \frac{1}{1\pm\bar{\beta}\cos\theta_j}\qquad\text{[\;$-$
($+$) if $j$ is attached to $p_3$ ($p_4$)\;]}\qquad \cos\theta_j\in
[-1,1]\nonumber\\
y_j & \propto
\begin{cases}
{1}/{y_j}\quad &\text{if $y_j \ge \varepsilon$}\\
{1}/{\varepsilon}\quad &\text{if $y_j < \varepsilon$}
\end{cases}
\qquad y_j\in [0,1]\,,
\label{eq:kfsvars}
\end{align}
and we build the light-like vectors $v_j$ and the vector $V$ defined above
Eq.~\eqref{eq:phspCRFbit2}. They allow to calculate
\begin{equation}
s_{34} = (p_3+p_4)^2 = \frac{4Q^2}{4 + V^2 + 4V^0}\,,
\label{eq:s34crf}
\end{equation}
which in turn is used to calculate the energy of $p_3$ and $p_4$ as
$E_{3} = E_4 = \frac{1}{2}\sqrt{s_{34}}$ and the final-state photon energies as
$\omega_j =\frac{\sqrt{s_{34}}}{2}\; \frac{y_j}{1-\sum_l y_l}$. Since we started by
assuming $p_3$ along the $z$ axis, we generate its true angles as
\begin{align}
\phi_3 & \propto 1\qquad\phi_3\in [0,2\pi]\nonumber\\
\cos\theta_3 & \propto
\begin{cases}
g(\cos\theta_3)\quad &\text{if $c_\text{min}\le \cos\theta_3\le c_\text{max}  $}\\
\bar{a}\quad &\text{if $\cos\theta_3 < c_\text{min}  $}\\
\bar{b}\quad &\text{if $\cos\theta_3 > c_\text{max}  $}
\end{cases}
\qquad \cos\theta_3\in [-1,1]\,,
\label{eq:p3varcrf}
\end{align}
where $g(c) = 1+c^2$ for muons and $g(c) = 1-c^2$ for pions. 
Finally we align $p_3$ in the direction defined by $\phi_3$ and
$\cos\theta_3$ and we rotate also $p_4$ and the $n_F$ photons in the
same rotated
coordinate system. Once all the momenta are known in the CRF frame we
boost them back in the c.m. reference frame.

One comment is in order concerning Eq.~\eqref{eq:p3varcrf}:
typically, at generator level the angles of $p_3$ and $p_4$, or
equivalently their cosines, are restricted to lie in a user-defined range
$[\theta_\text{min},\theta_\text{max}]$ or $[c_\text{min},c_\text{max}]$,
of course in the laboratory frame (which coincides here with the c.m.
frame). Since the boost going to the CRF is unknown
until all momenta are generated, we decide in the CRF to span all the
range $[-1,1]$ sampling as $g(c)$ inside $[c_\text{min},c_\text{max}]$
and flat outside it, in proportion to two constants $\bar{a}$ and
$\bar{b}$ tuned for efficiency. This ensures that we cover all the phase-space, being
understood that if in the c.m. frame $p_3$ or $p_4$ fall
outside the allowed ranges we reject the event, or equivalently we set
its weight to zero. The same comment applies to the sampling of $y_j$
in Eq.~\eqref{eq:kfsvars}: since we want all the photons in the c.m.
frame with energy $\omega >\varepsilon\;\frac{\sqrt{s}}{2}$, if, once boosted
back in the c.m. frame, any of the final-state photons does not fulfill
this condition the event is rejected.

We stress that, in the spirit of the multi-channel importance
 sampling, the cascade of samplings from
Eq.~\eqref{eq:kisangles} to Eq.~\eqref{eq:p3varcrf} aims to flatten as
much as possible one of the terms $\mathcal{N}_\lambda$ in the numerator
 $\mathcal{N}$ of Eq.~\eqref{eq:much1}. The flattening of the differential cross section
 is then taken care of by the presence of the denominator $\mathcal{D}$.

We conclude by describing the sampling of the photons' multiplicity
$n$. In LL approximation, the emission of extra photons from a hard
scattering process is regulated by a Poisson
distribution. With this feature in mind, we sample the discrete
variable $h = n -1$ (we use $h\ge 0$ because our hard scattering already has a photon
in the final state) as
\begin{align}
h & = n - 1  \propto \frac{\mu^h}{h!}e^{-\mu}\qquad h\in [0,\infty]\nonumber\\
\mu &
= \frac{\alpha}{\pi}\left[\log\frac{s}{m_e^2}\int_{0}^{1-\varepsilon}\dd
x\; P_f(x) + \log\frac{s}{m_F^2}\int_{0}^{1-\varepsilon}\dd x\;
P_r(x)\right]\quad
\begin{array}{ll}
&\text{$r=f$ for muons}\\
&\text{$r=s$ for pions}
\end{array}
\,.
\label{eq:nsampl}
\end{align}

\section{Clustering and mapping algorithms}
\label{sec:mapping}
We provide a brief description of the phase-space mapping which is 
implemented in the new version of the 
\textsc{BabaYaga@NLO} code. As shown in the main text, the PS algorithm requires mapping the $e^+e^-\to X^+X^- n \, \gamma $ kinematics into $e^+e^-\to X^+X^- m \, \gamma $, with $m<n$. Our goal is to retain as much as possible the soft and collinear limit for the photon emissions, in order to preserve the regime of the LL approximation. We use the same convention of Eq.~\eqref{eq:momenta_born} to indicate the charged particles momenta, while we label with $k_i$ the momentum of the $i-$th photon. By means of some algorithm,  we identify the $m$ \textit{hard} photons which we aim to describe with the exact $2\to 2+m\gamma$ matrix element. The remaining $n-m$ photons are then \textit{clustered} to the parent charged particle, in principle identified by means of an algorithm which can be different from the one used to pinpoint the \textit{hard} photons.
The CKKW algorithm, as detailed in 
Section~\ref{sec:PS}, is used 
for both purposes in the present implementation.
In Tab.~\ref{ALG:clustering} we give an operative description of the clustering algorithm.

\begin{table}[htbp]
\centering
\begin{tabularx}{\textwidth}{X}
\toprule
\textbf{Clustering algorithm}
Out of $n$ photons, cluster $n-m$ photons on the respective closest charged leg according to a CKKW-like algorithm\\
\midrule
\vspace{-5mm}
\setlength{\abovedisplayskip}{4pt}
\setlength{\belowdisplayskip}{4pt}
\setlength{\abovedisplayshortskip}{2pt}
\setlength{\belowdisplayshortskip}{2pt}

\newpage
  \begin{itemize}[itemsep=8pt,leftmargin=0pt]
  \item[] \textbf{Input}   \begin{tabular}[t]{@{}l l@{}}
    & $\{k_i\}_{i=1}^n$ momenta of the photons \\
    & $\{p_j\}_{j=1}^4$ momenta of the charged legs \end{tabular}
  \begin{itemize}[itemsep=8pt]
    \item[] \textbf{For} each photon $i=1,\dots,n$ and charged leg $j=1,\dots,4$
      \begin{itemize}[itemsep=4pt]
    \item[]\textbf{Compute} the CKKW-like distance between photon $i$ and charged particle $j$
        \[
          y_{ij}=\min\{k_{0,i}^2,p_{0,j}^2\}(1-\cos\theta_{ij})
        \]
        where $\theta_{ij}$ is the relative angle between $\vec{k}_i$ and $\vec{p}_j$
        
         \item[] \textbf{Determine} the minimum distance
        \[
          d(k_i)=\min_j \{{y_{i j}}\} 
         \]
         and the parent parton of the $i$-th photon defined as the charged leg for which $y_{i j}$ is minimised
  
      \end{itemize}
    \item[] \textbf{Order} photon indices $i$ such that
    \[
     d(k_1)\ge d(k_2) \ge \dots\ge d(k_m)\geq \dots\geq d(k_n)\]
     
     \item[] \textbf{Identify}
     \begin{tabular}[t]{@{}l l@{}}
    & $\{K\}=\{k_i\}_{i=1}^m$  the first $m$ photons as \textit{hard}\\
    & $\{k_j\}_{j=m+1}^n$ the remaining $n-m$ photons as \textit{soft}
  \end{tabular}
   \item[] \textbf{For} each leg $J$ 
   \begin{itemize}[itemsep=4pt]
      
       \item[] \textbf{Count} the number $n_{J}$ of \textit{soft} photons which share $p_J$ as parent parton
       
       \textbf{Define} the cluster momentum as
\[
      P_J = p_J +\delta_J\sum_{\lambda=1}^{n_J} k_{\alpha_\lambda} 
    \]       
    where $\delta_J=\pm1$ for final/initial state particles respectively and $k_{\alpha_\lambda} $ denote the set of soft photon momenta whose parent parton is $J$.

\end{itemize}
   \end{itemize}
    \item[] \textbf{Output}
   \begin{tabular}[t]{@{}l l@{}}
    & $\{K_I\}_{I=1}^m$ momenta of the \textit{hard} photons \\
    & $\{P_J\}_{J=1}^4$ momenta of the \textit{clusters}
  \end{tabular}
  \vspace{5mm}
\end{itemize}\\
\bottomrule
\end{tabularx}
\caption{Pseudo-algorithm describing the implementation of the clustering
procedure in the latest version of \textsc{BabaYaga@NLO}. As described in the text, the CKKW-like minimal distance $d(k_i)$ is used both as an order parameter and to identify the parent parton for each photon.}

\label{ALG:clustering}
\end{table}
After being grouped, the momenta $\{P_J\}$ of the clusters and the $\{K_i\}$ of the $m$ photons are $mapped$, in order to have a consistent $2\to 2+m\gamma$ kinematics that respects both on-shellness of external legs and momentum conservation. The algorithm is designed to retain the {\em angles} between clusters, so that kinematical invariants built out of clusters momenta are
conserved up to term of orders $k^2_\perp/E_J^2$, $k_\perp$ being the transverse momentum of the photon relative to the $J$ particle
with which it is clustered. We show the necessary steps to construct a reduced kinematics in Tab.~\ref{ALG:mapping}.
\begin{table}[htbp]
\centering
\begin{tabularx}{\textwidth}{X}
\toprule
\textbf{Mapping algorithm.}
Enforce on-shell and momentum conservation on clustered momenta.\\
\midrule
\vspace{-5mm}
\setlength{\abovedisplayskip}{4pt}
\setlength{\belowdisplayskip}{4pt}
\setlength{\abovedisplayshortskip}{2pt}
\setlength{\belowdisplayshortskip}{2pt}
  \begin{itemize}[itemsep=8pt,leftmargin=0pt]
    \item[] \textbf{Input} \begin{tabular}[t]{@{}l l@{}}
    & $\{K_I\}_{I=1}^m$ momenta of the \textit{hard} photons \\
    & $\{P_J\}_{J=1}^4$ momenta of the \textit{clusters}
  \end{tabular}
    \begin{itemize}[itemsep=8pt]
    \item[] \textbf{Boost} all momenta to the $(P_1+P_2)$ rest frame
    \[
    P_J \to \bar P_J\,\,\,(\bar P_J^2\neq m_J^2), \qquad K_I\to \bar K_I
    \]
    \item[] \textbf{Impose} on-shell conditions
 \[
\bar P_J \to \tilde P_J =
\left\{
\begin{array}{@{}l@{\qquad}l@{}}
\left(\bar{E}_J,\sqrt{E_J^2-m_J^2}\,\,\vec{u}_{\bar P_J}\right)
& J=1,2 \quad \text{initial state} \\[8pt]
\left(\sqrt{m_J^2+\bigl|\vec{\bar P}_J\bigr|^2},\,\vec{\bar P}_J\right)
& J=3,4 \quad \text{final state}
\end{array}
\right.
\]
where $\vec{u}_{\bar P_J}$ is the unit vector in the direction of $\vec{\bar{P}}_J$
    
    \item[] \textbf{Rescale} final-state charged particles $J=3,4$ and photons $I=1,\dots,m$
    \[
      \tilde P_J \to \tilde p_J =
\left(\sqrt{m_J^2+\bigl|y\,\vec{\bar{ P}}_J\bigr|^2},\,y\,\vec{\bar{P}}_J\right),
      \qquad
      \bar K_I \to \tilde k_I = y\,\bar K_I
    \]
    \item[] \textbf{Determine} the scaling factor $y$ by solving
    \[
      (\bar P_1+\bar P_2)^2 =
      \left(\tilde p_3+\tilde p_4+\sum_{j=1}^m \tilde k_j\right)^2
    \]
    \end{itemize}
\item[] \textbf{Output}  \begin{tabular}[t]{@{}l l@{}}
    & $\{\tilde k\}$ mapped momenta of the \textit{hard} photons \\
    & $\{\tilde p\}=\{\tilde p_{1,2}\equiv \tilde{P}_{1,2},\tilde{p}_{3,4}\}$ mapped momenta of the charged legs
  \end{tabular}
\vspace{5mm}
  \end{itemize}\\
\bottomrule
\end{tabularx}
\caption{Pseudo-algorithm describing the implementation of the phase space
mapping in the latest version of \textsc{BabaYaga@NLO}. The building of clusters momenta is described in table \ref{ALG:clustering}. }

\label{ALG:mapping}
\end{table}

\FloatBarrier

\newpage
\bibliographystyle{JHEP}
\bibliography{radiative}

\end{document}